\documentclass[12pt]{article}
\usepackage{amsmath}
\usepackage{graphicx,psfrag,epsf}
\usepackage{enumerate}
\usepackage{natbib}
\usepackage{url} % not crucial - just used below for the URL 

\newcommand{\blind}{0}

\usepackage{graphicx}
\usepackage{subfig}
\usepackage[pdflinkmargin=5pt,pdfstartview={FitBH -32768},plainpages=false,colorlinks=true,citecolor=blue]{hyperref}
\usepackage{multirow}
\usepackage{bbm}
\usepackage{nicefrac}
\usepackage{amsmath}
\usepackage{float}
\usepackage{xcolor}
\usepackage[normalem]{ulem}
\usepackage{amsmath,dsfont}
\usepackage{amssymb,amsmath,amsthm}
\renewcommand{\P}{\mathbb{P}}

\newcommand{\E}{\mathbb{E}}
\newcommand{\N}{\mathbb{N}}
\newcommand{\R}{\mathbb{R}}
\newcommand{\1}{\mathbbm{1}}
\newcommand{\KLEINO}{{\scriptstyle{\mathcal{O}}}}
\DeclareMathAccent{\verywidehat}{\mathord}{largesymbols}{'144}
\newcommand{\var}{\mathbb{V}\hspace*{-0.05cm}\textnormal{a\hspace*{0.02cm}r}}
\newcommand{\Cov}{\mathbb{C}\textnormal{O\hspace*{0.02cm}V}}
\newcommand{\cov}{\mathbb{C}\textnormal{o\hspace*{0.02cm}v}}

\DeclareMathOperator{\AVAR}{\mathbf{AVAR}}

\allowdisplaybreaks[3]
\newtheorem{remark}{Remark}[section]
\newtheorem{theo}{Theorem}[section]

\newtheorem{prop}{Proposition}[section]
\newtheorem{lem}{Lemma}

\newtheorem{cor}[prop]{Corollary}

\begin{document}

\def\spacingset#1{\renewcommand{\baselinestretch}%
{#1}\small\normalsize} \spacingset{1}

%%%%%%%%%%%%%%%%%%%%%%%%%%%%%%%%%%%%%%%%%%%%%%%%%%%%%%%%%%%%%%%%%%%%%%%%%%%%%%

\if0\blind
{
  \title{\bf Modeling and Forecasting Realized Volatility with Multivariate Fractional Brownian Motion}
  \author{Markus Bibinger\\
    Faculty of Mathematics and Computer Science, University of W\"urzburg\\
    and \\
    Jun Yu \\
    Faculty of Business Administration, University of Macau\\
    and \\
    Chen Zhang \\
   Lingnan College, Sun Yat-sen University}
  \maketitle
} \fi

\if1\blind
{
  \bigskip
  \bigskip
  \bigskip
   \bigskip
    \bigskip
 % \title{\bf Modeling and Forecasting Realized Volatility with Multivariate Fractional Brownian Motion}
  \begin{center}
     {\Large \bf Modeling and Forecasting Realized Volatility with Multivariate Fractional Brownian Motion}
 \end{center}
  \medskip
} \fi

\bigskip
\begin{abstract}
A multivariate fractional Brownian motion (mfBm) with component-wise Hurst exponents is used to model and forecast realized volatility (RV). We investigate the interplay between correlation coefficients and Hurst exponents and propose a novel method to estimate model parameters, establishing its consistency and asymptotic normality. Additionally, we develop a time-reversibility test, which is typically not rejected by RV data. When the data generating process is a time-reversible mfBm, we derive optimal forecasting formulae and analyze their properties. A key insight is that an mfBm with different Hurst exponents and non-zero correlations can reduce forecasting errors compared to a one-dimensional model. Consistent with this theory, out-of-sample forecasts using the time-reversible mfBm show improvements over univariate fBm, particularly when the estimated Hurst exponents differ significantly. Empirical results demonstrate that mfBm outperforms HAR and its variants in terms of out-of-sample forecast.
\end{abstract}

\noindent%
\vfill

\newpage
\spacingset{1.25} % DON'T change the spacing!
\section{Introduction\label{sec:1}}
Modeling and forecasting financial asset volatility play a crucial role in numerous areas of finance. In the univariate context, several stylized facts about volatility dynamics --- particularly long memory --- have been well established since the seminal work of \cite{ding1993}. Subsequent research has introduced multiple generations of long-memory volatility models and their approximations, including contributions by \cite{BAILLIE19963, comte1996, corsi2009}. \cite{andersen1997} further elucidated the mechanisms generating long memory in volatility. A recent advancement is the rough fractional stochastic volatility (RFSV) model of \cite{Gatheral-Jaisson-Rosenbaum-2018}, which employs fractional Brownian motion (fBm) with a Hurst exponent $H<0.5$. This model has been linked to long-memory ARFIMA processes \citep{shiyu2023, li2025, wangyu2022} and has demonstrated superior forecasting performance, as evidenced by \cite{Gatheral-Jaisson-Rosenbaum-2018} and \cite{forecast}.

It is well-documented that financial volatilities exhibit strong co-movement across assets and markets. This stylized fact has spurred extensive research on multivariate volatility modeling, including surveys of multivariate GARCH and stochastic volatility models \citep{Bauwen2006, Asai2006}. These frameworks effectively capture not only individual asset volatilities but also cross-asset correlations, which are often even more critical for applications such as portfolio optimization and risk diversification. At the same time, they enhance the accuracy of volatility forecasts for individual assets. In contrast, the existing literature on fractional models predominantly concentrates on univariate approaches, neglecting the correlation structure and probably resulting in inefficiencies.  

This paper proposes modeling and forecasting realized volatility (RV) using a multivariate fBm (mfBm) with component-wise Hurst exponents. The model extends the univariate fractional Brownian motion (fBm) framework, building on foundational work by \cite{mfBm} and \cite{coeurjolly2013wavelet}, which itself derives from the theoretical contributions of \cite{Lavancier-Philippe-Surgailis-2009, Lavancier-Philippe-Surgailis-2010}. It retains the appealing properties of univariate fBm --- which has seen growing application across various domains --- while addressing its limitations in capturing cross-sectional dynamics. The inclusion of component-specific Hurst exponents introduces significant mathematical complexity. This leads to a fundamental theoretical question that motivates our current investigation:
\begin{align}\label{Q}\notag&\text{Can two real-valued processes exhibit markedly different autocorrelation structures and}\\
&\text{ persistence patterns while maintaining strong contemporaneous correlation?}\tag{{\bf{Q}}}
\end{align}

\begin{figure}[t]
\begin{center}
	\includegraphics[width=16cm]{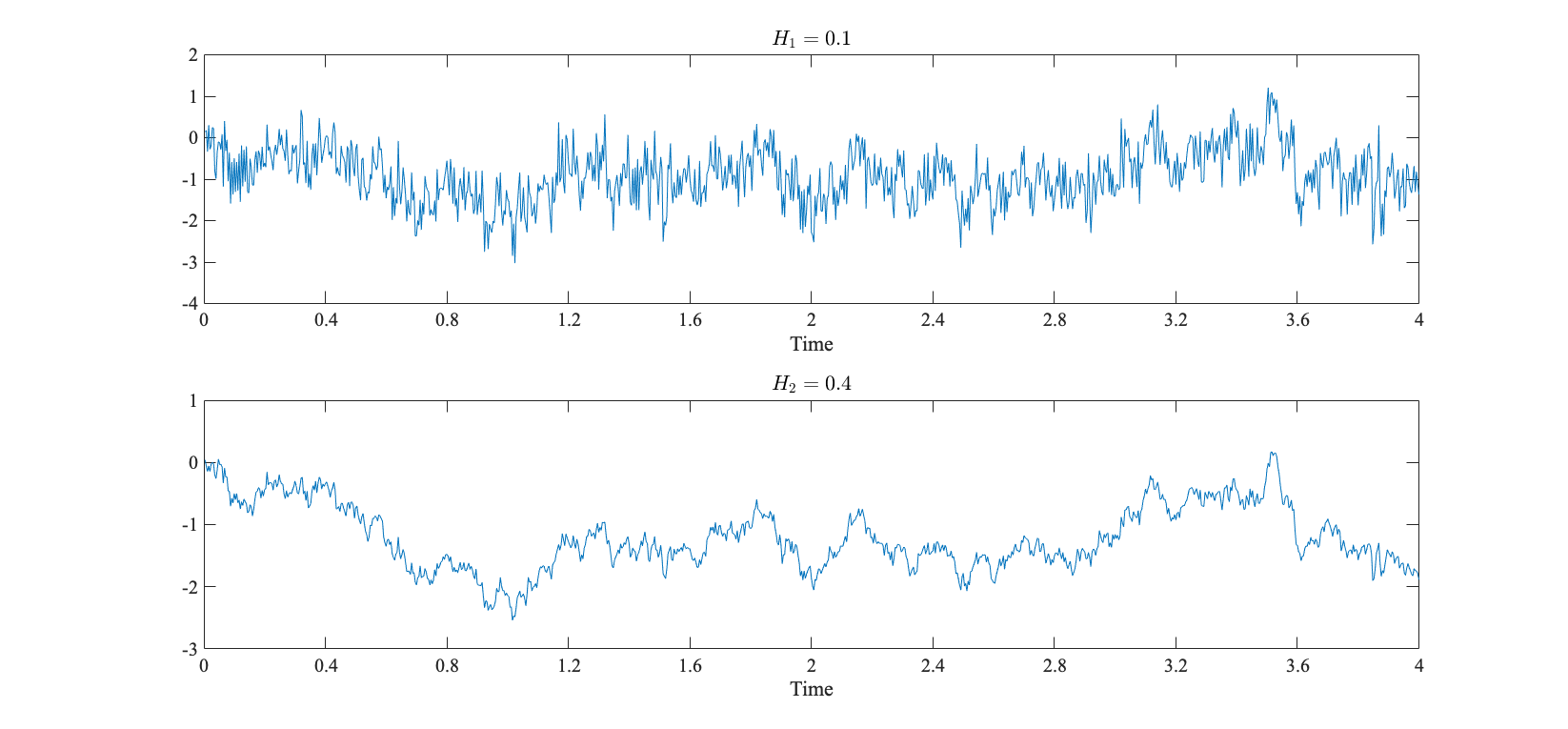}
	\vspace{-0.3cm}
	\caption{Simulated sample paths of fBm with $H_1=0.1$ (top), and $H_2=0.4$ (bottom), which are correlated with $\rho=0.8$ and $\eta_{1,2}=0$. The sample interval is $1/250$.}
\label{mfbm_sp3}
\end{center}
\end{figure}

The mfBm framework allows for distinct Hurst exponents across components while capturing their interdependencies through correlation parameters. Each component possesses its own scale parameter, and cross-component covariances incorporate both correlation and asymmetry parameters --- the latter capturing lead-lag effects between series. As illustrated in Figure \ref{mfbm_sp3}, these features provide preliminary evidence addressing question \eqref{Q}, with a comprehensive analysis developed in the following section. Given the well-documented strong correlation among RVs \citep{Ding2025}, we investigate the forecasting implications of multivariate rough fractional volatility modeling.

We propose a novel moment-based estimation method for all model parameters. The Hurst exponents and scale parameters are estimated component-wise, while correlation and asymmetry parameters are estimated pairwise. Under mild regularity conditions, we establish the consistency of these estimators. The new estimators are compared to the existing estimators proposed by \cite{amblard2011identification}. We demonstrate that our approach yields substantially improved estimation of asymmetry parameters while maintaining comparable performance for other parameters. Furthermore, we derive the asymptotic distribution of our estimators, enabling formal statistical inference. For forecasting applications, we focus on the time-reversible specification of mfBm (with zero asymmetry parameters), as our proposed hypothesis test fails to reject this null hypothesis for most assets at the 1\% significance level. 

Our theoretical analysis of time-reversible mfBm for RV forecasting yields several key insights. First, forecasting improvements emerge when components exhibit heterogeneous Hurst exponents. Second, stronger cross-asset correlations further enhance forecast accuracy. Contrary to traditional wisdom that correlation enhances forecasting accuracy, the benefits of time-reversible mfBm depend critically on its Hurst exponents. We implement this framework to model multivariate log RV series while producing univariate volatility forecasts. By partitioning the out-of-sample period into distinct windows, we identify significant efficiency gains from mfBm precisely when estimated Hurst exponents diverge across assets. Conversely, when Hurst exponents are similar, mfBm forecasts converge to their univariate fBm counterparts. These empirical findings confirm our theoretical predictions about the conditions under which multivariate modeling provides forecasting advantages.

We empirically examine whether incorporating additional information through multivariate modeling necessarily improves forecasting efficiency by comparing the heterogeneous autoregressive (HAR) and vector HAR models, where the univariate HAR serves as our benchmark. While theory suggests that correctly specified multivariate models should outperform their univariate counterparts by leveraging additional information, our analysis reveals two key findings: (1) contrary to theoretical expectations, the vector HAR model does not consistently deliver superior forecasting accuracy despite its richer information set; (2) when utilizing identical information sets, mfBm consistently generates more precise forecasts than HAR-type models. We further compare the mfBm with other variants of HAR models, including vector HAR models with a common factor \citep{Bauwens2023, Ding2025} and vector log-HAR models, and find that the mfBm remains superior. These results demonstrate that mfBm's mathematical structure provides distinct advantages in  extracting forecasting value from multivariate information.

The remainder of this paper is organized as follows. Section \ref{sec:2} introduces the model and explains its covariance structure and probabilistic properties. Section \ref{sec:3} develops a novel estimation method and its asymptotic theory and proposes a test for time-reversibility. Section \ref{sec:4} develops a formula for the optimal forecast with time-reversible mfBm. Section \ref{sec:6} conducts an empirical study where mfBm is fitted to log RV data and forecasting performance from alternative models is compared. Section \ref{sec:7} concludes. Appendix \ref{sec:est_amblard_coeurjolly} reviews the method of \cite{amblard2011identification} and the second-order-increment-based estimator, collects proofs of theoretical results, reports the Monte Carlo study and additional empirical results, and proposes an optimized GMM method for the correlation parameter. Throughout the paper, $\stackrel{p}{\to}$ denotes convergence in probability, $\stackrel{d}{\to}$ convergence in distribution, $\sim$ asymptotic equivalence for real-valued sequences, $\overset{d}{=}$ equality in distribution, and $\log$ the natural logarithm.

\section{Model, Correlation Structure and Properties\label{sec:2}}
We begin with a review of univariate fBm and its key properties. A univariate fBm $(B^H_t)_{t\in \R}$ is a continuous-time Gaussian process with almost surely continuous sample paths satisfying: (1) zero-mean: $\E[B^H_t]=0\; \forall t$; (2) covariance structure: $\cov(B^{H}_{t+h},B^{H}_t)=w(t,h,H)$ where
\begin{align}\label{covshort}w(t,h,H)=\frac12\big(|t+h|^{2H}+|t|^{2H}-|h|^{2H}\big)\, \mbox{ for } t,h\in\R.
\end{align}
The process exhibits two fundamental characteristics: (1) self-similarity: for any scaling factor $\lambda>0$, $\left(B^H_{\lambda t} \right)_{t\in \R} \overset{d}{=}\left(\lambda^{H} B^H_t\right)_{t\in \R}$; (2) stationary increments: for any $s,t\in \R$, $B^H_t-B^H_s\sim \mathcal{N}(0,|t-s|^{2H})$. The Hurst exponent
$H\in \left( 0,1\right)$ governs both the covariance structure and the self-similarity properties. Three important cases emerge. (1) When $H=0.5$, fBm reduces to standard Brownian motion. (2) When $H\neq 0.5$, fBm is neither a (semi-)martingale nor Markovian. (3) The sample paths exhibit H\"older continuity of order $\gamma$ for any $\gamma<H$. For discrete observations at intervals $\Delta>0$, the increment process $(\Delta_i B^H)_{1\le i \le n}$ defined by $\Delta_i B^H=B^H_{i\Delta}-B^H_{(i-1)\Delta}$ is a stationary Gaussian process whose autocovariance is
\begin{align}\notag
\cov\left( \Delta_i B^H ,\Delta_j B^H\right) & =\frac{\Delta^{2H}}{2}\left(
(i-j+1)^{2H}+(i-j-1)^{2H}-2(i-j)^{2H}\right) \text{, \,for any } i\geq j,  \label{acffgn}\\
& \sim \Delta^{2H} H(2H-1)(i-j)^{2H-2},\,\text{ for large } i-j.
\end{align}
Equation \eqref{acffgn} shows that for $H\in (0.5,1)$, $(\Delta_i B^H)_{1\le i\le n}$ is serially dependent with positive autocorrelations and its autocovariances are not absolutely summable. That is, it has a long memory for $H>0.5$. In contrast, if $H\in (0,0.5)$, $(\Delta_i B^H)_{1\le i \le n}$ has negative autocorrelations and 
\begin{equation*}
\sum\limits_{i-j=-\infty }^{+\infty }\cov\left( \Delta_i B^H ,\Delta_j B^H\right)  =0,
\end{equation*}%
such that the spectral density is zero at the origin. In this case, $(\Delta_i B^H)_{1\le i \le n}$ is an antipersistent process. 

The mfBm extends the univariate fBm to a multivariate setting such that its components are fBms (with possibly different Hurst exponents) but are possibly cross-correlated.\footnote{It should not be confused with \emph{multifractional} or \emph{mixed fractional Brownian motion}, which are univariate processes. The former has a time-varying Hurst function, while the latter is a linear combination of a Brownian motion and an independent fBm.} The mfBm is defined by \cite{mfBm} with its covariance introduced in a pairwise manner. For simplicity, we focus on a bivariate setting in the sequel. The bivariate fBm $(B_t)= \big((B^{(1)}_t, B^{(2)}_t)^{\top}\big)_{t\in \R}$ is constructed as a Gaussian process satisfying self-similarity:
\begin{align}\label{selfsim} \big((B^{(1)}_{\lambda t}, B^{(2)}_{\lambda t})^{\top})_{t\in \mathbbm{R}} \overset{d}{=}  \left(\big(\lambda^{H_1} B^{(1)}_{ t}, \lambda^{H_2}  B^{(2)}_{ t}\big)^{\top}\right)_{t\in \R}, ~\mbox{for}~\lambda>0\,,\end{align}
where $(H_1,H_2)^{\top}\in(0,1)^2$. Each component is an fBm with Hurst exponent $H_i\in(0,1)$ and scaling parameter $\sigma_i>0$. Their covariance function is
\[ \cov \left( B^{(i)}_{ t}, B^{(i)}_{ s} \right)=\frac{\sigma_i^2}{2}\left( \left\vert s\right\vert^{2H_i}+ \left\vert t\right\vert^{2H_i}-\left\vert s-t\right\vert^{2H_i} \right)=\sigma_i^2\,w(t,s-t,H_i),~i=1,2, \]
where $\sigma_i^2 =\var(B^{(i)}_1)$ and $w(\cdot,\cdot,\cdot)$ defined in \eqref{covshort}. The cross-covariances of $B^{(1)}_{ s}$ and $B^{(2)}_{ t}$ given in \cite[Prop.\ 3]{mfBm} are more complex and divided into two cases: the general case with $H_1+H_2\neq 1$ and the specific case with $H_1+H_2= 1$.\begin{enumerate}
\item If $H_1+H_2\neq 1$, there exists $(\rho,\eta_{1,2})\in [-1, 1]\times  \R$, such that
\begin{align}
 \cov \left(B^{(1)}_{ s}, B^{(2)}_{ t} \right) =&\frac{\sigma_1 \sigma_2}{2} \left( (\rho+\eta_{1,2} \,\text{sign}(s)) \left\vert s\right\vert^{H_1+H_2}+(\rho-\eta_{1,2}\,\text{sign}(t)) \left\vert t\right\vert^{H_1+H_2}\right. \notag\\
 & \left. -(\rho-\eta_{1,2}\,\text{sign}(t-s)) \left\vert t-s \right\vert^{H_1+H_2} \right)\,.
 \end{align}
\item  If $H_1+H_2= 1$, there exists $(\tilde{\rho},\tilde{\eta}_{1,2})\in [-1, 1]\times  \R$, such that
\[ \cov \left(B^{(1)}_{ s}, B^{(2)}_{ t} \right) =\frac{\sigma_1 \sigma_2}{2} \big(\tilde{\rho}\,(\left\vert s \right\vert+\left\vert t \right\vert- \left\vert s-t\right\vert)+\tilde{\eta}_{1,2}(t\log  \left\vert t \right\vert - s\log  \left\vert s \right\vert - (t-s)\log \left\vert t-s\right \vert )\big)\,.\]
\end{enumerate}
%\enlargethispage*{.25cm}
As the notation suggests, $\rho=\operatorname{corr}(B^{(1)}_{1},B^{(2)}_{1})$ is a correlation parameter. The second parameter carries a subscript to reflect its asymmetry: $\eta_{1,2}=-\eta_{2,1}$ and $\tilde{\eta}_{1,2}=-\tilde{\eta}_{2,1}$. This asymmetry arises when swapping the roles of the two components, i.e., transitioning from $\cov(B^{(1)}_{ s}, B^{(2)}_{ t} )$ to $\cov (B^{(1)}_{ t}, B^{(2)}_{ s})$. These asymmetry parameters are defined by 
\begin{align*}\eta_{1,2}&=\frac{\E[B^{(1)}_{1}B^{(2)}_{-1}]-\E[B^{(2)}_{1}B^{(1)}_{-1}]}{\sigma_1\sigma_2(2-2^{H_1+H_2})},~\mbox{if}~H_1+H_2\neq 1\,,\\
\tilde\eta_{1,2}&=\frac{\E[B^{(1)}_{1}B^{(2)}_{-1}]-\E[B^{(2)}_{1}B^{(1)}_{-1}]}{\sigma_1\sigma_2 2\log(2)},~\mbox{if}~H_1+H_2=1\,.
\end{align*}

A key observation is that the cross-covariance function depends on the Hurst exponents only through their sum. While the general covariance structure is complex and potentially difficult to interpret, an important special case is the symmetric scenario where $B_t\stackrel{d}{=}B_{-t}$, which is particularly relevant for financial applications. This defines the bivariate time-reversible fBm. In this simplified case, when asymmetry parameters vanish, the spectral density becomes real, and the covariance turns symmetric. That is, $\cov(B^{(1)}_{s}, B^{(2)}_{t} )=\cov (B^{(1)}_{t}, B^{(2)}_{s})$. This is the process we consider in the forecasting application. It has the following cross-covariance function
\begin{equation} \cov\big(B^{(1)}_{s}, B^{(2)}_{t}\big)=\frac{\rho\sigma_1 \sigma_2 }{2} \left( \left\vert s\right\vert^{2H}+ \left\vert t \right\vert^{2H}- \left\vert t-s\right\vert^{2H} \right)= \rho\sigma_1 \sigma_2\,w(t,s-t,H)\,, \label{mcov}
\end{equation}
where $H = (H_1 + H_2)/2$ denotes the cross Hurst exponent. This process possesses a covariance structure where the cross-covariances mirror the form of univariate fBm autocovariances, incorporating both a correlation coefficient $\rho$ and the cross Hurst exponent $H$. Importantly, the value of $H$ determines the nature of interdependence. $H<1/2$ results in a short-range interdependent process, while $H>1/2$ leads to long-range interdependency.

A multivariate process with asymmetric covariances --- distinct from the well-known asymmetry in return distributions and Zumbach effects --- can induce lead-lag effects between components.  Specifically, one component influences another with a delayed reaction. Although this phenomenon is relevant in certain applications \citep[p. 147]{wackernagel2003multivariate}, it appears less critical for volatility time series. For RV, estimated asymmetry parameters are typically close to zero. In Section \ref{sec:3}, we develop a hypothesis test for time-reversibility. Empirical results show that the null hypothesis cannot be rejected in most cases. Consequently, we focus on the time-reversible mfBm in our forecasting exercise.

\begin{figure}[t]
%\begin{framed}
%\hspace*{-.35cm}
\centering
\includegraphics[width=10cm]{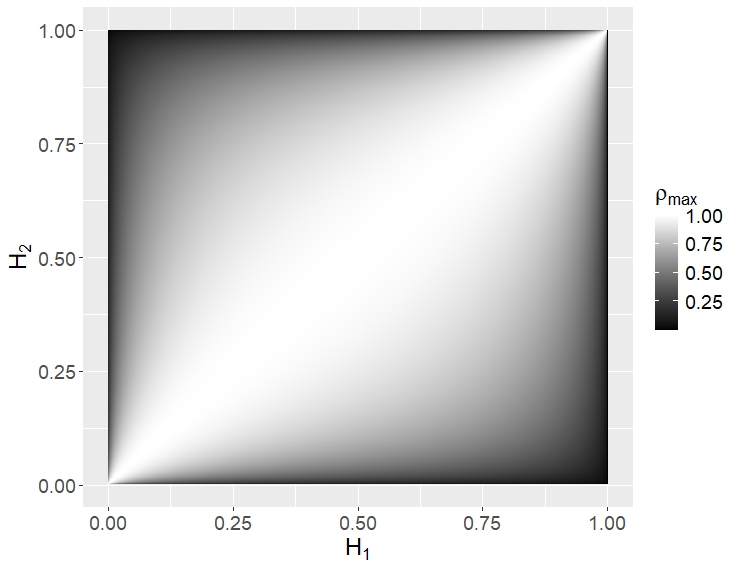}
%\end{framed}
\caption{\label{Fig:rhomax}{Maximal possible (absolute) correlation $\rho_{max}$ of the bivariate time-reversible fBm with Hurst exponents $H_1$ and $H_2$.}}
\end{figure}

To ensure the covariance kernel of mfBm remains positive semi-definite, certain parameter restrictions must hold. In this section, we examine these constraints for the bivariate time-reversible fBm case. The more general setting, which incorporates asymmetry, leads to greater complexity and less intuitive parameter coherence; we refer to \cite[Fig.\ 1]{mfBm} for illustrative examples. In the bivariate case, the correlation is bounded above by a maximum value
\begin{align*}\rho_{max}(H_1,H_2)=\frac{\sqrt{\sin(\pi H_1) \sin(\pi H_2)\Gamma(2H_1+1)\Gamma(2H_2+1)}}{\sin(\pi H)\Gamma(2H+1)}.\end{align*}
For correlations $\rho$, with $|\rho|\le \rho_{max}$, time-reversible fBm exists; see \cite[page 10]{mfBm}. Due to the symmetry between positive and negative correlations, i.e., $\rho_{min}=-\rho_{max}$, we only display the maximal absolute correlation $\rho_{max}$ in Figure \ref{Fig:rhomax}.

The plot resembles a heatmap, showing that smaller differences $|H_1-H_2|$ permit higher correlations between components. The range within which $\rho$ can vary also becomes larger when $H_1,H_2,H$ are closer to 1/2. In the special case where $H=H_1=H_2$ --- which we term unifractional --- there are no restrictions on $\rho\in[-1,1]$. Here, the covariance structure simplifies, as the same Hurst exponent $H$ governs all autocovariances, both marginals and cross-component. Only in this scenario can mfBm be represented as a linear transformation of a vector of independent fBms, a property we exploit in certain proofs. 

However, when Hurst exponents differ significantly, the constraints tighten. For example, if $H_1=0.2$, $H_2=0.8$, the maximal correlation drops to $\rho_{max}(0.2,0.8)\approx 0.662$. This answers our initial question \eqref{Q}: while not all combinations are feasible, the mfBm framework still accommodates such cases. Remarkably, even when the components exhibit starkly contrasting properties --- one being rough and short-range dependent ($H_1=0.2$), the other smooth and long-range dependent ($H_1=0.8$) --- a substantial correlation (up to 
$\sim 2/3$) is achievable. For more extreme disparities, such as $H_1=0.1$ and $H_2=0.9$, the maximum correlation further declines to $\rho_{max}(0.1,0.9)\approx 0.383$. Sample paths illustrating such a behavior, with $H_1=0.1$, $H_2=0.4$, $\rho=0.8$, are shown in Figure \ref{mfbm_sp3}.

\begin{figure}[t]
%\begin{framed}
\centering
\includegraphics[width=5cm]{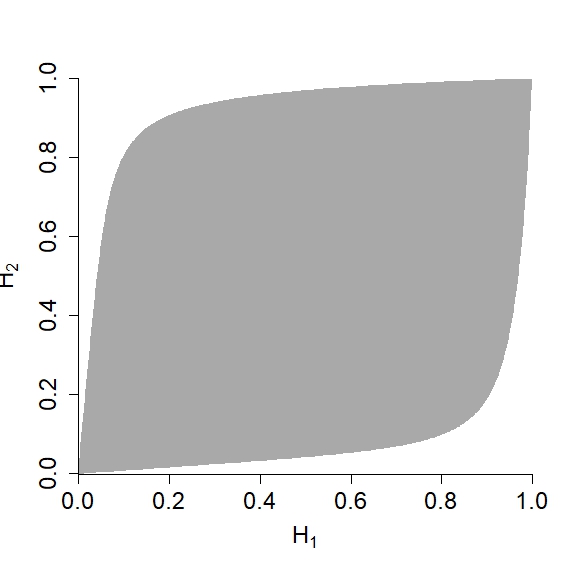}\hspace{0.5cm}\includegraphics[width=5cm]{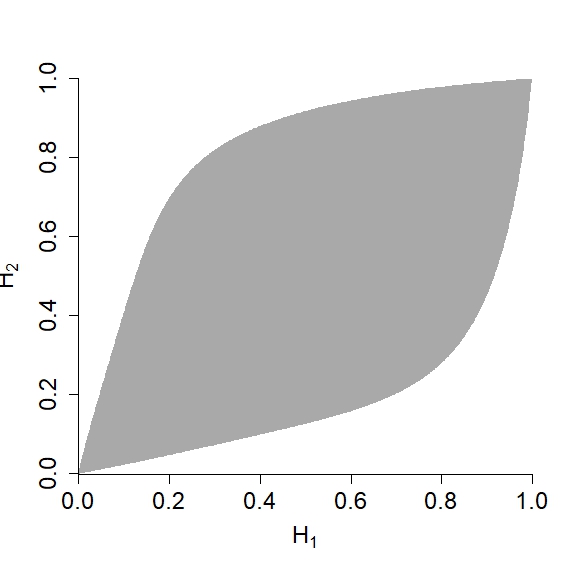}\hspace{0.5cm}\includegraphics[width=5cm]{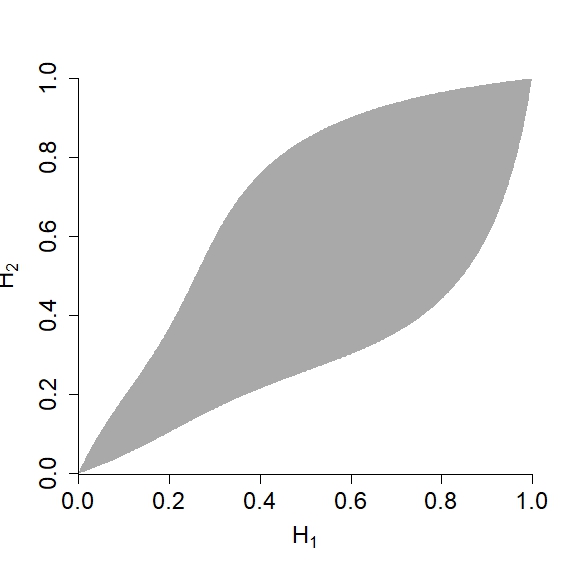}
%\end{framed}
\caption{\label{Fig:Exist}Parameter values for which bivariate time-reversible fBm exists (gray), and combinations that are not possible (white). The left-hand side is for $\rho=0.5$, the middle for $\rho=0.75$, and the right-hand side for $\rho=0.9$.}
\end{figure}

In Figure \ref{Fig:Exist}, we examine three fixed correlation values: $\rho=0.5$, $\rho=0.75$, and $\rho=0.9$, shown in separate panels. The gray region in each panel represents the set of Hurst exponent pairs $(H_1,H_2)$, for which bivariate time-reversible fBm exists. These admissible regions exhibit symmetry about the diagonal, a property that holds for all $\rho\in[-1,1]$. Notably, the diagonal itself always lies entirely within the admissible parameter space. The left panel demonstrates that $\rho=0.5$ is achievable for nearly all $(H_1,H_2)$ combinations, with exclusion limited to cases where $|H_1-H_2|$ approaches 1. The middle panel shows that $\rho=0.75$ remains feasible for most parameter pairs, though the admissible region begins to narrow. This restriction becomes particularly evident in the right panel ($\rho=0.9$), where the gray area contracts significantly, indicating stronger constraints on permissible Hurst exponent combinations.
%\begin{remark}
The discussed minimal conditions on the parameters are not only required for a mfBm, but necessary for a well-defined covariance (i.e., positive semi-definite). For any such parametrization, the existence of the process can be established through a spectral integral representation \citep[Sec.\ 3]{mfBm}.
%\end{remark}

\section{Statistics for mfBm \label{sec:3}}
\subsection{Observation model\label{sec:3a}}
Fractional time series models have demonstrated strong performance in volatility forecasting applications, as evidenced by \citep{andersen2003,halbleib} and related literature. Recent work, building on the framework established by \cite{Gatheral-Jaisson-Rosenbaum-2018}, has expanded this line of research to include fractional continuous-time models with Hurst exponents $H<1/2$. These approaches, grounded in fBm, have shown particular promise for modeling and forecasting log RV.

Traditionally, fBm and its increment process --- fractional Gaussian noise --- were primarily employed to model long memory phenomena. Given the well-documented presence of long memory in volatility, \cite{comte1998} proposed using fBm with Hurst exponent $H>1/2$ capturing this persistence property. However, recent empirical studies have consistently found that when modeling volatility using fBm, estimated Hurst exponents typically fall below $1/2$ \citep{Gatheral-Jaisson-Rosenbaum-2018, Wang-Xiao-Yu-2023-JoE, Bolko-Christensen-Pakkanen-Veliyev-2023-GMM, Bibinger-Sonntag-2023}. This is evident even from simple autocorrelation plots of the data, which typically show negative autocorrelations in the increments. Therefore, the rough fractional stochastic volatility model from \cite{Gatheral-Jaisson-Rosenbaum-2018} has become increasingly popular and has been utilized already for options pricing \citep{Bayer-Friz-Gatheral-2016, Garnier-Solna-2018}, %Livieri-2018, 
variance swaps \citep{Bayer-Friz-Gatheral-2016}, portfolio choice \citep{Fouque-Hu-2018}, designing trading strategies \citep{Glasserman-He-2020}, and dynamic hedging \citep{Euch-Rosenbaum-2018}. 

Akin to the approach of fractional time series in \cite{andersen2003}, \cite{Wang-Xiao-Yu-2023-JoE} employs discrete observations from fBm (or generalized fractional processes) to model log RV --- an observable quantity. In contrast, recent theoretical work has focused on latent spot volatility derived from discrete price recordings, providing evidence of roughness in spot volatility \citep{Chong2025, Bolko-Christensen-Pakkanen-Veliyev-2023-GMM}. Since our objective, like in \cite{andersen2003,Wang-Xiao-Yu-2023-JoE}, is to forecast RV, we adopt a modeling framework for this observable measure. Specifically, we treat available time series as discrete observations of fractional processes, where $B^{(i)}_{t}$ is the log RV of asset $i$ at time $t$. This approach ensures that any empirical findings we establish pertain strictly to RV, not spot volatility.

Consistent with the existing literature that employs univariate fBm for modeling and forecasting log RV, we adopt a model based on equi-spaced high-frequency observations over a fixed time interval, such as $[0,1]$. While the univariate framework provides valuable insights, extending the statistical inference methods to the multivariate case offers greater practical utility. We therefore consider observations specified by
\[\Big(B_{j\Delta}\Big)= \Big(B^{(1)}_{j\Delta}, B^{(2)}_{j\Delta}, ..., B^{(d)}_{j\Delta}\Big)^{\top},~0\le j\le n,~\text{with}~\Delta=1/n\,,\]
where $d$ denotes the dimension of $B_{j\Delta}$ and asymptotic results are established under the asymptotic scheme where $\Delta\to 0$, $n\to\infty$.\footnote{The same high-frequency asymptotics have been employed in the existing literature \citep{Gatheral-Jaisson-Rosenbaum-2018, Wang-Xiao-Yu-2023-JoE}. Our theory and proofs leverage self-similarity and hence, extend readily to the fixed frequency settings (i.e., $T\to\infty$ with $\Delta$ fixed where $\Delta=T/n$). However, this long-span asymptotic scheme does not universally apply to univariate fractional models with a drift term.}

\subsection{Our estimation method\label{sec:3c}}
Like the first and only available parameter estimation method for mfBm \citep{amblard2011identification}, our new estimation method is also based on moments and hence, is method-of-moments (MM). The estimators of $H_j$ and $\sigma^2_j$ are based on the marginal distributions of mfBm, while correlation and asymmetry parameters are estimated pairwise. Recall the notation for increments $(\Delta_k B^{(j)})_{1\le k \le n}$, defined by $\Delta_k B^{(j)}=B^{(j)}_{k\Delta}-B^{(j)}_{(k-1)\Delta}$, which we use in the sequel with the components $B^{(j)},1\le j\le d$.

For a simple notation, consider estimation of a bivariate fBm with components $B^{(1)}$, and $B^{(2)}$, and with Hurst exponents $H_1,H_2$, correlation $\rho$, variance parameters $\sigma_1^2,\sigma_2^2$, and asymmetry parameter $\eta_{1,2}$. %Hence, we do not directly restrict to the time-reversible case. 
The Hurst exponent is component-wise and is estimated by
\begin{align}\label{Hest}\hat H_j=\frac{1}{2\log(2)}\log\bigg(\frac{\sum_{k=1}^{{{n-1}}}\big(\Delta_{k,2} B^{(j)}\big)^2}{\sum_{k=1}^{{{n}} }\big(\Delta_k B^{(j)}\big)^2}\bigg),~j=1,2\,,\end{align}
where $\Delta_{k,2} B^{(j)}={{B^{(j)}_{(k+1)\Delta}-B^{(j)}_{(k-1)\Delta}}}$ for $j=1,2$ stands for lag-2-increments. Then, the variance parameters can be estimated component-wise by
\begin{align}\label{sigmaest}\hat \sigma_j^2=\frac{\sum_{k=1}^n\big(\Delta_k B^{(j)}\big)^2}{n\Delta^{2\hat H_j}},~j=1,2\,.\end{align}

To estimate $\rho$ and $\eta_{1,2}$, as in \cite{amblard2011identification}, we restrict to the case that $H_1+H_2\ne 1$, which is relevant for modeling RV, as prior studies (e.g., \citet{Gatheral-Jaisson-Rosenbaum-2018, Wang-Xiao-Yu-2023-JoE}) and our own estimates of roughness (reported in subsequent sections) find the estimated $H_j$ below 0.5. Our proposed estimators of $\rho$ and $\eta_{1,2}$ are
\begin{subequations}
\begin{align}\label{rhohat}
\hat\rho&=\frac{\sum_{k=1}^n\Delta_k B^{(1)} \Delta_k B^{(2)}}{\sqrt{\sum_{k=1}^n\big(\Delta_k B^{(1)}\big)^2 \sum_{k=1}^n\big(\Delta_k B^{(2)}\big)^2}}\,,\\
\label{etahat}
\hat\eta_{1,2}&=\frac{\sum_{k=1}^{n-1}\Big(\Delta_{k+1} B^{(2)} \Delta_k B^{(1)}-\Delta_{k+1} B^{(1)} \Delta_k B^{(2)}\Big)}{\sqrt{\sum_{k=1}^{n-1} %\big(B^{(1)}_{(k+2)\Delta}-B^{(1)}_{k\Delta}\big)^2 \sum_{k=1}^{n-2} \big(B^{(2)}_{(k+2)\Delta}-B^{(2)}_{k\Delta}\big)^2}
\big(\Delta_{k,2} B^{(1)}\big)^2\sum_{k=1}^{n-1}\big(\Delta_{k,2} B^{(2)}\big)^2}-2\sqrt{\sum_{k=1}^n\big(\Delta_k B^{(1)}\big)^2 \sum_{k=1}^n\big(\Delta_k B^{(2)}\big)^2}}\,.
\end{align}
\end{subequations}
The estimators constitute proper statistics that depend solely on increments. For $\hat{\rho}$, the identical normalizing factors $\Delta^{H_1+H_2}$ required for convergence in both the numerator and the denominator cancel out. For $\hat\eta_{1,2}$, the covariance structure introduces an additional factor $2^{H_1+H_2}-2$ in the numerator's expectation. This explains the inclusion of lag-2-increments in the denominator --- their specific form ensures cancellation of this factor. The self-normalization of $\hat{\rho}$ and $\hat\eta_{1,2}$ turn out to be important for ensuring good finite-sample performance. Both estimators are empirical counterparts of the theoretical identities that define the parameters and can be computed at no cost. 

In Appendix \ref{generalestimation} we propose an optimal GMM estimator of $\rho$. When comparing $\hat\rho$ with the optimal GMM estimator of $\rho$, we find that GMM yields slight efficiency gains over $\hat\rho$. In contemporaneous and independent work, \cite{dugo2024} propose a GMM approach to estimate all parameters of a multivariate fractional Ornstein–Uhlenbeck process. Due to computational difficulties, they implement non-optimal GMM. A fully general GMM approach for all parameters in mfBM is also computationally demanding and does not provide much statistical improvement to our MM method in the unreported experiments.

\subsection{Asymptotic properties of MM and test for time-reversibility\label{sec:3d}}

The proposed estimators (\ref{Hest}, \ref{sigmaest}, \ref{rhohat}, \ref{etahat}) are consistent for general bivariate fBm under mild regularity conditions. Tables~\ref{dj30_corr_7}  reports the estimated parameters from our dataset. The Hurst exponents are all below 0.5, the correlations of increments across components are approximately 0.3, and the asymmetry parameters are close to zero. The presence of nonzero correlations supports the adoption of a multivariate modeling framework, while the near-zero asymmetry parameters justify a focus on the time-reversible case. We derive the asymptotic distribution for the bivariate time-reversible fBm and develop a formal hypothesis test for \( H_0: \eta_{1,2} = 0 \). The test results provide strong statistical evidence in favor of the time-reversible modeling assumption. Our asymptotic theory  requires the additional restriction $H<3/4$, a condition satisfied by empirical estimates of roughness in RV, as discussed earlier.

%However, their asymptotic distributions may depend on the model specification, e.g. the estimator for the correlation parameter $\rho$ is sensitive to asymmetry. 

%while the component-wise estimators exhibit asymptotic variances that are invariant to asymmetry, 

%and statistically indistinguishable from zero given 
%supporting the adequacy of time-reversible mfBm specification.  This empirical evidence motivates our theoretical focus on the time-reversible case. 
 %Additionally, we develop a formal hypothesis test for $H_0: \eta_{1,2} = 0$ to statistically validate this modeling assumption.

%\textcolor{red}{Our asymptotic theory  require the additional restriction $H<3/4$, a condition satisfied by empirical estimates of realized volatility roughness, as discussed earlier.} Furthermore, Table \ref{dj30_eta_7} reveals that the estimated asymmetry parameters for our dataset are statistically indistinguishable from zero, supporting the adequacy of time-reversible mfBm specification. This empirical evidence motivates our theoretical focus on the time-reversible case. Additionally, we develop a formal hypothesis test for $H_0: \eta_{1,2} = 0$ to statistically validate this modeling assumption.

\begin{theo}\label{propstat}For a general bivariate fBm satisfying $\max(H_1,H_2)<3/4$, $H_1+H_2\ne 1$, the component-wise estimators obey the following asymptotic properties. For each $j$, as $n\rightarrow \infty$,  
\begin{equation}\label{cltHhat}\hat H_j\stackrel{p}{\rightarrow}H_j,\; \sqrt{n}\big(\hat H_j-H_j\big)\stackrel{d}{\rightarrow}\mathcal{N}\big(0,\text{AVAR}_{\hat H_j}\big)\,,
\end{equation}
where 
{\scriptsize
\begin{align*}
\text{AVAR}_{\hat H_j}&=\frac{1}{4\log^2(2)}\Big(4+\sum_{r=1}^{\infty}\Big(|r+1|^{2H_j}+|r-1|^{2H_j}-2|r|^{2H_j}\Big)^2+2^{-4H_j}\sum_{r=1}^{\infty}\Big(|r+2|^{2H_j}+|r-2|^{2H_j}-2|r|^{2H_j}\Big)^2\\
&-2^{1-2H_j}\sum_{r=1}^{\infty}\Big(|r+1|^{2H_j}+|r-2|^{2H_j}-|r|^{2H_j}-|r-1|^{2H_j}\Big)^2\Big)\,.
\end{align*}}
%$\text{AVAR}_{\hat H_j}$  is given by \eqref{cltHhat3} in Appendix.
%\begin{align*}
%\text{AVAR}_{\hat H_j}&=\frac{1}{4\log^2(2)}\Big(4+\sum_{r=1}^{\infty}\Big(|r+1|^{2H_j}+|r-1|^{2H_j}-2|r|^{2H_j}\Big)^2\\
%& +2^{-4H_j}\sum_{r=1}^{\infty}\Big(|r+2|^{2H_j}+|r-2|^{2H_j}-2|r|^{2H_j}\Big)^2\\
%&-2^{1-2H_j}\sum_{r=1}^{\infty}\Big(|r+1|^{2H_j}+|r-2|^{2H_j}-|r|^{2H_j}-|r-1|^{2H_j}\Big)^2\Big)\,.
%\end{align*}
Under the same conditions, for each $j$, as $n\rightarrow \infty$,
\begin{equation}\label{cltsigmaest}\hat \sigma_j^2\stackrel{p}{\rightarrow}\sigma_j^2,\; \frac{\sqrt{n}}{\log(n)}\big(\hat \sigma_j^2-\sigma_j^2\big)\stackrel{d}{\rightarrow}\mathcal{N}\big(0,\text{AVAR}_{\hat H_j}\cdot 4\sigma_j^4\big)\,.
\end{equation}
\end{theo}

\begin{remark}
The component-wise estimators $\hat{H}_j$ and $\hat{\sigma}_j$, for $j = 1, 2$, are unaffected by asymmetry. Their estimation results remain invariant regardless of whether the underlying model is a general bivariate fBm or a bivariate time-reversible fBm.
\end{remark}

Estimator $\hat{H}_j$ achieves consistency with the optimal parametric convergence rate of $\sqrt{n}$ when $H_j<3/4$. This condition mirrors the requirement for limit theory of normalized realized variances in univariate fBm (see, e.g., \cite{nourdin2010central}). While general asymptotic theory for univariate fBm estimators based on first-order increments exists (e.g., \cite{Coeurjolly-2001}), we provide a complete proof for \eqref{cltHhat} to derive explicit variance formulas for $\hat{H}_j$. 

A second-order-increment-based estimator is given in \eqref{rhohatso} in Appendix \ref{generalestimation}. It satisfies a central limit theorem for all $H\in(0,1)$. Statistics based on second-order increments have been studied in the literature (e.g., \cite{begyn2007functional}, \cite{kubilius}, \cite{Wang-Xiao-Yu-2023-JoE}, \cite{Bibinger-Sonntag-2023}). However, our estimator \eqref{Hest} offers two key advantages: (1) a simpler asymptotic variance structure, and (2) uniformly smaller variance across $H\in(0,3/4)$ compared to second-order-increment-based estimators. These properties make \eqref{Hest} particularly suitable for rough volatility series with small Hurst exponents. The convergence rate in \eqref{cltsigmaest}, including its logarithmic factor, is also optimal \citep[Section 4.1]{brouste2018local}. Notably, the relationship between asymptotic variances in \eqref{cltHhat} and \eqref{cltsigmaest} parallels that of the second-order-increment-based estimator reported in Theorem 4.1 of \cite{Wang-Xiao-Yu-2023-JoE}.

\begin{theo}\label{theostat2}
For a general bivariate fBm with $\max(H_1,H_2)<3/4$, $H_1+H_2\ne 1$, as $n\rightarrow \infty$, we have 
\[\hat\eta_{1,2}\stackrel{p}{\rightarrow}\eta_{1,2},\; \hat\eta_{1,2}-\eta_{1,2}=\mathcal{O}_{\P}\big(n^{-1/2}\big).\]
In the time-reversible case, it satisfies a central limit theorem with an asymptotic variance $\text{AVAR}_{\hat \eta_{1,2}}(H_1,H_2,\rho)$ given by \eqref{avareta} in Appendix \ref{sec:proof_theorem}.
\end{theo}

Based on the developed asymptotic theory for $\hat \eta_{1,2}$, in the following corollary, we propose a test for time reversibility.

\begin{cor}\label{etatest}Given a bivariate fBm with $\max(H_1,H_2)<3/4$, $H_1+H_2\ne 1$, the test rejects the null hypothesis $\eta_{1,2}=0$, in favor of the alternative hypothesis that $\eta_{1,2}\ne 0$, when
\begin{align} \sqrt{n} \frac{\big|\hat \eta_{1,2}\big|}{\sqrt{\text{AVAR}_{\hat \eta_{1,2}}(\hat H_1,\hat H_2,\hat \rho)}}>\Phi^{-1}(1-\alpha/2)\,,\end{align}
where $\alpha\in(0,1)$, and $\Phi$ is the cumulative distribution function of the standard normal. This test attains the asymptotic size $\alpha$ and the asymptotic power $1$.
\end{cor}
%For discretely observed data, let $\Delta X (m)$\ denote the increment of the discretely observed mfBm, that is, 
 %\[\Delta X (m) = X (m\Delta) -X\left((m-1)\Delta\right)=\left(\Delta X_1 (m),...,\Delta X_p (m) \right).\] 
 %The  covariance for the increments, $(\Delta X_1(m), \Delta X_2(n))$, of observations take the form of 
%\begin{align}
 %Cov\left(\Delta X_i (m), \Delta X_j (n)\right)=\Sigma_{ij} \Delta^{H_i}\Delta^{H_j}\gamma_{H_i,H_j}\left(|m-n|\right), 
 %\end{align}
%where
%\[ \gamma_{H_i,H_j}\left(k\right) = \frac{1}{2}\left(|k+1|^{H_i+H_j}+|k-1|^{H_i+H_j}-2|k|^{H_i+H_j}\right). \]

\begin{remark}
There are several advantages in our estimators relative to those proposed by \cite{amblard2011identification}. First and foremost, our estimators have closed-form expressions, facilitating the computation. Second, our estimators have simpler asymptotic variance structures that are more straightforward to estimate. Third, our asymmetry parameter estimator shows superior performance compared to that of \cite{amblard2011identification}, as demonstrated in Table \ref{sec:finite_per_c}. This improvement is notable given \cite{amblard2011identification}'s own observation that estimating $\eta_{1,2}$ was ``very difficult to estimate, at least with the method adopted here''. Our approach successfully overcomes this challenge.
\end{remark}

\begin{theo}\label{theostat}
For a bivariate time-reversible fBm with $\max(H_1,H_2)<3/4$, $|\rho|<1$, $H_1+H_2\ne 1$, as $n\rightarrow \infty$, we have 
\[n^{-1}\Delta^{-H_1-H_2}\sum_{k=1}^n\Delta_k B^{(1)}\cdot \Delta_k B^{(2)}\stackrel{p}{\rightarrow} \varsigma=\rho\sigma_1\sigma_2,\]
\begin{equation}\label{cltcov}\sqrt{n}\Big(n^{-1}\Delta^{-H_1-H_2}\sum_{k=1}^n\Delta_k B^{(1)}\cdot \Delta_k B^{(2)}-\varsigma\Big)\stackrel{d}{\rightarrow}\mathcal{N}\Big(0,\sigma_1^2\sigma_2^2\big(1+\rho^2+\rho^2\upsilon_1+\upsilon_2\big)\Big)\,,
\end{equation}
where
{\scriptsize
\begin{subequations}
\begin{align}
\upsilon_1&=\upsilon_1(H_1,H_2)=\frac12 \sum_{r=1}^{\infty}\big(|r+1|^{H_1+H_2}+|r-1|^{H_1+H_2}-2|r|^{H_1+H_2}\big)^2,\\
\upsilon_2&=\frac12 \sum_{r=1}^{\infty}\big(|r+1|^{2H_1}+|r-1|^{2H_1}-2|r|^{2H_1}\big)\big(|r+1|^{2H_2}+|r-1|^{2H_2}-2|r|^{2H_2}\big)\,.
\end{align}
\end{subequations}}
Under the same set of assumptions, 
the estimator $\hat\rho$ defined in \eqref{rhohat} satisfies, 
\begin{align}\label{cltrho}\hat\rho\stackrel{p}{\rightarrow}\rho,\; \sqrt{n}\big(\hat\rho-\rho\big)\stackrel{d}{\rightarrow}\mathcal{N}\Big(0,\text{AVAR}_{\hat \rho}\Big)\,,\end{align}
where
{\scriptsize
\begin{eqnarray*}
  &\text{AVAR}_{\hat \rho} = (1-\rho^2)^2+\rho^2\Big((1+\rho^2)\upsilon_1(H_1,H_2)+\frac{\upsilon_1(H_1,H_1)}{2}+\frac{\upsilon_1(H_2,H_2)}{2}-\upsilon_3(H_1,H_2) -\upsilon_3(H_2,H_1)\Big)+\upsilon_2,
\end{eqnarray*}
\[\upsilon_3(H_1,H_2)=\sum_{r=1}^{\infty}\big(|r+1|^{2H_1}+|r-1|^{2H_1}-2|r|^{2H_1}\big)\big(|r+1|^{H_1+H_2}+|r-1|^{H_1+H_2}-2|r|^{H_1+H_2}\big).\]}
\end{theo}

\begin{remark}
$\hat\rho-\rho=\mathcal{O}_{\P}\big(n^{-1/2}\big)$ can be generalized to general bivariate fBm. A non-zero asymmetry parameter will just influence the asymptotic variance, which we compute for the time-reversible case of interest.
\end{remark}

\begin{remark}
When we set all series coefficients $(\upsilon_1,\upsilon_2,\upsilon_3)$ in Theorem \ref{theostat} to zero, the resulting asymptotic variances correspond exactly to those of a multivariate i.i.d.\ model --- or equivalently, a multivariate Brownian motion. Specifically: (1) for the empirical covariance in \eqref{cltcov}, the asymptotic variance reduces to $(1+\rho^2)\sigma_1^2\sigma_2^2$; (2) for the empirical correlation in \eqref{cltrho}, the asymptotic variance becomes $(1-\rho^2)^2$.
\end{remark}

We examine the finite-sample performance of the proposed estimators through Monte Carlo studies. The proposed method performs well and the empirical standard deviation is accurately predicted by the asymptotic theory. Our estimator for $\rho$ shows performance comparable to that of \cite{amblard2011identification}, while our estimator for $\eta_{1,2}$ demonstrates substantially improved accuracy, particularly when $\eta_{1,2} = 0$.  Additionally, consistent with theoretical expectations, the power of the test for time-reversibility grows monotonically as the true $\eta_{1,2}$ value deviates further from zero and gets larger for a larger sample size. More results can be found in Appendix \ref{sec:monte_carlo}.

\section{Optimal Forecast\label{sec:4}}
Given the empirical evidence showing only weak asymmetry in multivariate volatility dynamics, we recommend employing time-reversible mfBm for modeling log realized volatility. In what follows, we examine the forecasting advantages offered by this modeling approach.

The conditional expectation represents the optimal forecast in that it minimizes the mean squared forecast error (MSFE). In practice, almost always discrete observations over multiple periods are available. Consequently, the optimal forecast of the mfBm is obtained through conditional expectation based on the multivariate normal distribution\footnote{When continuous records are available, there exist formulas for generating the optimal forecast. For example, in an idealized continuous-time setting where a univariate fBm is observed over the infinite past $(-\infty,t]$, \cite[Eq.\ (34)]{Nuzman_Poor_2000} derive the conditional expectation for the process at future time $t + h$. However, this result cannot be directly applied in practice due to two key limitations: (1) real-world observations are available only at discrete time points, and (2) historical data spans a finite, rather than infinite, time window. To address these constraints, \cite{Gatheral-Jaisson-Rosenbaum-2018} propose necessary modifications to the original formula. As comprehensively reviewed in \cite{forecast}, the method proposed by \cite{Gatheral-Jaisson-Rosenbaum-2018} performs worse than the conditional expectation based on the multivariate normal distribution.}.  Consider one has equispaced discrete observations $\mathcal{X}_{t,d}=\left\{ B_{\Delta},\ldots, B_{t\Delta}\right\} $ from a $d$-dimensional mfBm $B=(B^{(1)},\ldots,B^{(d)})^{\top}$, where $\Delta$ is the sampling interval and $t$ is the number of periods where the mfBm is observed. In the volatility literature, the length of a year is conventionally normalized to 1. Hence, for $\Delta =1/12,1/52,1/252$, $\mathcal{X}_{t,d}$ represents monthly, weekly, and daily observations accordingly. We stack the data $\mathcal{X}_{t,d}$ into a column vector, such that 
\[\mathcal{X}_{t,d}=(B^{(1)}_{\Delta},\ldots,B^{(d)}_{\Delta},\ldots,B^{(1)}_{t\Delta},\ldots,B^{(d)}_{t\Delta})^{\top }\,.\] 
The optimal $h$-step-ahead forecast of $B^{(j)}_{(t+h) \Delta }$, for any $j\in\{1,\ldots,d\}$, is again the conditional mean and has a closed-form expression of the form
\begin{align}
\hat{B}^{(j)}_{(t+h) \Delta |(1:t)\Delta} = \E\Big[{B}^{(j)}_{(t+h) \Delta}  |\mathcal{X}_{t,d}\Big]=\big( \gamma_{t,h }^{j}\big)^{\top }\Sigma _{t,d}^{-1} \,\mathcal{X}_{t,d}\,,\label{opfore}
\end{align}
where $\Sigma _{t,d}$ denotes the covariance matrix of $\mathcal{X}_{t,d}$, whose elements are readily obtained from \eqref{mcov} and $\gamma _{t, h}^{j}$ is a vector that contains the covariance of ${B}^{(j)}_{(t+h)\Delta}$ with each entry of $\mathcal{X}_{t,d}$, which are also readily obtained from \eqref{mcov}.

Since the explicit formula (\ref{opfore}) is generally complex, the forecasting advantages of this modeling approach over the univariate framework are not immediately apparent. To develop key insights into forecasting within the mfBm model, we begin with simplified scenarios—specifically, the optimal forecast based on a single observation: (1) single-period forecasts in bivariate settings, and (2) their multivariate extensions, for which explicit expressions for the weights in (\ref{opfore}) are available. The general case, involving multiple observations of discretized paths, will be treated later. This approach is inspired by known results for univariate fBm, where optimal forecasts assign maximal weight to the most recent observation and nearly all weight to only a few recent data points—a pattern demonstrated in \cite{forecast}.

\subsection{Optimal forecast with observations in one period}
\label{sec:4.1}

In this subsection, we first assume that a single observation in a bivariate setting is the only available data. For simplicity, we ignore the sampling frequency $\Delta$ and denote the observation by $(B_t^{(1)},B_t^{(2)})$. Recall that $H=(H_1+H_2)/2$.
 
\begin{prop}\label{forecastprop}
Consider a bivariate time-reversible fBm $(B_t^{(1)}, B_t^{(2)})^{\top}$. The optimal $h$-step-ahead forecast of $B_{t+h}^{(1)}$, conditional on $B_t^{(1)},B_t^{(2)}$, is given by
\begin{align}\label{forecast}\hat B_{t+h|t}^{(1)}=w^{11}_{t+h|t}\cdot B_t^{(1)}+w^{12}_{t+h|t}\cdot B_t^{(2)}\,,\end{align}
for any $\rho \in (-1,1)$, with weight functions
%{\scriptsize
\begin{equation*}
w^{11}_{t+h|t}=\frac{1}{1-\rho^2}\Big(\frac{w(t,h,H_1)}{t^{2H_1}}-\rho^2 \frac{ w(t,h,H)}{t^{2H}}\Big),\;
w^{12}_{t+h|t}=\frac{\rho}{1-\rho^2}\frac{\sigma_1}{\sigma_2}\Big(\frac{w(t,h,H)}{t^{2H_2}}-\frac{ w(t,h,H_1)}{t^{2H}}\Big)\,,
\end{equation*}
%}
where $w(t,h,H)$ is defined in \eqref{covshort}. Its MSFE is
%{\scriptsize
\begin{align*}\E\Big[\big(\hat B_{t+h|t}^{(1)}-B_{t+h}^{(1)}\big)^2\Big]&=\sigma_1^2(t+h)^{2H_1}-\frac{\sigma_1^2}{1-\rho^2}\frac{(w(t,h,H_1))^2}{t^{2H_1}}\\
&\quad +\frac{\sigma_1^2\rho^2}{1-\rho^2}\bigg(\frac{2 w(t,h,H_1) w(t,h,H)}{t^{2H}}-\frac{(w(t,h,H))^2}{t^{2H_2}}\bigg)\,.\end{align*}
%}
\end{prop}

The roles of both components are symmetric and can be interchanged. We now examine important implications arising from Proposition \ref{forecastprop}.  Incorporating $B_t^{(2)}$ consistently improves forecasting accuracy by reducing the MSFE, or at worst maintains equivalent performance. This improvement becomes evident when we express the term $(1-\rho^2)^{-1}$ as $1+\rho^2/(1-\rho^2)$ and represent the reduction in forecasting error relative to the univariate case as a squared quantity. Regarding the role of each parameter, the MSFE of $\hat B_{t+h|t}^{(1)}$ does not depend on $\sigma_2$; only the Hurst exponents and correlation coefficients are relevant. We highlight several key special cases related to these parameters. (1) If $H_1=H_2=H$, then $w^{12}_{t+h|t}=0$ for any $\rho$. Thus, in the unifractional case, the optimal forecast coincides with the optimal forecast conditional on $B_t^{(1)}$ only. That is, we do not attain any gains from observing the second component $B_t^{(2)}$. The MSFE is $\sigma_1^2((t+h)^{2H_1}-(w(t,h,H_1))^2t^{-2H_1})$, and $w^{11}_{t+h|t}=w(t,h,H_1)t^{-2H_1}$.  (2) If $\rho=0$, then $w^{12}_{t+h|t}=0$ for any $H_2$. Hence, the optimal forecast coincides with the optimal forecast conditional on $B_t^{(1)}$ only.  Contrary to traditional wisdom that correlation enhances forecasting accuracy, the benefits of time-reversible mfBm depend critically on its Hurst exponents. Additionally, for arbitrary parameter values, the weights satisfy the following limiting behavior as the forecast horizon approaches zero: $w^{11}_{t+h|t}\to 1$ and $w^{12}_{t+h|t}\to 0$.

\begin{figure}[t]
%\begin{framed}
\centering
\includegraphics[width=4.45cm]{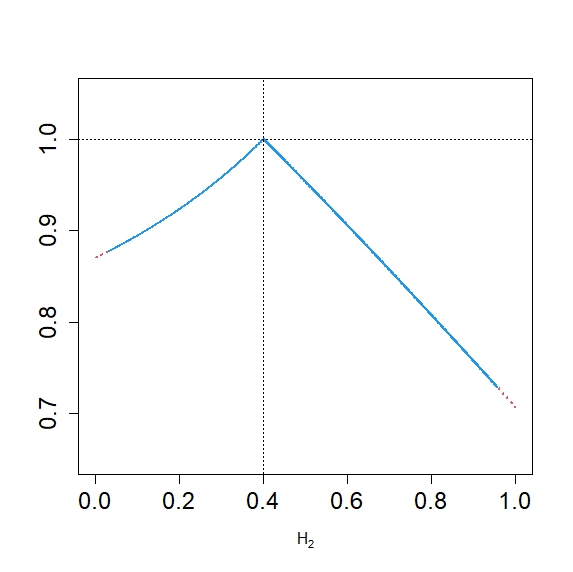}\hspace{.4cm} \includegraphics[width=4.45cm]{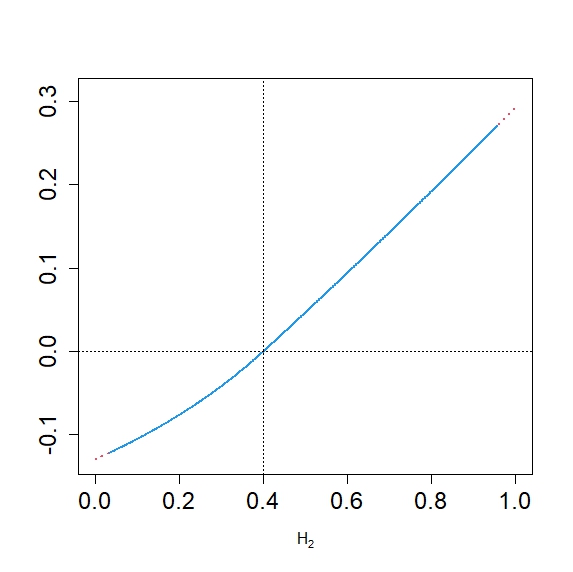} \hspace{.4cm}\includegraphics[width=4.45cm]{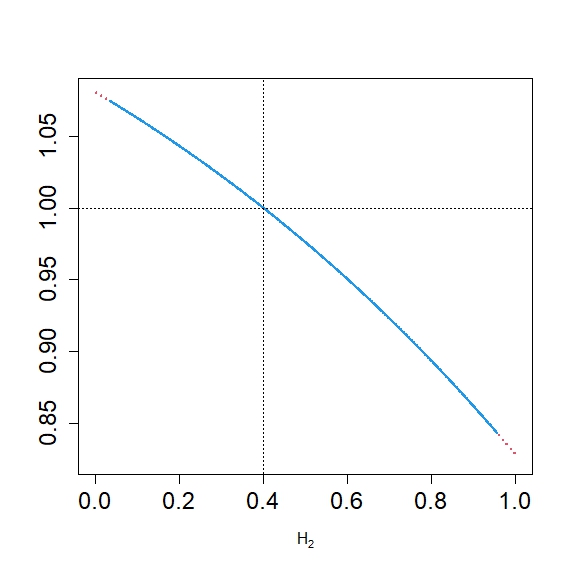}
%\end{framed}
\caption{\label{Fig:weights}Relative weights $w^{11}_{2|1}/(|w^{11}_{2|1}|+|w^{12}_{2|1}|)$ (left) and $w^{12}_{2|1}/(|w^{11}_{2|1}|+|w^{12}_{2|1}|)$ (middle) and $w^{11}_{2|1}t^{2H_1}/w(1,1,H_1)$ (right), for parameters $\rho=1/2$, $\sigma_1=\sigma_2=1$ and $H_1=0{.4}$ as functions of $H_2$.}
\end{figure}

Figure \ref{Fig:weights} illustrates the role of the Hurst exponents in determining the weights by fixing the moderate  and empirically relevant coefficients at $\rho = 1/2$ and  $H_1 = 0.4$, while allowing $H_2$ to vary. The other parameters are set to $\sigma_1 = \sigma_2 = 1$, $t = 1$, and $h = 1$. Although in this case the weight function is drawn over the whole support $H_2\in(0,1)$, some parameter configurations, especially when $H_2$ takes values close to the boundaries, are not possible. Over these sub-intervals the function is drawn with a dotted red line. The left panel presents the relative weight when the first component is forecasted, computed as $w^{11}_{2|1}/(|w^{11}_{2|1}|+|w^{12}_{2|1}|)$. The middle panel shows the relative weight on the other component, computed as $w^{12}_{2|1}/(|w^{11}_{2|1}|+|w^{12}_{2|1}|)$). The right panel plots $w^{11}_{2|1}t^{2H_1}/w(1,1,H_1)$, which is the ratio of the values of the first weight when $\rho=0.5$ and those when $\rho=0$, the latter coinciding with the factor of the univariate forecast based on observing $B_t^{(1)}$ only.

%Several observations emerge. In generally, the weights are not zero, which means forecasting with own past observation will lead to inefficiency. Hurst exponents have interesting implications on the weights. First, as our theory predicts, when $H_2=H_1=0.4$, all weight is on $B_t^{(1)}$ and $B_t^{(2)}$ is not used for the forecast even the  correlation coefficient, $\rho=1/2$, from these three panels. Second, the sign of weights depends on the relative magenitude of $H_2$ and $H_1$. the middle panels show that  the relative weight is a strictly increasing function, which equals zero at $H_2=0.4$, takes negative values when $H_2<0.4$, and approaches $0.3$ as $H_2$ nears $1$.  Third, the right panel shows that $w^{11}_{2|1}t^{2H_1}/w(1,1,H_1)$ is strictly decreasing, which means the univariate forecasting will from underweight its own past observation  to overweight as $H_2$ increases.

Several observations emerge. First, in general, the weights on the second component are nonzero, indicating that relying solely on the first component's own past observation leads to inefficiency in forecasting.  Second, as predicted by our theory, when $H_2 = H_1 = 0.4$, all the weight is placed on $B_t^{(1)}$, and $B_t^{(2)}$ is not used in the forecast, despite the moderate correlation coefficient $\rho = 1/2$, as shown in all three panels. Third, the sign of the weight depends on the relative magnitude of $H_2$ and $H_1$. The middle panel shows that the relative weight is a strictly increasing function: it equals zero when $H_2 = 0.4$, becomes negative when $H_2 < 0.4$, and approaches $0.3$ as $H_2$ nears $1$. Fourth, the right panel shows that $w^{11}_{2|1} t^{2H_1} / w(1,1,H_1)$ is strictly decreasing, implying that the univariate forecast shifts from underweighting to overweighting its own past observation as $H_2$ increases.

%Second, the left panel shows that for large differences between $H_1$ and $H_2$, the relative weight on the past observation of the component, which we aim to forecast, falls below $3/4$. The function is not symmetric with respect to the vertical dashed line. The middle panel show that the relative weight is a strictly increasing function, which equals zero at $H_2=0.4$, takes negative values when $H_2<0.4$, and approaches $0.3$ as $H_2$ nears $1$, demonstrating significant influence of the second component in these cases. Third, the right panel shows that $w^{11}_{2|1}t^{2H_1}/w(1,1,H_1)$ is strictly decreasing.  

%The left panel in Figure \ref{Fig:MSE} shows the relative MSFE when forecasting \textcolor{red}{$B^{(1)}_{2}$} based on \textcolor{red}{$(B_{1}^{(1)},B_{1}^{(2)})$} relative to the univariate benchmark (using only  $B_{10}^{(1)}$), normalized such that the maximum value equals 1, correspond to the parameter setting in Figure \ref{Fig:weights}.  All cases with $H_1\ne H_2$ show efficiency gains (values below 1) relative to univariate forecasting, yield meaningful, though limited, improvements. \sout{The MSFE remains invariant under proportional rescaling of both $t$ and $h$.} If we increase $\rho$ to be $0.9$ in the right panel, Higher correlation coefficients enable greater error reduction. 

\begin{figure}[t]
%\begin{framed}
\centering
\includegraphics[width=5cm]{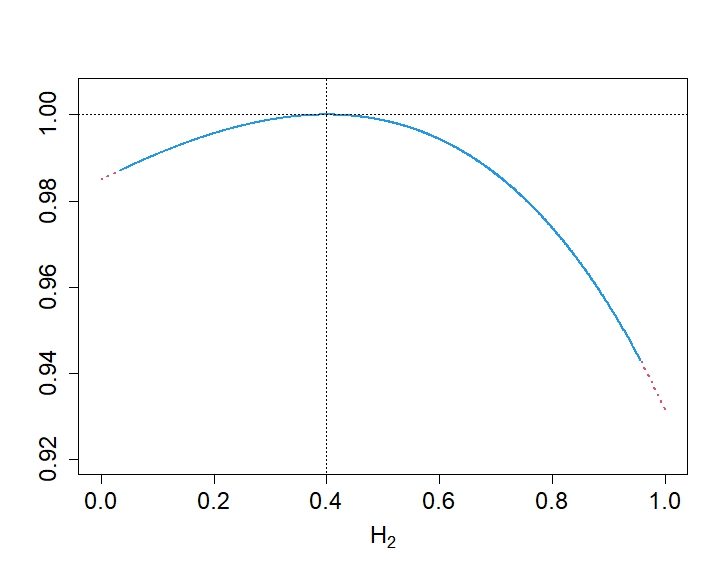}
\hspace{1.2cm}
\includegraphics[width=5cm]{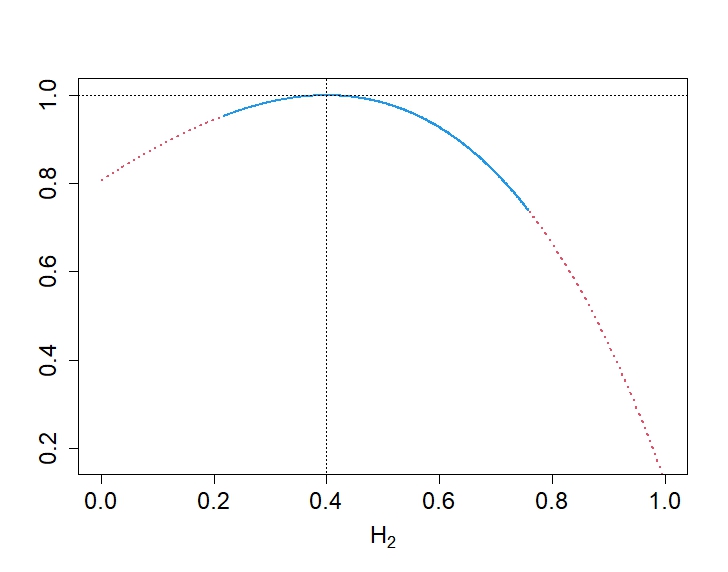}%\includegraphics[width=4.45cm]{RelMSE09015new.jpeg}
%\end{framed}
\caption{\label{Fig:MSE}Relative MSFEs as functions of $H_2$ for $t=1$, $h=1$, $\sigma_1=\sigma_2=1$, with $\rho=0.5$ and $H_1=0.4$ (left), with $\rho=0.9$ and $H_1=0.4$ (right).}
\end{figure}

The left panel of Figure~\ref{Fig:MSE} displays the relative MSFE for forecasting $B^{(1)}_{2}$ based on $(B_{1}^{(1)}, B_{1}^{(2)})$, relative to the univariate benchmark that uses only $B_{1}^{(1)}$, corresponding to the parameter settings in Figure~\ref{Fig:weights}. The values are hence normalized so that the maximum equals 1. In all cases where $H_1 \ne H_2$, we observe efficiency gains (values below 1) relative to the univariate forecast, with the improvements being  meaningful. If we increase the correlation coefficient to $\rho = 0.9$, as shown in the right panel, the forecast error reduction becomes more pronounced, highlighting the role of correlation in enhancing bivariate forecast.

%We fix $\sigma_1=\sigma_2=1$ in all cases, $\rho=0.5$ in the left panel, $\rho=0.9$ in the middle and right panels, and $H_1=0.4$ in the left and middle panel, $H_1=0.15$ in the right panel.   Three observations emerge. (1) All cases with $H_1\ne H_2$ show efficiency gains (values below 1) relative to univariate forecasting. (2) Higher correlation coefficients enable greater error reduction. (3) Even moderate correlations (left panel)cyield meaningful, though limited, improvements.

Propositions \ref{forecastprop2} and \ref{msedim} extend our bivariate results to the general $d$-dimensional setting. A key finding that persists in this broader context is the absence of forecasting gains from multivariate modeling when all Hurst exponents are equal.

\begin{prop}\label{forecastprop2}
Consider a time-reversible mfBm $(B_t^{(1)},\ldots, B_t^{(d)})^{\top}$, and denote $\Sigma=(\Sigma_{ij})_{1\le i,j\le d}\in\R^{d\times d}$ the positive definite covariance matrix of $(B_1^{(1)},\ldots, B_1^{(d)})^{\top}$, with inverse $\Sigma^{-1}=((\Sigma^{-1})_{ij})_{1\le i,j\le d}$. The optimal $h$-step-ahead forecast of $B_{t+h}^{(1)}$, given $B_t^{(1)},\ldots, B_t^{(d)}$, is given by
\begin{align}\label{forecastgen}\hat B_{t+h|t}^{(1)}&=\sum_{j=1}^d w^{1j}_{t+h|t}\cdot B_t^{(j)}\,,~\text{where}\\
\notag w^{1j}_{t+h|t}&=\sum_{k=1}^d\Sigma_{1k}(\Sigma^{-1})_{kj}\Big(\frac{w\big(t,h,(H_1+H_k)/2\big)}{t^{H_j+H_k}}\Big)~,1\le j\le d\,.
\end{align} 
\end{prop}

A key remaining question concerns how higher dimensions ($d > 2$) may further reduce forecasting errors when (1) correlations are non-zero and (2) Hurst exponents differ across components. The following specific result provides valuable insight through its tractable closed-form expressions, revealing how forecast accuracy improves with increasing dimensionality.

\begin{prop}\label{msedim}
Consider mfBm with dimension $d>2$, with $\sigma_j^2=1$ for all $j$, and all correlations equal to $\rho>0$. Assume $H_1$ is the Hurst exponent of $(B_{t}^{(1)})$, while $(B_{t}^{(j)})$ has Hurst exponent $H_j=H\ne H_1$, for all $2\le j\le d$.\footnote{We slightly abuse the notation here as $H$ is not the average of all Hurst exponents.} The optimal forecast for $B_{t+h}^{(1)}$, given $(B_{t}^{(1)},B_{t}^{(2)},\ldots, B_{t}^{(d)})^{\top}$, has the following expression for MSFE
%{\scriptsize
\begin{align*}\E\Big[\big(\hat B_{t+h|t}^{(1)}-B_{t+h}^{(1)}\big)^2\Big]&=(t+h)^{2H_1}-\frac{1}{1+(d-2)\rho-(d-1)\rho^2}\bigg((1+(d-2)\rho)\frac{(w(t,h,H_1))^2}{t^{2H_1}}\\
&\quad -2 (d-1)\rho^2\frac{w(t,h,H_1) w(t,h,H)}{t^{2H}}+(d-1)\rho^2\frac{(w(t,h,H))^2}{t^{2H_2}}\bigg)\,.\end{align*}
%}
In particular, the limit as $d\to\infty$ exists and remains strictly positive.
\end{prop}
\begin{figure}[t]
\centering
%\begin{framed}
\hspace*{.25cm}\includegraphics[width=7cm]{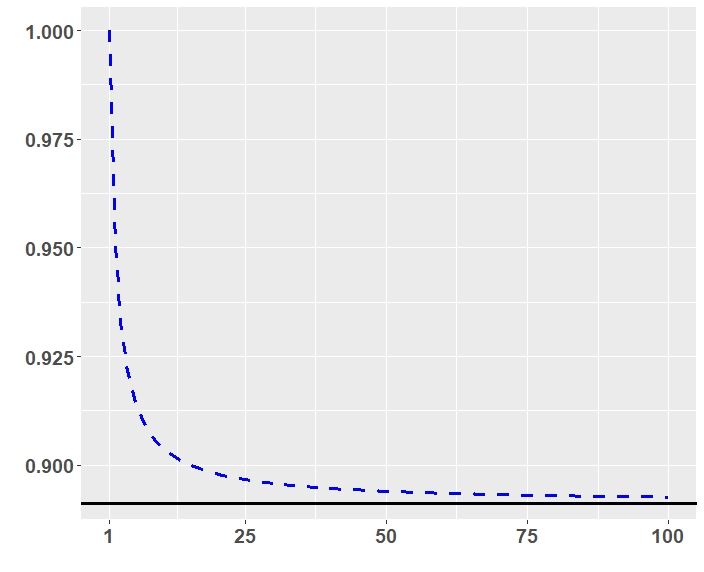}\hspace*{1cm}\includegraphics[width=7cm]{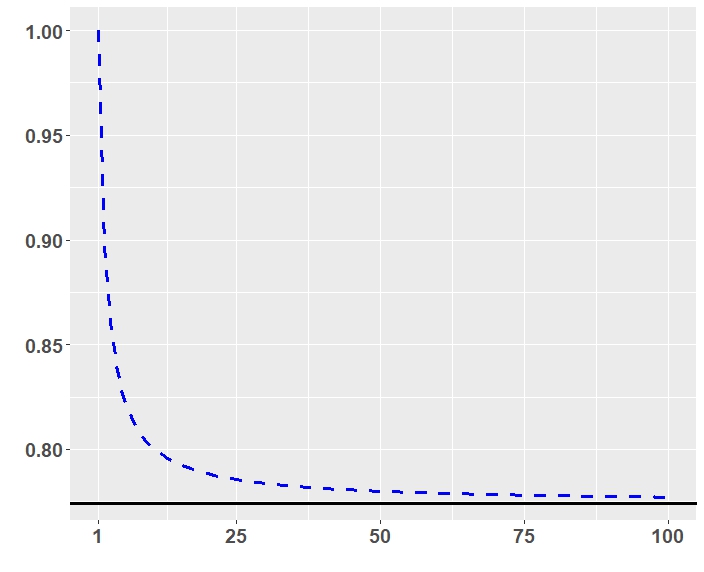}
%\end{framed}
\caption{\label{Fig:msedim}Ratios of MSFE in Prop.\ \ref{msedim}, with $\rho=0.8$, $H_1=0.4$, $H=0.1$, and that in the univariate model, for $t=1,h=1$ (left) and $t=10,h=1$ (right),  as a function of dimension $d$. The horizontal reference lines show the limits as $d\to\infty$.}
\end{figure}

Since the equal correlation condition requires $\rho\in(-(d-1)^{-1},1)$, we restrict our analysis to $\rho>0$ in Proposition \ref{msedim}, which is empirically relevant. Figure \ref{Fig:msedim} illustrates the results by plotting the relative MSFEs for $t=1$ and $t=10$, $h=1$, $\rho=0.8$, $H_1=0.1$ and $H=0.4$. The relative MSFE is normalized by the minimal univariate MSFE and evaluated across dimensions $d=1,\ldots,100$. The horizontal lines indicate the asymptotic limits as $d\to\infty$.

Compared to the bivariate case, incorporating additional components significantly reduces the forecast error. However, the risk does not vanish as $d$ increases. In fact, for the scenario in Figure \ref{Fig:msedim}, the MSFE improvement becomes almost negligible beyond $d\ge 20$. As $d\to\infty$ MSFE settles around 89\% of that of the minimal univariate case for $t=1$. Setting $t=10$ yields a smaller ratio of approximately 77.4\% (for much lager $t$, the ratio only decreases slightly). Thus, Proposition \ref{msedim} presents a mixed outcome for forecasting efficiency: while substantial gains are achievable by increasing dimensionality, these gains plateau early, and the forecasting risk remains bounded away from zero.

\subsection{Optimal forecast with all historical observations}
%\footnote{In a roll-window approach, we should note that $\Sigma _{t,d}$ depends on the variance of the first historical observations that is time varying. Moreover, the result in \ref{opfore} can be easily extended to cover non-equispaced discrete observations}
The conclusions on optimal forecasting drawn from the simple cases in Section \ref{sec:4.1} can be extended to the cases using all historical observations. Two additional propositions are presented in Appendix \ref{sec:prop_fore}: one for the bivariate time-reversible fBm in Proposition \ref{unifracprop} and the other for the $d$-dimensional case in Proposition \ref{unifracgen}. The corresponding Monte Carlo studies are reported in Appendix \ref{sec:monte_carlo}. In the Monte Carlo studies, we consider a realistic setting that explicitly accounts for estimation errors and find that mfBm continues to deliver more accurate forecasts when the correlation is nonzero and the Hurst exponents differ.

%Proposition \ref{unifracprop} establishes that, for a bivariate time-reversible fBm with identical Hurst exponents $H_1=H_2=H$, observing the historical path of the second component $B_t^{(2)}$ provides no additional forecasting benefit. This result extends naturally to the $d$-dimensional case in Proposition \ref{unifracgen}, provided all Hurst exponents coincide. The propositions are presented in Appendix~\ref{sec:prop_fore}, and the related Monte Carlo studies are reported in Appendix~\ref{sec:monte_carlo}.

\begin{table}[ht]                                                            
\centering            
\caption{Estimates of $H$, $\rho$  and $\eta$ for 5 Dow Jones 30 stocks between January 5, 2005 and January 14, 2025, with the asymptotic standard errors in parentheses.}           
\scalebox{0.5}[0.5]{                                                     
\begin{tabular}{l|c|ccccc|ccccc}            
\hline
\hline                                         
 Ticker& $\hat{H}$ & \multicolumn{5}{c}{Correlation estimates}& \multicolumn{5}{|c}{Asymmetry estimates}\\      
 \hline                 
AAPL &  0.2609 & 1 &       &       &      &    &0  &       &        &      \\
 &(0.0123)  & - &       &       &      &      & - &       &        &  \\
ALD  & 0.2050 & 0.3902 & 1 &       &      &    & 0.0950 &  0     &        &    \\
  & (0.0128) & (0.0131)  & -&       &      &     &  (0.0322)&   -     &        &          \\
AMGN & 0.1645 & 0.3074 & 0.3205 & 1 &       &     & 0.0123 & -0.0080 &      0  &       \\
 & (0.0131) & (0.0141) & (0.0142) & -&       &     & (0.0314) & (0.0292) &       - &       \\
AXP  & 0.2058 & 0.3721 & 0.4011 & 0.2893 & 1 &      & 0.0556 & -0.0593 & 0.0276 &   0   \\
  &  (0.0128)&(0.0133)  & (0.0132) & (0.0145)  & -&      & (0.0325) & (0.0298) & (0.0296)  &   -    \\
BA   & 0.2153 & 0.3602 & 0.3850 & 0.2696 & 0.3405& 1   & 0.0899 & -0.0937 & -0.0239 & -0.0336&0 \\
   & (0.0127) & (0.0134) & (0.0133)  &(0.0146)  &(0.0138)  &-     & (0.0332) &(0.0304)  & (0.0301) &(0.0310)&- \\
\hline  
\hline  
\end{tabular}}
\label{dj30_corr_7}                          
\end{table}

\section{Empirical Study\label{sec:6}}
Volatility modeling and forecasting have been subjects of significant interest for decades. Recent advances in continuous-time finance highlight the strong performance of rough fBm and rough fOU processes in volatility forecasting. In this study, we investigate the practical implications of using rough mfBm and its associated theoretical framework. Our analysis focuses on daily realized volatilities of the Dow Jones 30 (DJ30) constituents from January 5, 2005, to January 14, 2025—a span of 20 years—using data provided by Risk Lab.\footnote{See Dacheng Xiu's website: \href{https://dachxiu.chicagobooth.edu/\#risklab}{https://dachxiu.chicagobooth.edu/\#risklab}.}  Robustness check has been also considered for the Magnificent 7 stocks, which can be found in Appendix \ref{sec:m6}.

%%%%%%%%%%%%%%%%%%%%%%%%%%%%%%%%%%%%%%%%%%
\subsection{Estimation results}
We assume the true model is time-reversible mfBm and estimate parameters using the proposed MM approach.  For brevity, Table~\ref{dj30_corr_7} reports the point estimates of $H$, $\rho$, and $\eta_{1,2}$ with the asymptotic standard errors in parentheses, for the first five Dow Jones 30 stocks in alphabetical order.\footnote{Empirical results for other Dow Jones 30 stocks are reported in Tables \ref{dj30_corr30}-\ref{dj30_eta30} in Appendix \ref{sec:add_empirics}.}  From Table \ref{dj30_corr_7}, we can see that the null hypothesis $\eta_{1,2}=0$ cannot be rejected at the 1\% level in most cases, supporting a time-reversible mfBm specification.\footnote{This does not necessarily imply that the true (latent) volatility is time-reversible; see Appendix \ref{sec:error_on_eta_rho} for an analysis with estimation noises being considered.} Table \ref{dj30_corr_7}  also suggests that there are sizable differences among the estimated Hurst exponents and that the estimated correlation parameters are not close to zero.  According to our theory, mfBm should offer forecasting improvements over univariate fBm.\footnote{Since RV follows a log-normal distribution, adjustments for the $h\Delta $-period-ahead optimal forecast of RV are needed. For the univariate case, see \cite{bennedsen2017,forecast}.}

\subsection{Forecasting results}
For brevity, we focus on forecasting RV of AAPL,\footnote{We obtain similar results when log RV is forecasted, consistent with theoretical predictions.} examining whether incorporating additional cross-asset information enhances predictive accuracy. Using a two-year rolling window approach \citep{Gatheral-Jaisson-Rosenbaum-2018}, we compare forecasts derived from AAPL's historical data alone—as is standard in the literature—with those augmented by data from four additional stocks (ALD, AMGN, AXP, and BA), selected alphabetically. We evaluate five model specifications in terms of MSFE, progressively expanding the asset set: fBm (AAPL), bfBm (AAPL and ALD),  mfBm3 (AAPL, ALD and AMGN), mfBm4 (AAPL, ALD, AMGN and AXP) and  mfBm5 (AAPL, ALD, AMGN, AXP and BA). The forecast period is between January 3, 2007 and January 14, 2025.

The top panel in Table \ref{emp: mfbm_full} reports MSFEs and p-values of model confidence set (MCS) \citep{hansen2011} for five competing models for the full forecast period.\footnote{Implementation details can be found in Appendix \ref{sec:mcs}.}  A smaller MSFE or a larger MCS p-value indicates a better-performing model. It is clear that bfBm outperforms fBm and, in some cases mfBm processes outperform bfBm, consistent with the prediction of our theory.

%\begin{figure}[H]
%\caption{Rolling window estimates of $H$ for log realized volatilities}
%\label{fig:roll_h}\centering
%\subfloat[
%AAPL (red) and ALD (blue)]{\includegraphics[scale=0.28]{ALD.png}} 
%\hspace{0.6cm}
%\subfloat[
%AAPL (red) and AMGN (blue)]{\includegraphics[scale=0.28]{AMGN.png}} 
%\newline
%\subfloat[
%AAPL (red) and AXP (blue)]{\includegraphics[scale=0.28]{AXP.png}} 
%\hspace{0.6cm}
%\subfloat[
%AAPL (red) and BA (blue)]{\includegraphics[scale=0.28]{BA.png}} 
%\newline
%\end{figure}
%\vspace{-1cm}

\begin{figure}[t]
\centering
\makebox[\textwidth]{
	\includegraphics[width=15cm]{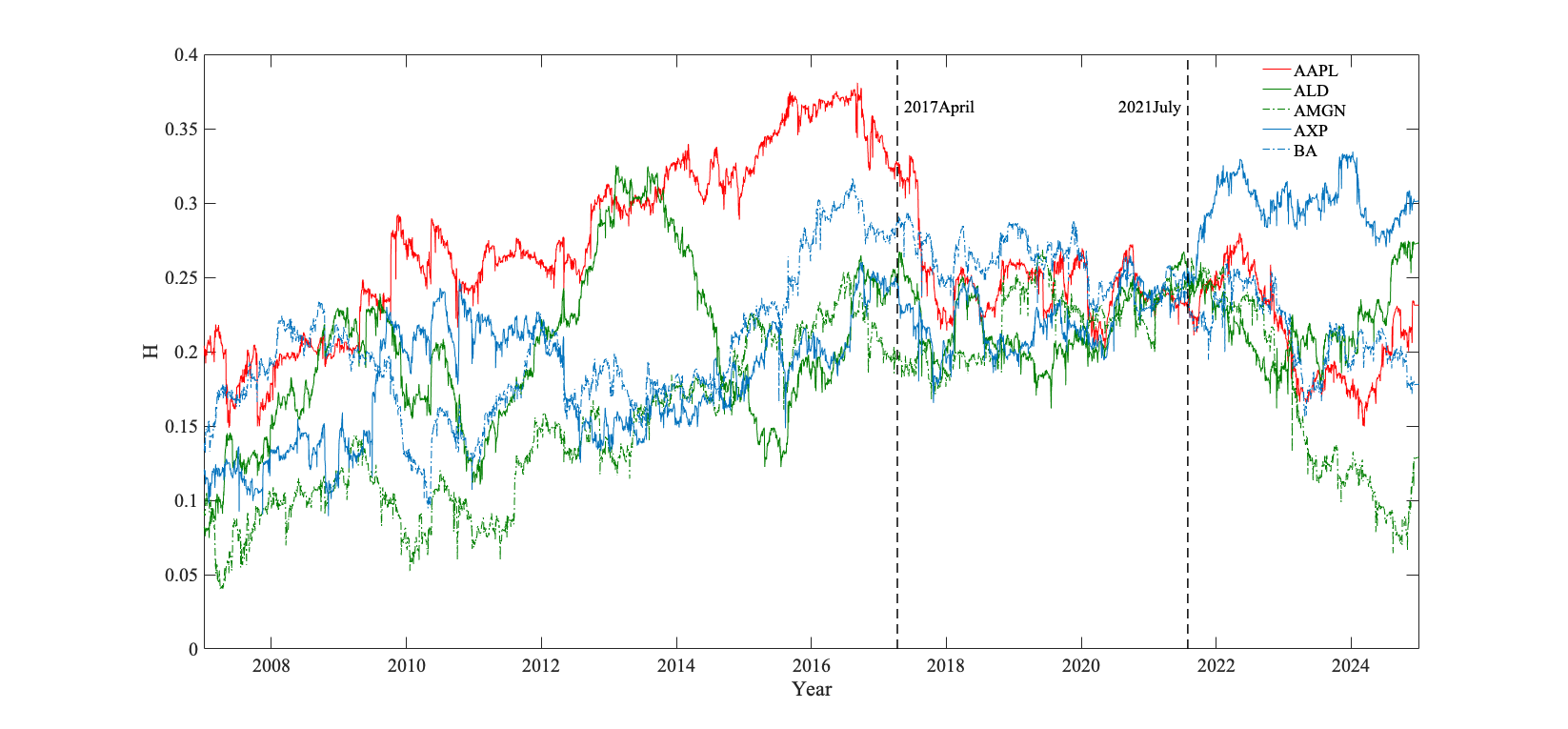}}
	\vspace{-0.3cm}
	\caption{Rolling window estimates of $H$ for log realized volatilities.}
\label{fig:roll_h}
\end{figure}

Figure \ref{fig:roll_h} displays the rolling window estimates of the Hurst exponents for each log realized volatility series. A key pattern emerges: over the period 2017–2021, the estimated Hurst exponents for AAPL and the other assets exhibit convergence, with differences shrinking toward zero. This finding motivates the consideration of two subperiods: January~3,~2007–April~11,~2017, and April~11,~2017–July~30,~2021. As the differences between the estimated Hurst exponents are large in the first forecast period, we expect substantial gains in using mfBm over fBm. By contrast, as the estimated Hurst exponents converge in the second forecast period, we expect no gains in using mfBm over fBm.

To validate our theory, we report MSFEs and MCS p-values in the middle and bottom panels of Table \ref{emp: mfbm_full} for the two forecast periods, respectively. In the first period, mfBm demonstrates superior forecasting performance compared to fBm. For example, for one-week-ahead forecasts, the improvement of mfBm5 over fBm increases from 1.2\% in the full sample to 1.8\% in the first subsample in terms of MSFE. For one-month-ahead forecasts, the improvement rises from 2.36\% to 3.31\%. Conversely, in the second period, the forecasts generated by mfBm and fBm become nearly identical. This empirical evidence strongly supports our analytical conclusions. The MCS p-values also favor the mfBm in both the full sample and the first subsample.

\begin{table}[H]
\centering
\caption{MSFEs of alternative models for $h$-day-ahead forecasts of RV , along with p-values from model confidence set (MCS) tests for each period,  during 3 forecast periods.}
\scalebox{0.6}{ 
\begin{tabular}{cccccccccccccccccc}
\hline \hline
&  \multicolumn{8}{c}{MSFE $\times$ 100}  & & \multicolumn{8}{c}{MCS p-value} \\
\hline
       & $h=1  $    &  $h=2 $     &  $h=3  $    &  $h=4 $     &  $h=5 $     &  $h=10 $    &  $h=15 $    &  $h=20  $  & & $h=1  $    &  $h=2 $     &  $h=3  $    &  $h=4 $     &  $h=5 $     &  $h=10 $    &  $h=15 $    &  $h=20  $ \\
                       \hline
\multicolumn{18}{c}{January 3, 2007 to January 14, 2025}      \\\hline
fBm   & 0.5003 & 0.6428 & 0.7296 & 0.7778 & 0.8523 & 1.0705 & 1.2456 & 1.3685&& 0.4016 & 0.3250 & 0.2032 & 0.2020 & 0.1000 & 0.0800 & 0.0810 & 0.0732\\
bfBm  & 0.4993 & 0.6405 & 0.7260 & 0.7730 & 0.8465 & 1.0611 & 1.2328 & 1.3532 &&0.6312 & 0.7382 & 0.4030 & 0.4882 & 0.2734 & 0.3828 & 0.3538 & 0.3592\\
mfBm3 & 0.4986 & 0.6397 & 0.7246 & 0.7716 & 0.8439 & 1.0592 & 1.2278 & 1.3471 &&0.6312 & 0.8352 & 0.4600 & 0.4882 & 0.2872 & 0.3876 & 0.3538 & 0.3592 \\
mfBm4 & 0.4984 & 0.6397 & 0.7235 & 0.7705 & 0.8419 & 1.0548 & 1.2206 & 1.3378 &&0.6312 & 0.8352 & 0.5318 & 0.4882 & 0.4700 & 0.4714 & 0.9082 & 0.8260\\
mfBm5 & 0.4979 & 0.6391 & 0.7230 & 0.7696 & 0.8415 & 1.0550 & 1.2195 & 1.3370&&1.0000 & 1.0000 & 1.0000 & 1.0000 & 1.0000 & 1.0000 & 1.0000 & 1.0000\\
\hline 
\multicolumn{18}{c}{January 3, 2007 to April 11, 2017}      \\\hline
fBm   & 0.6349 & 0.8028 & 0.8933 & 0.9288 & 1.0137 & 1.2629 & 1.4685 & 1.6195 &&0.2618 & 0.2862 & 0.2028 & 0.2160 & 0.1058 & 0.0908 & 0.0910 & 0.0788 \\
bfBm  & 0.6328 & 0.7985 & 0.8869 & 0.9202 & 1.0036 & 1.2474 & 1.4479 & 1.5957 && 0.5622 & 0.7330 & 0.4654 & 0.5952 & 0.3304 & 0.4104 & 0.3890 & 0.3368\\
mfBm3 & 0.6315 & 0.7973 & 0.8850 & 0.9184 & 0.9998 & 1.2449 & 1.4398 & 1.5859 &&0.5622 & 0.7492 & 0.4654 & 0.5952 & 0.3304 & 0.4104 & 0.3890 & 0.3368\\
mfBm4 & 0.6309 & 0.7970 & 0.8829 & 0.9166 & 0.9965 & 1.2379 & 1.4272 & 1.5698 &&0.5622 & 0.7492 & 0.5478 & 0.5952 & 0.4208 & 0.4104 & 0.8110 & 0.6426\\
mfBm5 & 0.6301 & 0.7960 & 0.8822 & 0.9152 & 0.9956 & 1.2378 & 1.4250 & 1.5676&&1.0000 & 1.0000 & 1.0000 & 1.0000 & 1.0000 & 1.0000 & 1.0000 & 1.0000\\
\hline
\multicolumn{18}{c}{April 12, 2017 to July 30 2021}     \\\hline
fBm   & 0.3842 & 0.5180 & 0.6119 & 0.7114 & 0.8080 & 1.0794 & 1.2913 & 1.4648 &&1.0000 & 1.0000 & 0.9274 & 0.9938 & 1.0000 & 0.8420 & 0.8442 & 0.8778\\
bfBm  & 0.3849 & 0.5186 & 0.6121 & 0.7115 & 0.8081 & 1.0782 & 1.2886 & 1.4600 &&0.2298 & 0.7328 & 0.9274 & 0.9938 & 0.9968 & 0.8420 & 0.8442 & 0.8884\\
mfBm3 & 0.3850 & 0.5185 & 0.6116 & 0.7113 & 0.8081 & 1.0781 & 1.2890 & 1.4604 &&0.1112 & 0.7328 & 1.0000 & 1.0000 & 0.9968 & 0.8918 & 0.8442 & 0.8884\\
mfBm4 & 0.3858 & 0.5191 & 0.6121 & 0.7116 & 0.8083 & 1.0775 & 1.2875 & 1.4582 &&0.0486 & 0.4882 & 0.9274 & 0.9538 & 0.9352 & 1.0000 & 1.0000 & 1.0000\\
mfBm5 & 0.3855 & 0.5191 & 0.6122 & 0.7115 & 0.8090 & 1.0793 & 1.2895 & 1.4612&&0.0556 & 0.4882 & 0.9274 & 0.9938 & 0.4800 & 0.8420 & 0.8128 & 0.7246\\
\hline
\hline
\end{tabular}}
\label{emp: mfbm_full}
\end{table}

\subsection{Does more information necessarily lead to efficiency gains?}
Our analysis naturally prompts two fundamental questions: First, what justifies our use of the mfBm framework over conventional multivariate time series models? Second, does incorporating additional information (e.g., moving from univariate to multivariate specifications) necessarily yield efficiency gains in forecasting?
These questions address both the methodological choice and the practical implications of model complexity.

\begin{table}[H]
\centering
\caption{MSFEs of HAR-type models for $h$-day-ahead forecasts of RV, along with p-values from model confidence set (MCS) tests,  between January 3, 2007 to January 14, 2025.}
\scalebox{0.6}{ 
\begin{tabular}{ccccccccccccccccccc}
\hline \hline
&  \multicolumn{8}{c}{MSFE $\times$ 100}  & & \multicolumn{8}{c}{MCS p-value} \\
\hline
              &       $h=1  $    &  $h=2 $     &  $h=3  $    &  $h=4 $     &  $h=5 $     &  $h=10 $    &  $h=15 $    &  $h=20$ &&$h=1  $    &  $h=2 $     &  $h=3  $    &  $h=4 $     &  $h=5 $     &  $h=10 $    &  $h=15 $    &  $h=20$  \\
                         \hline
HAR   & 0.5157 & 0.6513 & 0.7369 & 0.7870 & 0.8678 & 1.1109 & 1.2468 & 1.3072 && 1.0000 & 1.0000 & 1.0000 & 1.0000 & 1.0000 & 1.0000 & 1.0000 & 1.0000 \\
VHAR2  & 0.5316 & 0.6703 & 0.7701 & 0.8292 & 0.9307 & 1.2834 & 1.4560 & 1.5053&&0.0490 & 0.0790 & 0.0310 & 0.0424 & 0.0352 & 0.0450 & 0.0558 & 0.1044  \\
VHAR3 & 0.5384 & 0.6838 & 0.7863 & 0.8482 & 0.9563 & 1.3236 & 1.5065 & 1.5581&&0.0466 & 0.0244 & 0.0190 & 0.0204 & 0.0154 & 0.0450 & 0.0558 & 0.1044  \\
VHAR4 & 0.5457 & 0.6977 & 0.8074 & 0.8750 & 0.9852 & 1.3666 & 1.6630 & 1.9388&&0.0394 & 0.0232 & 0.0172 & 0.0174 & 0.0150 & 0.0450 & 0.0558 & 0.1044 \\
VHAR5 & 0.5580 & 0.7072 & 0.8158 & 0.8973 & 1.0088 & 1.3626 & 1.6406 & 1.8773&&0.0394 & 0.0018 & 0.0086 & 0.0012 & 0.0100 & 0.0450 & 0.0558 & 0.1044\\
\hline \hline
\end{tabular}}
\label{emp: HAR}
\end{table}

Given the HAR model's established role as a benchmark in volatility forecasting, we systematically compare the performance of univariate HAR and vector HAR (VHAR) models to evaluate whether increased information incorporation necessarily improves forecast accuracy.\footnote{The detailed specifications of HAR and VHAR can be found in Appendix \ref{sec:add_empirics}.} The VHAR model can be viewed as a multivariate approximation of a long-memory vector autoregressive fractionally integrated moving average (VARFIMA) process \citep{halbleib}, in a manner similar to the HAR model. To do so, we forecast AAPL's realized volatility using the same set of information sources (AAPL, ALD, AMGN, AXP, and BA) across the same five progressively expanded model specifications as before. Table \ref{emp: HAR} reports MSFEs and MCS p-values for HAR and four VHAR models.

By examining Table \ref{emp: HAR} and comparing it with Table \ref{emp: mfbm_full}, we draw the following conclusions. First, unlike the mfBm framework, which benefits from incorporating additional information, the vector HAR model fails to improve forecasting accuracy, contrary to conventional expectations. Second, when using the same information, the mfBm generally delivers more accurate forecasts than HAR-type models. Given the strong performance of fBm-based models, the mfBm emerges as a powerful tool for leveraging extra information to enhance predictions. These findings underscore a key advantage of the mfBm framework: its unique capacity to effectively utilize additional information where traditional multivariate approaches fall short.\footnote{We also examine other variants of HAR, including the vector HAR model with a single common factor \citep{Bauwens2023, Ding2025} and vector log-HAR models. These models remain inferior to our mfBm. In addition to MSFE, QLIKE \citep{bollerslev2016} is also used as a loss function. Detailed model specifications and related results are provided in Appendix \ref{sec:add_empirics}.}
%%%%%%%%%%%%%%%%%%%%%%%%%%%%%%%%%%
\section{Conclusion\label{sec:7}}
This paper introduces the first multivariate rough fractional volatility model, proposing a time-reversible multivariate fBm to jointly model a panel of realized volatilities. We develop the moment-based estimation method and investigate its asymptotic properties. We establish theoretical results on optimal forecasts, demonstrating that the multivariate model outperforms its univariate counterpart when correlations are non-zero and Hurst exponents differ across components. Empirically, the model delivers significant efficiency gains in volatility forecasting compared to both univariate and benchmark approaches.

During the revision of this paper, we became aware of an independent study by \cite{dugo}, where a multivariate fractional Ornstein-Uhlenbeck (mfOU) process is proposed. While this work primarily explores probabilistic properties rather than feasible estimation and forecasting, their follow-up paper \cite{dugo2024} conducts empirical analysis of the model. Their empirical analysis mostly confirms our findings, and they also report evidence of volatility spillover effects from the U.S. market to the Asian market, resulting in asymmetry. Thus, there are some cases in which volatility data can violate time-reversibility. This highlights the importance of our statistical test.

Our study opens several promising avenues for future research: (1) asymmetric forecasting theory—developing a general theoretical framework for forecasting with asymmetry, which could broaden the model's applicability across different financial contexts; (2) mfOU modelling and forecasting --- investigating the modelling and forecasting performance of mfOU processes and assessing their potential practical advantages over existing approaches; (3) efficient estimation methods --- exploring improved estimation techniques, such as maximum likelihood estimation, and quantifying the expected efficiency gains relative to current methods; (4) structural break analysis --- extending the model to incorporate change-point detection, as suggested by the regime-switching behavior observed in AAPL’s Hurst exponent, to better identify and interpret structural shifts in volatility dynamics.

%\section*{Supplementary Materials}
%The supplementary material contains R code that replicates the simulations in Section 3. An online supplement contains the proofs of all theoretical results, along with the simulation studies and robustness checks.

%\section*{Acknowledgements}
%The authors are grateful to the editor, the associate editor, and the two anonymous reviewers for their insightful comments, which significantly improved the quality of this paper.

%\section*{Disclosure Statement}
%The authors report there are no competing interests to declare.

\section*{Funding}
 Yu acknowledges support from the University of Macau Development Foundation.
 
%\bibliographystyle{Chicago}
%\bibliography{Bibliography-MM-MC}
%\bibliography{Bibliography-Main}

\newpage
\appendix
\begin{center}
\textbf{\Large Appendix}
\end{center}

\section{Existing estimation method}
\label{sec:est_amblard_coeurjolly}
%\label{sec:3b}
The first and only available parameter estimation method for mfBm was introduced in \cite{amblard2011identification}. Let \( B_{k \Delta}^m \) denote the time series obtained by applying the \( m \)-th dilated version of a filter to the \( d \)-dimensional \( B_{k \Delta} \), where \( m \) is an integer and\footnote{\( a \) is in the set of filters,
\[ \mathcal{A}_{l,q} = \left\{ (a_t)_{t \in \mathbb{Z}} : a_t = 0, \forall t \in \mathbb{Z}^{-} \cup \{l+1, ..., +\infty\} \text{ and } \sum_{t \in \mathbb{Z}} t^l a_t = 0, \forall l = 0, ..., q-1 \right\},\]
where $\mathbb{Z}^{-}=\{...,-2,-1,0\}$, $l$ and $q$ be two positive integers. Typical examples are the difference filter and its compositions, Daubechies wavelet filters, and any known wavelet filter with compact support and a sufficient number of vanishing moments.} 
\vspace{-0.2cm}
\[
a^m_t = 
\begin{cases}
a_{t/m}, & \text{if } t \in m\mathbb{Z} \\
0, & \text{if } t \notin m\mathbb{Z}
\end{cases},
\] 
with $\mathds{Z}=\{0,1,2,...\}$. To be more specific, the $i$-th component of \( B_{k \Delta}^m \), denoted by \( B_{k\Delta}^{(i),m} \), is
\[
B_{k\Delta}^{(i),m} = \sum_{t \in \mathbb{Z}} a^m_t B_{(k-t)\Delta}^{(i)}.
\]

The estimation principle for the case \( H_i + H_j \neq 1 \) in \cite{amblard2011identification} relies for a given \( m \geq 1 \) on the theoretical covariance 
\begin{align}
\Cov^{m}_{ij}(h\Delta) :=& \Cov(B_{k\Delta}^{(i),m},B_{(k+h)\Delta}^{(j),m})\notag \\
=& -\frac{\sigma_i \sigma_j}{2} \sum_{t,l \in \mathbb{Z}} a_t a_l \left( \rho_{ij} - \eta_{ij} \text{sign}(h + m(t-l)) \right) |(h + m(t - l))\Delta|^{H_i + H_j},\notag
\end{align}
for all \( i, j = 1, \dots, d \), and its empirical estimator
\[
C_{ij}^{m}(h\Delta) = \frac{1}{n-ml-h} \sum_{k=ml+1}^{n-h} B_{k\Delta}^{(i),m} B_{(k+h)\Delta}^{(j),m}.
\]
Subsequently, the differences between the theoretical covariances and their empirical counterparts are
\begin{align}
\epsilon_{v_{i}}^{m} =& \log C_{ii}^{m}(0) - \log \Cov^{m}_{ii}(0),\notag \\
\epsilon_{c_{ij}}^{m} =& \log |C_{ij}^{m}(0)| - \log | \Cov^{m}_{ij}(0)|,\notag \\
\epsilon_{d_{ij}}^{m} =& \log \left( 0.5 |C_{ij}^{m}(ml\Delta) - C_{ji}^{m}(ml\Delta)| \right) - \log \left( 0.5 |\Cov^{m}_{ij}(ml\Delta) -  \Cov^{m}_{ij}(ml\Delta)| \right).\notag
\end{align}

\cite{amblard2011identification} proposed to estimate the Hurst exponents, scaling, correlation and asymmetry parameters by minimizing the following weighted mean squared error over all values of \( m \) taken from a discrete set \( \mathcal{M} \):
\[  \sum_{m \in \mathcal{M}} \left( w_v \sum_{i=1}^{d} (\epsilon_{v_i}^m)^2 + w_c \sum_{i=1,j>i}^{d} (\epsilon_{c_{ij}}^m)^2 + w_{dd} \sum_{i=1,j>i}^{d} (\epsilon_{d_{ij}}^m)^2 \right).\]
 
\cite{amblard2011identification} stated that in a typical setting one can choose the weights \( w_c = w_{dd} = 0 \), which is what we adopt in the present paper. Moreover, \cite{amblard2011identification} proved that the estimator is consistent regardless of the choice of weights. They also provide the condition under which the estimator is asymptotically normally distributed. However, the complexity of the asymptotic variance expression poses significant challenges for constructing asymptotic confidence intervals in practice.

\section{Propositions for all historical observations}
\label{sec:prop_fore}

\begin{prop}\label{unifracprop}
Consider a bivariate time-reversible fBm with a positive definite covariance and $H_1=H_2=H$. The optimal forecast for $B_{t+h}^{(1)}$, given $t$ discrete observations $(B^{(1)}_{\Delta},B^{(2)}_{\Delta}),\ldots, (B^{(1)}_{t\Delta},B^{(2)}_{t\Delta})$, does not depend on the component of $(B^{(2)})$ and coincides with the optimal forecast based on univariate fBm.
\end{prop}

The proof of Proposition \ref{unifracprop} benefits from a more explicit and transparent representation of the conditional expectation in the bivariate case compared to the general $d$-dimensional setting. Nevertheless, the overarching result in Proposition \ref{unifracgen} and its proof strategy follow a similar approach.

\begin{prop}\label{unifracgen}
Consider a time-reversible mfBm with a positive definite covariance and all component-wise Hurst exponents equal. The optimal forecast for $B_{t+h}^{(1)}$, given $t$ discrete observations $(B^{(1)}_{\Delta},\ldots,B^{(d)}_{\Delta}),\ldots, (B^{(1)}_{t\Delta},\ldots,B^{(d)}_{t\Delta})$, does only depend on the first component $(B^{(1)})$ and coincides with the optimal forecast based on univariate fBm.
\end{prop}

We conduct Monte Carlo experiments to evaluate forecasting performance in realistic empirical settings: 1) investigating how forecast accuracy varies with inter-component correlation; 2) analyzing how accuracy depends on differences in Hurst exponents; 3) evaluating accuracy when incorporating additional mfBm components. Overall, simulation results are consistent with analytical predictions. More results can be found in  Online Supplement \ref{sec:monte_carlo}.

\section{Proof of Theoretical Results}
\label{sec:proof_theorem}
\subsection{Proof of Theorem \ref{propstat}}
For components $(p,q)\in\{1,\ldots,d\}^2$, the covariance of the $k$th and $j$th increments, $(j,k)\in\{1,\ldots,n\}^2$, of the time-reversible mfBm yields
\begin{align}\label{covincr}\cov\big(\Delta_k B^{(p)},\Delta_j B^{(q)}\big)=\rho_{p,q}\sigma_p\sigma_q\Delta^{H_p+H_q}\gamma_{p,q}(k-j)\,,\end{align}
where the dependence on time, or the lag $l\in\mathds{Z}$, is via the function
\begin{align}\label{gamma}\gamma_{p,q}(l)=\frac12\Big(|l+1|^{H_p+H_q}+|l-1|^{H_p+H_q}-2|l|^{H_p+H_q}\Big)\,.\end{align}
The function is symmetric in the sense that $\gamma_{p,q}(l)=\gamma_{p,q}(-l)$. For the correlation parameters, we use the notation $\rho_{p,p}=1$, for $p=q$. For $p=q$, we further write shortly $\gamma_{p}$ instead of $\gamma_{p,p}$. It holds $\gamma_{p,q}(0)=1$, for any $p,q$. Having Gaussian processes, a crucial ingredient to compute variances and covariances of our statistics, is obtained from the Isserlis' theorem (\cite{isserlis}):
\begin{align}\label{isserlis}\cov\big(\Delta_k B^{(p)}\Delta_j B^{(q)},\Delta_i B^{(r)}\Delta_l B^{(s)}\big)&=\cov\big(\Delta_k B^{(p)},\Delta_i B^{(r)}\big)\cov\big(\Delta_j B^{(q)},\Delta_l B^{(s)}\big)\\
&\notag\quad +\cov\big(\Delta_k B^{(p)},\Delta_l B^{(s)}\big)\cov\big(\Delta_j B^{(q)},\Delta_i B^{(r)}\big),\end{align}
for all $(p,q,r,s)\in\{1,\ldots,d\}^4$ and $(j,k,l,i)\in\{1,\ldots,n\}^4$.
We obtain the following variances of suitably normalized realized volatilities: 
\begin{align*}
\var\Big(\frac{\Delta^{-2H_p}}{\sqrt{n}}\sum_{k=1}^n\big(\Delta_kB^{(p)}\big)^2\Big)&=\frac{\Delta^{-4H_p}}{n}\sum_{1\le k,l\le n}\cov\big(\big(\Delta_kB^{(p)}\big)^2,\big(\Delta_lB^{(p)}\big)^2\big)\\
&\stackrel{\eqref{isserlis}}{=}\frac{\Delta^{-4H_p}}{n}\sum_{1\le k,l\le n}2\cov^2\big(\Delta_kB^{(p)},\Delta_lB^{(p)}\big)\\
&\stackrel{\eqref{covincr}}{=}\frac{2\sigma_p^4}{n}\sum_{1\le k,l\le n}\gamma_p^2(k-l)\\
&=\frac{2\sigma_p^4}{n}\Big(n\gamma_p^2(0)+2\sum_{r=1}^n(n-r)\gamma_p^2(r)\Big)\\
&=2\sigma_p^4\Big(1+2\sum_{r=1}^n\frac{n-r}{n}\gamma_p^2(r)\Big)\,.
\end{align*}
With dominated convergence, we obtain that
\begin{align*}\lim_{n\to\infty} \var\Big(\frac{\Delta^{-2H_p}}{\sqrt{n}}\sum_{k=1}^n\big(\Delta_kB^{(p)}\big)^2\Big)&=2\sigma_p^4\Big(1+2\sum_{r=1}^{\infty}\gamma_p^2(r)\Big)\\
&\stackrel{\eqref{gamma}}{=}\sigma_p^4\Big(2+\sum_{r=1}^{\infty}\Big(|r+1|^{2H_p}+|r-1|^{2H_p}-2|r|^{2H_p}\Big)^2\Big)\,.
\end{align*}
Applying twice the mean value theorem for differentiation with the functions $x\mapsto x^{2H_p}$, and $x\mapsto x^{2H_p-1}$, we find that
\begin{align*}(r+1)^{2H_p}+(r-1)^{2H_p}-2(r)^{2H_p}&=2H_p\big(\xi_r^{2H_p-1}-\xi_{r-1}^{2H_p-1}\big)\\
&=2H_p(2H_p-1)\zeta_r^{2H_p-2}\,,\end{align*}
with some $\xi_r\in(r,r+1),r\in\N$, and $\zeta_r\in(\xi_{r-1},\xi_r)\subset(r-1,r+1)$. This suffices to see (based on a majorant and minorant) that the series $\sum_{r=1}^{\infty}\gamma_p^2(r)$ converges if and only if $H_p\in(0,3/4)$. Similar arguments apply to upcoming series expressions in the proofs. Univariate central limit theorems 
\begin{align*}%\label{cltqv}
\sqrt{n}\Big(\frac{\Delta^{-2H_p}}{n}\sum_{k=1}^n\big(\Delta_kB^{(p)}\big)^2-\sigma_p^2\Big)\stackrel{d}{\rightarrow}\mathcal{N}\Big(0,\sigma_p^4\Big(2+\sum_{r=1}^{\infty}\big(|r+1|^{2H_p}+|r-1|^{2H_p}-2|r|^{2H_p}\big)^2\Big)\Big)\end{align*}
are known to hold in this case by an application of the celebrated limit theorem by \cite{breuermajor} with the Hermite polynomial $x\mapsto x^2-1$ of Hermite rank 2. For its rank, it suffices to see that $\E[(X^2-1)X]= 0$ and $\E[(X^2-1)^2]=2\ne 0$, for a normalized centred normal random variable $X$.

Next, consider lag-2-increments, which equal (telescoping) sums of increments:
\[\Delta_{k,2} B^{(p)}=B_{(k+1)\Delta}^{(p)}-B_{(k-1)\Delta}^{(p)}=\Delta_k B^{(p)}+\Delta_{k+1}B^{(p)},~k\in\{1,\ldots,n-1\}\,.\]
It holds that
\begin{align*}
\cov(\Delta_{k,2} B^{(p)},\Delta_{l,2} B^{(p)})&=\cov(\Delta_k B^{(p)}+\Delta_{k+1}B^{(p)},\Delta_l B^{(p)}+\Delta_{l+1}B^{(p)})\\
&=\Delta^{2H_p}\sigma_p^2\big(2\gamma_p(k-l)+\gamma_p(k+1-l)+\gamma_p(k-l-1)\big)\,
\end{align*}
by linearity. Note the relation $2^{2H_p}=(2\gamma_p(1)+2)$ for the variances. This relation and applying the Isserlis formula directly to lag-2-increments, what is possible, yields a simple computation of the variance
\begin{align*}
&\var\Big(\frac{\Delta^{-2H_p}}{\sqrt{n}}\sum_{k=1}^{n-1}\big(\Delta_{k,2}B^{(p)}\big)^2\Big)=\frac{\Delta^{-4H_p}}{n}\sum_{1\le k,l\le n-1}2\cov^2(\Delta_{k,2} B^{(p)},\Delta_{l,2} B^{(p)})\\
&\quad=\frac{2\sigma_p^4}{n}\Big((n-1)(2+2\gamma_p(1))^2+2\sum_{r=1}^{n-2}(n-1-r)\big(2\gamma_p(r)+\gamma_p(r+1)+\gamma_p(r-1)\big)^2\Big)\\
&\quad=\sigma_p^4\Big(2\cdot 2^{4H_p}\frac{n-1}{n}+4\sum_{r=1}^{n-2}\frac{n-1-r}{n}\big(2\gamma_p(r)+\gamma_p(r+1)+\gamma_p(r-1)\big)^2\Big)\,.
\end{align*}
With dominated convergence and elementary transformations, we obtain that
\begin{align*}\lim_{n\to\infty} \var\Big(\frac{\Delta^{-2H_p}}{\sqrt{n}}\sum_{k=1}^{n-1}\big(\Delta_{k,2}B^{(p)}\big)^2\Big)&=\sigma_p^4\Big(2\cdot 2^{4H_p}+4\sum_{r=1}^{\infty}\big(2\gamma_p(r)+\gamma_p(r+1)+\gamma_p(r-1)\big)^2\Big)\\
&\stackrel{\eqref{gamma}}{=}\sigma_p^4\Big(2\cdot 2^{4H_p}+\sum_{r=1}^{\infty}\Big(|r+2|^{2H_p}+|r-2|^{2H_p}-2|r|^{2H_p}\Big)^2\Big)\,.
\end{align*}
Since we sum over all $k$, and do not only use half of the observations, this is not obvious from the above asymptotics of the realized volatility. A central limit theorem follows analogously as for the realized volatility. For the covariance of both statistics, we use that
\begin{align*}
\cov\big(\Delta_k B^{(p)},\Delta_{l,2} B^{(p)}\big)&=\cov\big(\Delta_k B^{(p)},\Delta_{l} B^{(p)}+\Delta_{l+1} B^{(p)}\big)\\
&=\Delta^{2H_p}\sigma_p^2\big(\gamma_p(k-l)+\gamma_p(k-l-1)\big),~1\le k\le n,1\le l\le n-1\,.
\end{align*}
We derive that
\begin{align*}
&\cov\Big(\frac{\Delta^{-2H_p}}{\sqrt{n}}\sum_{k=1}^n\big(\Delta_k B^{(p)}\big)^2,\frac{\Delta^{-2H_p}}{\sqrt{n}}\sum_{l=1}^{n-1}\big(\Delta_{l,2}B^{(p)}\big)^2\Big)=\frac{\Delta^{-4H_p}}{n}\sum_{\substack{1\le k\le n\\1\le l\le n-1}}2\cov^2\big(\Delta_k B^{(p)},\Delta_{l,2} B^{(p)}\big)\\
&=\frac{2\sigma_p^4}{n}\sum_{\substack{1\le k\le n\\1\le l\le n-1}}\big(\gamma_p(k-l)+\gamma_p(k-l-1)\big)^2\\
&= \frac{2\sigma_p^4}{n}\Big(2(n-1)\big(\gamma_p(0)+\gamma_p(1)\big)^2+\sum_{r=1}^{n-2}(n-r-1)\big(\gamma_p(r)+\gamma_p(r+1)\big)^2+\sum_{r=2}^{n-1}(n-r)\big(\gamma_p(r)+\gamma_p(r-1)\big)^2\Big)\\
&= 4\sigma_p^4\sum_{r=1}^{n-1}\frac{n-r}{n}\big(\gamma_p(r)+\gamma_p(r-1)\big)^2\,,
\end{align*}
such that with dominated convergence 
\begin{align*}&\lim_{n\to\infty} \cov\Big(\frac{\Delta^{-2H_p}}{\sqrt{n}}\sum_{k=1}^n\big(\Delta_k B^{(p)}\big)^2,\frac{\Delta^{-2H_p}}{\sqrt{n}}\sum_{l=1}^{n-1}\big(\Delta_{l,2}B^{(p)}\big)^2\Big)=4\sigma_p^4\sum_{r=1}^{\infty}\big(\gamma_p(r)+\gamma_p(r-1)\big)^2\\
&\quad \stackrel{\eqref{gamma}}{=}\sigma_p^4\sum_{r=1}^{\infty}\Big(|r+1|^{2H_p}+|r-2|^{2H_p}-|r|^{2H_p}-|r-1|^{2H_p}\Big)^2\,.
\end{align*}
If we show for scalar products $\langle \,\cdot,\,\cdot\,\rangle$, with any $\alpha\in\R^2$, that
\begin{align}\label{bicltsc}
\sqrt{n}\Big\langle\alpha,\Big(\frac{\Delta^{-2H_p}}{n}\sum_{k=1}^n(\Delta_k B^{(p)})^2-\sigma_p^2,\frac{\Delta^{-2H_p}}{n}\sum_{k=1}^{n-1}(\Delta_{k,2} B^{(p)})^2-2^{2H_p}\sigma_p^2\Big)^{\top}\Big\rangle\stackrel{d}{\rightarrow}\langle \alpha, Z\rangle\,,
\end{align}
with the bivariate normal
\begin{align*}
Z\sim\mathcal{N}\bigg(\bigg(\begin{array}{c}0 \\[-.25cm] 0\end{array}\bigg),\bigg(\begin{array}{cc}\text{AVAR}_{11}& \text{AVAR}_{12}\\[-.25cm] \text{AVAR}_{12}&\text{AVAR}_{22}\end{array}\bigg)\bigg)\,,
\end{align*}
where
\begin{subequations}
\begin{align}\label{bicltavar1}
\text{AVAR}_{11}&=\sigma_p^4\Big(2+\sum_{r=1}^{\infty}\Big(|r+1|^{2H_p}+|r-1|^{2H_p}-2|r|^{2H_p}\Big)^2\Big) \,,\\
\label{bicltavar2}\text{AVAR}_{12}&=\sigma_p^4 \sum_{r=1}^{\infty}\Big(|r+1|^{2H_p}+|r-2|^{2H_p}-|r|^{2H_p}-|r-1|^{2H_p}\Big)^2\,,\\
\label{bicltavar3}\text{AVAR}_{22}&=\sigma_p^4\Big(2\cdot 2^{4H_p}+\sum_{r=1}^{\infty}\Big(|r+2|^{2H_p}+|r-2|^{2H_p}-2|r|^{2H_p}\Big)^2\Big)\,,
\end{align}
\end{subequations}
the Cram\'{e}r-Wold device implies the bivariate central limit theorem
\begin{align}\label{bicltqv}\sqrt{n}\left(
\begin{array}{c}\frac{\Delta^{-2H_p}}{n}\sum_{k=1}^n(\Delta_k B^{(p)})^2-\sigma_p^2\\ \frac{\Delta^{-2H_p}}{n}\sum_{k=1}^{n-1}(\Delta_{k,2} B^{(p)})^2-2^{2H_p}\sigma_p^2\end{array}\right)\stackrel{d}{\rightarrow} Z\,.
\end{align}
By self-similarity $B_{k\Delta}^{(p)}\stackrel{d}{=}\Delta^{H_p}B_k^{(p)}$, such that (instead of a triangular array) we consider the $\R^2$-valued stationary, mean-zero Gaussian sequence 
\[(X_1,\ldots,X_{n-1}),~\mbox{with}~X_j=\big(X_j^{(1)},X_j^{(2)}\big)^{\top}=\big(\sigma_p(B_{j+1}^{(p)}-B_j^{(p)}),\sigma_p(B_{j+1}^{(p)}-B_{j-1}^{(p)})\big)^{\top}\,,\]
and for $\alpha=(\alpha_1,\alpha_2)^{\top}$ the function $f:\R^2\to\R,f(x,y)=\alpha_1x^2+\alpha_2y^2$. Since Hermite polynomials $H_r,r\in\N$, satisfy $\E[H_{r_1}(X_j^{(1)})H_{r_2}(X_j^{(2)})]=\delta_{r_1,r_2}r_1!c_{12}$, with (covariance) constant $c_{12}$, i.e., zero for $r_1\ne r_2$, and non-zero for $r_1=r_2$, $f$ has Hermite rank 2 in the sense of Eq.\ (2.2) in \cite{arcones}; see \eqref{rank} also. Neglecting the first increment in the first component, which is asymptotically negligible, the limit theorem from \cite{arcones} thus readily yields \eqref{bicltsc} and we conclude \eqref{bicltqv}. A (more direct) proof based on moment generating functions can be worked out similar to Theorem 4 of \cite{kubilius}. We have that
\begin{align*}\hat H_p=\frac{1}{2\log(2)}\log\Big(\frac{T}{S}\Big),~\mbox{with}~S=\frac{\Delta^{-2H_p}}{n}\sum_{k=1}^n(\Delta_k B^{(p)})^2\,,~\mbox{and}~T=\frac{\Delta^{-2H_p}}{n}\sum_{k=1}^{n-1}(\Delta_{k,2} B^{(p)})^2\,.
\end{align*}
From \eqref{bicltqv}, we can therefore derive a central limit theorem for $\hat H_p$ based on the multivariate $\Delta$-method applied to the function $g:\R^2\to\R$, $g(x,y)=\log(y/x)/(2\log(2))$. Since $g\big(\sigma_p^2,2^{2H_p}\sigma_p^2\big)=H_p$, the $\Delta$-method yields
\begin{align*}
\sqrt{n}\big(\hat H_p-H_p\big)
\stackrel{d}{\rightarrow}\mathcal{N}\bigg(0,\big(\nabla g\big(\sigma_p^2,2^{2H_p}\sigma_p^2\big)\big)^{\top}\bigg(\begin{array}{cc}\text{AVAR}_{11}& \text{AVAR}_{12}\\[-.2cm] \text{AVAR}_{12}&\text{AVAR}_{22}\end{array}\bigg)\nabla g\big(\sigma_p^2,2^{2H_p}\sigma_p^2\big)\bigg)\,,
\end{align*}
with $\big(\nabla g(x,y)\big)^{\top}=(-x^{-1},y^{-1})^{\top}/(2\log(2))$, such that
\begin{align*}
&\nabla g\big(\sigma_p^2,2^{2H_p}\sigma_p^2\big)\big)^{\top}\bigg(\begin{array}{cc}\text{AVAR}_{11}& \text{AVAR}_{12}\\[-.2cm] \text{AVAR}_{12}&\text{AVAR}_{22}\end{array}\bigg)\nabla g\big(\sigma_p^2,2^{2H_p}\sigma_p^2\big)\\
&\quad=\frac{1}{4\log^2(2)\sigma_p^4}\Big(\text{AVAR}_{11}-2\cdot2^{-2H_p}\text{AVAR}_{12}+2^{-4H_p}\cdot \text{AVAR}_{22}\Big)\,.
\end{align*}
We conclude with \eqref{bicltavar1}-\eqref{bicltavar3} for $H_p\in(0,3/4)$ that
\begin{align}\label{cltHhat2}
\sqrt{n}\big(\hat H_p-H_p\big)
\stackrel{d}{\rightarrow}\mathcal{N}\big(0,\text{AVAR}_{\hat H_p}\big)\,,
\end{align}
with asymptotic variance as in \eqref{cltHhat}
\begin{align}\label{cltHhat3}
\text{AVAR}_{\hat H_p}&=\frac{1}{4\log^2(2)}\Big(4+\sum_{r=1}^{\infty}\Big(|r+1|^{2H_p}+|r-1|^{2H_p}-2|r|^{2H_p}\Big)^2 \notag \\
&\hspace*{2.5cm} +2^{-4H_p}\sum_{r=1}^{\infty}\Big(|r+2|^{2H_p}+|r-2|^{2H_p}-2|r|^{2H_p}\Big)^2 \notag \\
&\hspace*{2.5cm}-2\cdot 2^{-2H_p}\sum_{r=1}^{\infty}\Big(|r+1|^{2H_p}+|r-2|^{2H_p}-|r|^{2H_p}-|r-1|^{2H_p}\Big)^2\Big)\,.
\end{align}
This asymptotic variance depends on $H_p$, but not on $\sigma_p^2$. Based on the expansion
\[\Delta^{2(H_p-\hat H_p)}=\exp\big(2(\hat H_p-H_p)\log(\Delta^{-1})\big)=1+2(\hat H_p-H_p)\log(n)+\mathcal{O}_{\P}\big(n^{-1}\log^2(n)\big)\,,\]
we obtain that
\begin{align*}
&\frac{\sqrt{n}}{\log(n)}\big(\hat\sigma_p^2-\sigma_p^2\big)=\frac{\sqrt{n}}{\log(n)}\Big(\frac{\Delta^{-2H_p}}{n}\sum_{k=1}^n\big(\Delta_k B^{(p)}\big)^2\Delta^{2(H_p-\hat H_p)}-\sigma_p^2\Big)\\
&\quad=\frac{\sqrt{n}}{\log(n)}\Big(\frac{\Delta^{-2H_p}}{n}\sum_{k=1}^n\big(\Delta_k B^{(p)}\big)^2-\sigma_p^2\Big)+2\sqrt{n}\big(\hat H_p-H_p\big)\frac{\Delta^{-2H_p}}{n}\sum_{k=1}^n\big(\Delta_k B^{(p)}\big)^2+\mathcal{O}_{\P}\big(n^{-1/2}\log^2(n)\big)\\
&\quad=2\sqrt{n}\big(\hat H_p-H_p\big)\sigma_p^2+\KLEINO_{\P}(1)\stackrel{d}{\rightarrow}\mathcal{N}\big(0,4\sigma_p^4\text{AVAR}_{\hat H_p}\big)\,,
\end{align*}
by Slutsky's theorem and \eqref{cltHhat2}. This proves \eqref{cltsigmaest} and completes the proof of Theorem \ref{propstat}.

\subsection{Proof of Theorem \ref{theostat2}}
For the general, bivariate fBm with asymmetry parameter $\eta_{1,2}=-\eta_{2,1}$, the cross-covariances of increments generalize from \eqref{covincr} to 
\begin{align}\label{covincr2}\cov(\Delta_k B^{(1)},\Delta_j B^{(2)})\negthinspace =\negthinspace \Delta^{\negthinspace H_1+H_2}\sigma_{\negthinspace 1}\sigma_{\negthinspace 2}\big(\rho_{1,2}\gamma_{1,2}(k\negthinspace -\negthinspace j)+\eta_{1,2}\tilde\gamma_{1,2}(k\negthinspace -\negthinspace j)\big),\end{align}
%\begin{align}\label{covincr2}\cov\big(\Delta_k B^{(1)},\Delta_j B^{(2)}\big)=\Delta^{H_1+H_2}\sigma_{1}\sigma_{2}\big(\rho\gamma_{1,2}(k - j)+\eta_{1,2}\tilde\gamma_{1,2}(k\negthinspace -\negthinspace j)\big)\,,\end{align}
for $(j,k)\in\{1,\ldots,n\}^2$, and with $\gamma_{1,2}$ from \eqref{gamma} and
\begin{align*}\tilde \gamma_{1,2}(r)=\tfrac12({{\operatorname{sgn}(r+1)}}|r\negthinspace+\negthinspace 1|^{H_1+H_2}+{\operatorname{sgn}(r\negthinspace-\negthinspace 1)}|r\negthinspace -\negthinspace 1|^{H_1+H_2}\negthinspace-\negthinspace2\,{\operatorname{sgn}(r)}|r|^{H_1+H_2}).\end{align*}
The functions $\gamma_{1,2}$ and $\tilde\gamma_{1,2}$ are closely related with $\gamma_{1,2}(r)=\tilde \gamma_{1,2}(r)$, for $r\ge 1$, but $\tilde\gamma$ is anti-symmetric while $\gamma$ is symmetric.
Covariances of increments of one component are not affected by the asymmetry. This implies 
\begin{align*}\E\big[\Delta_{k+1} B^{(1)}\Delta_k B^{(2)}-\Delta_{k} B^{(1)}\Delta_{k+1} B^{(2)}\big]&=\E\big[\Delta_{k+1} B^{(1)}\Delta_k B^{(2)}\big]-\E\big[\Delta_{k} B^{(1)}\Delta_{k+1} B^{(2)}\big]\\
&\stackrel{\eqref{covincr2}}{=}\sigma_1\sigma_2\Delta^{H_1+H_2}\gamma_{1,2}(1)\big(\big(\rho_{1,2}+\eta_{1,2})-(\rho_{1,2}+\eta_{2,1})\big)\\
&=2\eta_{1,2}\sigma_1\sigma_2\Delta^{H_1+H_2}\gamma_{1,2}(1)\,,
\end{align*}
using $\gamma_{1,2}(1)=\tilde\gamma_{1,2}(1)$, with $2\gamma_{1,2}(1)=2^{H_1+H_2}-2$. With a simple bound as $\big(\rho_{1,2}+\eta_{1,2}\operatorname{sign}{(k-j)}\big)^2\le 2(\rho_{1,2}^2+\eta_{1,2}^2)$, the above analysis for the covariation readily yields that
\[\frac{\Delta^{-H_1-H_2}}{n}\sum_{k=1}^{n-1}\Big(\Delta_{k+1} B^{(2)}\cdot \Delta_k B^{(1)}-\Delta_{k+1} B^{(1)}\cdot \Delta_k B^{(2)}\Big)-\eta_{1,2}\sigma_1\sigma_2(2^{H_1+H_2}-2)=\mathcal{O}_{\P}(n^{-1/2})\,.\]
The asymptotics of the rescaled realized variances, with increments and lag-2-increments, i.e., the above central limit theorems, imply that
\begin{align*}
\frac{\Delta^{-2H_j}}{n}\sum_{k=1}^{n-1}\big(\Delta_{k,2} B^{(j)}\big)^2-2^{2H_j}\sigma_j^2=\mathcal{O}_{\P}(n^{-1/2}),~j=1,2,\\
\frac{\Delta^{-2H_j}}{n}\sum_{k=1}^{n}\big(\Delta_{k} B^{(j)}\big)^2-\sigma_j^2=\mathcal{O}_{\P}(n^{-1/2}),~j=1,2,
\end{align*}
such that elementary computations yield that 
\[\hat\eta_{1,2}-\eta_{1,2}=\mathcal{O}_{\P}(n^{-1/2})\,.\]
Since we aim to construct an asymptotic test for $\eta_{1,2}=0$ vs.\ $\eta_{1,2}\ne 0$, we work out the central limit theorem for $\hat\eta_{1,2}$ under the null hypothesis that $\eta_{1,2}=0$. Therefore, the following derivation of the asymptotic variance may restrict to covariances \eqref{covincr} instead of \eqref{covincr2}, what simplifies some terms. We deduce that
\begin{align*}
&\var\bigg(\frac{\Delta^{-H_1-H_2}}{\sqrt{n}}\sum_{k=1}^{n-1}\Big(\Delta_{k+1} B^{(2)}\cdot \Delta_k B^{(1)}-\Delta_{k+1} B^{(1)}\cdot \Delta_k B^{(2)}\Big)\bigg)\\
&\quad =\frac{\Delta^{-2(H_1+H_2)}}{n}\sum_{1\le i,j\le n-1}\cov\Big(\Delta_{i+1} B^{(2)} \Delta_i B^{(1)}-\Delta_{i+1} B^{(1)} \Delta_i B^{(2)},\Delta_{j+1} B^{(2)} \Delta_j B^{(1)}-\Delta_{j+1} B^{(1)} \Delta_j B^{(2)}\Big)\\
&\quad \stackrel{\eqref{isserlis}}{=}\frac{\sigma_1^2\sigma_2^2}{n}\sum_{1\le i,j\le n-1}\Big(2\big(\gamma_1(i-j)\gamma_2(i-j)+\rho^2\gamma_{1,2}(i+1-j)\gamma_{1,2}(i-j-1)\big)\\
&\hspace*{1.5cm}-\gamma_1(i+1-j)\gamma_2(i-j-1)-\gamma_1(i-j-1)\gamma_2(i+1-j)-2\rho^2\gamma_{1,2}^2(i-j)\Big)\\
&\quad =\sigma_1^2\sigma_2^2\Big(\frac{n-1}{n}\big(2-2\gamma_1(1)\gamma_2(1)-2\rho^2\big(1-\gamma_{1,2}^2(1)\big)\big)\\
&\hspace*{1.5cm}+2\sum_{r=1}^{n-1}\frac{n-1-r}{n}\big(2\gamma_1(r)\gamma_2(r)-\gamma_1(r+1)\gamma_2(r-1)-\gamma_1(r-1)\gamma_2(r+1)\big)\\
&\hspace*{1.5cm}-4\rho^2\sum_{r=1}^{n-1}\frac{n-1-r}{n}\big(\gamma_{1,2}^2(r)-\gamma_{1,2}(r+1)\gamma_{1,2}(r-1)\big)\Big)\,.
\end{align*}
With dominated convergence, we conclude under the null, $\eta_{1,2}=0$, the asymptotic variance
\begin{align}
&\notag\lim_{n\to\infty}n\cdot\var\big(\hat\eta_{1,2}\big)=(2^{H_1+H_2}-2)^{-2}\Big(2\big(1-\gamma_1(1)\gamma_2(1)\big)+2\rho^2\big(\gamma_{1,2}^2(1)-1\big)\\
&\notag\quad-\rho^2\sum_{r=1}^{\infty}\Big(\big(|r+1|^{H_1+H_2}+|r-1|^{H_1+H_2}-2|r|^{H_1+H_2}\big)^2\\
&\notag\hspace*{1cm}-\big(|r+2|^{H_1+H_2}+|r|^{H_1+H_2}-2|r+1|^{H_1+H_2}\big)\big(|r|^{H_1+H_2}+|r-2|^{H_1+H_2}-2|r-1|^{H_1+H_2}\big)\Big)\\
&\notag\quad+\frac{1}{2}\sum_{r=1}^{\infty}\Big(2\big(|r+1|^{2H_1}+|r-1|^{2H_1}-2|r|^{2H_1}\big)\big(|r+1|^{2H_2}+|r-1|^{2H_2}-2|r|^{2H_2}\big)\\
&\notag\hspace*{2cm}-\big(|r+2|^{2H_1}+|r|^{2H_1}-2|r+1|^{2H_1}\big)\big(|r|^{2H_2}+|r-2|^{2H_2}-2|r-1|^{2H_2}\big)\\
&\notag\hspace*{2cm}-\big(|r+2|^{2H_2}+|r|^{2H_2}-2|r+1|^{2H_2}\big)\big(|r|^{2H_1}+|r-2|^{2H_1}-2|r-1|^{2H_1}\big)\Big)\Big)\\
&\hspace*{2cm}=\text{AVAR}_{\hat \eta_{1,2}}\label{avareta}\,.
\end{align}
For the proof of Theorem \ref{theostat2}, we use the asymptotic variance of $\hat\eta_{1,2}$ from \eqref{avareta} and show a central limit theorem. By self-similarity \eqref{selfsim}, it holds that $\big(B_{k\Delta}^{(1)},B_{k\Delta}^{(2)}\big)^{\top}\stackrel{d}{=}\big(\Delta^{H_1}B_k^{(1)},\Delta^{H_2}B_k^{(2)}\big)^{\top}$. To prove asymptotic normality of the numerator in \eqref{etahat}, multiplied with $\Delta^{-H_1-H_2}/\sigma_1\sigma_2$, we apply Theorem 4 from \cite{arcones} to prove asymptotic normality of $\sum_{k=1}^{n-1}f(X_k)$, with the $\R^4$-valued, stationary, mean-zero Gaussian sequence $(X_1,\ldots,X_{n-1})$, with
\[X_j=\big(X_j^{(1)},X_j^{(2)},X_j^{(3)},X_j^{(4)}\big)^{\top}=\big((B_{j}^{(1)}-B_{j-1}^{(1)}),(B_{j}^{(2)}-B_{j-1}^{(2)}),(B_{j+1}^{(1)}-B_{j}^{(1)}),(B_{j+1}^{(2)}-B_{j}^{(2)})\big)^{\top},\]
where the entries are already standardized with variances equal to one, and the function 
\[f:\R^4\to\R,f(x^{(1)},x^{(2)},x^{(3)},x^{(4)})=x^{(1)}\cdot x^{(4)}-x^{(2)}\cdot x^{(3)}\,.\]
It suffices to show that the Hermite rank of $f$ with respect to $X_1$ is 2 to conclude with Theorem 4 from \cite{arcones}. The Hermite rank is defined by
\begin{align}\label{rank}\operatorname{rank}(f):=\inf\Big\{\tau\ge 1|\sum_{j=1}^4l_j=\tau~\text{and}~\E\big[\big(f(X_1)-\E[f(X_1)]\big)\prod_{k=1}^4 H_{l_k}(X_1^{(k)})\big]\ne 0\Big\}\,,\end{align}
with $H_l,l\ge 0$, the Hermite polynomials. We have $\operatorname{rank}(f)>1$, since with $H_1(x)=x$ and $H_0=1$, it suffices to show
\[\E\big[\big(f(X_1)-\E[f(X_1)]\big)X_1^{(k)}\big]=0,~k=1,2,3,4.\]
For all $k=1,2,3,4$, the proof is similar, we consider without loss of generality $k=1$:
\begin{align*}\E\big[\big(f(X_1)-\E[f(X_1)]\big)X_1^{(1)}\big]&=\E\big[\big(X_1^{(1)}\cdot X_1^{(4)}-X_1^{(2)}\cdot X_1^{(3)}-\E[f(X_1)]\big)X_1^{(1)}\big]\\
&=\E\big[(X_1^{(1)})^2\cdot X_1^{(4)}\big]-\E\big[X_1^{(1)}\cdot X_1^{(2)}\cdot X_1^{(3)}\big]=0\,,
\end{align*}
where the last identity follows since $-X_1\stackrel{d}{=}X_1$, tantamount with the (trivial) odd case in Isserlis' theorem. To show $\operatorname{rank}(f)=2$, consider
\begin{align*}&\E\big[\big(f(X_1)-\E[f(X_1)]\big)\,X_1^{(1)}X_1^{(4)}\big]\\
&=\E\big[(X_1^{(1)}\cdot X_1^{(4)})^2\big]-\E\big[X_1^{(1)}\cdot X_1^{(2)}\cdot X_1^{(3)}\cdot X_1^{(4)}\big]-\Big(\E\big[X_1^{(1)}\cdot X_1^{(4)}\big]\Big)^2+\E\big[X_1^{(1)}\cdot X_1^{(4)}\big]\E\big[X_1^{(2)}\cdot X_1^{(3)}\big]\\
&\stackrel{\eqref{isserlis}}{=}1+\Big(\E\big[X_1^{(1)}\cdot X_1^{(4)}\big]\Big)^2-\E\big[X_1^{(1)}\cdot X_1^{(3)}\big]\E\big[X_1^{(2)}\cdot X_1^{(4)}\big]-\E\big[X_1^{(1)}\cdot X_1^{(2)}\big]\E\big[X_1^{(3)}\cdot X_1^{(4)}\big]\\
&=1+\rho^2\gamma_{1,2}^2(1)-\rho^2-\gamma_1(1)\gamma_2(1)\,.
\end{align*}
We make a case differentiation. If $\rho=0$, the above yields $1-\gamma_1(1)\gamma_2(1)>0$. If $H_1=H_2$, we have that $\gamma_1(1)=\gamma_2(1)=\gamma_{1,2}(1)=\gamma(1)$, what yields $(1-\rho^2)(1-\gamma^2(1))>0$. For the other case, it suffices to consider with $H_2(x)=x^2-1$ the expectation
\begin{align*}&\E\big[\big(f(X_1)-\E[f(X_1)]\big)\big(\big(X_1^{(1)}\big)^2-1\big)\big]\\
&\quad\quad=\E\big[(X_1^{(1)})^3\cdot X_1^{(4)}-X_1^{(1)}\cdot X_1^{(4)}+X_1^{(2)}\cdot X_1^{(3)}-(X_1^{(1)})^2\cdot X_1^{(2)}\cdot X_1^{(3)}\big]\\
&\quad\quad=3\E\big[(X_1^{(1)})^2\big]\E\big[X_1^{(1)} X_1^{(4)}\big]-\E\big[  X_1^{(1)} X_1^{(4)}\big]\\ &\quad\quad\quad+\E\big[X_1^{(2)} X_1^{(3)}\big]-2\E\big[X_1^{(1)} X_1^{(2)}\big]\E\big[X_1^{(1)} X_1^{(3)}\big]-\E\big[(X_1^{(1)})^2\big]\E\big[X_1^{(2)} X_1^{(3)}\big]\\
&\quad\quad=2\E\big[X_1^{(1)} X_1^{(4)}\big]-2\E\big[X_1^{(1)} X_1^{(2)}\big]\E\big[X_1^{(1)} X_1^{(3)}\big]\\
&\quad\quad=2\rho\big(\gamma_{1,2}(1)-\gamma_1(1)\big)\,,
\end{align*}
what is non-zero if $\rho\ne 0$, and $H_1\ne H_2$. The central limit theorem for $\hat\eta_{1,2}$, under the null hypothesis $\eta_{1,2}=0$, is hence implied by Slutsky's lemma. Since the map $(\rho,H_1,H_2)\mapsto \text{AVAR}_{\hat \eta_{1,2}}$ is continuous, what can be seen by dominated convergence, Corollary \ref{etatest} is hence implied by the convergence of the normalized realized volatilities in the denominator, continuous mapping for convergence in probability and Slutsky's lemma.

\subsection{Proof of Theorem \ref{theostat}}
Consider the variances of suitably normalized realized covariances
\begin{align*}
&\var\Big(\frac{\Delta^{-H_p-H_q}}{\sqrt{n}}\sum_{k=1}^n\Delta_kB^{(p)}\Delta_kB^{(q)}\Big)=\frac{\Delta^{-2(H_p+H_q)}}{n}\sum_{1\le k,l\le n}\cov\big(\Delta_kB^{(p)}\Delta_kB^{(q)},\Delta_lB^{(p)}\Delta_lB^{(q)}\big)\\
&\quad\stackrel{\eqref{isserlis}}{=}\frac{\Delta^{-2(H_p+H_q)}}{n}\sum_{1\le k,l\le n}\Big(\cov\big(\Delta_kB^{(p)},\Delta_lB^{(p)}\big)\cov\big(\Delta_kB^{(q)},\Delta_lB^{(q)}\big)\\
&\hspace*{4.2cm}+\cov\big(\Delta_kB^{(p)},\Delta_lB^{(q)}\big)\cov\big(\Delta_kB^{(q)},\Delta_lB^{(p)}\big)\Big)\\
&\quad\stackrel{\eqref{covincr}}{=}\frac{1}{n}\sum_{1\le k,l\le n}\Big(\sigma_p^2\sigma_q^2\gamma_p(k-l)\gamma_q(k-l)+\rho_{p,q}^2\sigma_p^2\sigma_q^2\gamma_{p,q}^2(k-l) \Big)\\
&\quad=\sigma_p^2\sigma_q^2\Big(\gamma_p(0)\gamma_q(0)+2\sum_{r=1}^n\frac{n-r}{n}\gamma_p(r)\gamma_q(r)+\rho_{p,q}^2\big(\gamma_{p,q}^2(0)+2\sum_{r=1}^n\frac{n-r}{n}\gamma^2_{p,q}(r)\big)\Big)\,,
\end{align*}
such that with dominated convergence, we obtain that
\begin{align*}\lim_{n\to\infty} \var\Big(\frac{\Delta^{-H_p-H_q}}{\sqrt{n}}\sum_{k=1}^n\Delta_kB^{(p)}\Delta_kB^{(q)}\Big)&=
\sigma_p^2\sigma_q^2(1+\rho_{p,q}^2)+2\sigma_p^2\sigma_q^2\sum_{r=1}^{\infty}\gamma_p(r)\gamma_q(r)\\
&\hspace*{2.75cm}+2\rho_{p,q}^2\sigma_p^2\sigma_q^2\sum_{r=1}^{\infty}\gamma_{p,q}^2(r)\,.
\end{align*}
The Hermite rank of $G:\R^2\to\R,G(x,y)=x\cdot y$, defined in general in Eq.\ (2.2) of \cite{arcones} as in \eqref{rank}, is 2. This is checked using that, for $(X,Y)^{\top}$ jointly centred normal, $\E[X^2Y^2]\ne 0$, while $\E[X^2Y]=0$. Therefore, the central limit theorem 
\begin{align*}%\label{cltqv}
\sqrt{n}\Big(\frac{\Delta^{-(H_p+H_q)}}{n}\sum_{k=1}^n\Delta_kB^{(p)}\Delta_kB^{(q)}-\rho_{p,q}\sigma_p\sigma_q\Big)\stackrel{d}{\rightarrow}\mathcal{N}\Big(0,\upsilon\Big)\end{align*}
is implied by \cite{arcones}, where
\begin{align*}\upsilon&=\sigma_p^2\sigma_q^2(1+\rho_{p,q}^2)+\frac{\sigma_p^2\sigma_q^2}{2}\Big(\rho^2_{p,q}\sum_{r=1}^{\infty}\big(|r+1|^{H_p+H_q}+|r-1|^{H_p+H_q}-2|r|^{H_p+H_q}\big)^2\\
&\hspace*{.5cm}+\sum_{r=1}^{\infty}\big(|r+1|^{2H_p}+|r-1|^{2H_p}-2|r|^{2H_p}\big)\big(|r+1|^{2H_q}+|r-1|^{2H_q}-2|r|^{2H_q}\big)\Big)\,.\end{align*}
Based on similar computations, we deduce the asymptotic covariances
\begin{align*}\lim_{n\to\infty} \cov\Big(\frac{\Delta^{-H_p-H_q}}{\sqrt{n}}\sum_{k=1}^n\Delta_kB^{(p)}\Delta_kB^{(q)},\frac{\Delta^{-2H_p}}{\sqrt{n}}\sum_{k=1}^n\big(\Delta_kB^{(p)}\big)^2\Big)&=2\sigma_p^3\sigma_q\rho\Big(1+2\sum_{r=1}^{\infty}\gamma_p(r)\gamma_{p,q}(r)\Big)
\end{align*}
between normalized realized variance and covariance, and between realized variances:
\begin{align*}\lim_{n\to\infty} \cov\Big(\frac{\Delta^{-2H_p}}{\sqrt{n}}\sum_{k=1}^n\big(\Delta_kB^{(p)}\big)^2,\frac{\Delta^{-2H_q}}{\sqrt{n}}\sum_{k=1}^n\big(\Delta_kB^{(q)}\big)^2\Big)&=2\rho^2\sigma_p^2\sigma_q^2\Big(1+2\sum_{r=1}^{\infty}\gamma_{p,q}^2(r)\Big)\,.
\end{align*}
Based on \cite{arcones} and Cram\'{e}r-Wold, analogously as above, we obtain a multivariate central limit theorem
\begin{align}\label{cltmatrix}\sqrt{n}\left(\begin{array}{c}\frac{\Delta^{-2H_1}}{n}\sum_{k=1}^n\big(\Delta_kB^{(1)}\big)^2-\sigma_1^2\\[.1cm] \frac{\Delta^{-2H_2}}{n}\sum_{k=1}^n\big(\Delta_kB^{(2)}\big)^2-\sigma_2^2\\[.1cm] \frac{\Delta^{-(H_1+H_2)}}{n}\sum_{k=1}^n\Delta_kB^{(1)}\Delta_kB^{(2)}-\rho\sigma_1\sigma_2\end{array}\right)\stackrel{d}{\rightarrow}\mathcal{N}\big(0,\operatorname{AV}\big)\,,
\end{align}
with the $(3\times 3)$ asymptotic variance-covariance matrix $\operatorname{AV}$, which contains the above given limiting variances and covariances. Since
\[\hat\rho=\frac{\frac{\Delta^{-(H_1+H_2)}}{n}\sum_{k=1}^n\Delta_kB^{(1)}\Delta_kB^{(2)}}{\sqrt{\frac{\Delta^{-2H_1}}{n}\sum_{k=1}^n\big(\Delta_kB^{(1)}\big)^2\frac{\Delta^{-2H_2}}{n}\sum_{k=1}^n\big(\Delta_kB^{(2)}\big)^2}}\,,\]
we derive \eqref{cltrho} from \eqref{cltmatrix} with the multivariate $\Delta$-method. With the function $h:(x,y,z)^{\top}\mapsto z/\sqrt{xy}$, $\nabla h(x,y,z)=\big(-yz/(2(xy)^{3/2}),-xz/(2(xy)^{3/2}),(xy)^{-1/2}\big)^{\top}$, we obtain
\begin{align*}\text{AVAR}_{\hat \rho}&=\big(\nabla h(\sigma_1^2,\sigma_2^2,\rho\sigma_1\sigma_2)\big)^{\top}\cdot\operatorname{AV}\cdot \nabla h(\sigma_1^2,\sigma_2^2,\rho\sigma_1\sigma_2)\\
&=\frac{\rho^2}{4\sigma_1^4}\operatorname{AV}_{\negthinspace 11}+\frac{\rho^2}{4\sigma_2^4}\operatorname{AV}_{\negthinspace 22}+\frac{\operatorname{AV}_{\negthinspace 33}}{\sigma_1^2\sigma_2^2}+\frac{\rho^2\operatorname{AV}_{\negthinspace 12}}{2\sigma_1^2\sigma_2^2}-\frac{\rho\operatorname{AV}_{\negthinspace 13}}{\sigma_1^3\sigma_2}-\frac{\rho\operatorname{AV}_{\negthinspace 23}}{\sigma_1\sigma_2^3}.
\end{align*}
Inserting the above limiting variances and covariances and simplifying the terms yields \eqref{cltrho}. This finishes the proof of Theorem \ref{theostat}.
 %the Hermite expansion of the square function \eqref{hermite} with Hermite rank 2, see also \cite[(A.6)]{power}

\section{Proof of Propositions} %(not for publication)}
\label{sec:propositions}
\subsection{Proof of Proposition \ref{forecastprop}}
Consider a multivariate normally distributed random vector
 \[ X = \left(\begin{array}{c}Y\\ Z\end{array}\right)\sim \mathcal{N}\bigg(
      \bigg(\begin{array}{c}\mu_Y\\[-.2cm] \mu_Z\end{array}\bigg), \bigg(\begin{array}{cc}\Sigma_Y & \Sigma_{Y,Z} \\[-.2cm] \Sigma_{Y,Z}^{\top} & \Sigma_Z \end{array}\bigg)\bigg) = \mathcal{N}(\mu, \Sigma),\]
which is $ \R^{d}$-valued with sub-vectors $Y$ taking values in  $ \R^{p}$ and $Z$ taking values in  $ \R^{d-p}$, $p,(d-p)\in\N$,
with marginals $Y \sim\mathcal{N}(\mu_Y, \Sigma_Y)$, $Z\sim \mathcal{N}(\mu_Z, \Sigma_Z)$, where \(\mu_Y \in \R^p\), \(\mu_Z \in \R^{d-p}\), \(\Sigma_Y \in \R^{p \times p}\), \(\Sigma_Z \in \R^{(d-p) \times (d-p)}\) and \(\Sigma_{Y,Z} \in \R^{p \times (d-p)}\). 
In this case, it is well known\footnote{See, for instance, \cite[Proposition 3.13]{eaton}.} that the conditional distribution of $Y$ given $Z=z$ is multivariate normal:
  \begin{align}\label{condexp}
    (Y \mid Z =  z) \sim \mathcal{N}\big(\mu_Y + \Sigma_{Y,Z} \Sigma_Z^{-1} (z -  \mu_Z),\Sigma_Y - \Sigma_{Y,Z} \Sigma_Z^{-1} {\Sigma^{\top}_{Y,Z}}\big).\end{align}
Having a Gaussian process, we apply this general result with $d=3$, $p=1$, to $Y=B_{t+h}^{(1)}$, with $\mu_Y=0$, $\Sigma_Y=\sigma_1^2(t+h)^{2H_1}$, and $Z=(B_t^{(1)},B_t^{(2)})^{\top}$, with $\mu_Z=(0,0)^{\top}$, and 
\[\Sigma_Z=\left(\begin{array}{cc}\sigma_1^2t^{2H_1} & \rho\sigma_{1}\sigma_2t^{2H} \\ \rho\sigma_{1}\sigma_2t^{2H} & \sigma_2^2t^{2H_2} \end{array}\right)\,,\]
with $H$ the cross Hurst exponent from \eqref{mcov}. The determinant 
\[\det(\Sigma_Z)=\sigma_1^2\sigma_2^2(1-\rho^2)t^{4H}\]
depends on $(H_1,H_2)$ only via $H$. We obtain that
\[\Sigma_Z^{-1}=\left(\begin{array}{cc}\big(\sigma_1^2(1-\rho^2)t^{2H_1}\big)^{-1} & -\rho\big(\sigma_1\sigma_2(1-\rho^2)t^{2H}\big)^{-1}  \\ -\rho\big(\sigma_1\sigma_2(1-\rho^2)t^{2H}\big)^{-1} & \big(\sigma_2^2(1-\rho^2)t^{2H_2}\big)^{-1}\end{array}\right)\,,\]
such that with $\Sigma_{Y,Z}=(\sigma_1^2 w(t,h,H_1),\rho\sigma_1\sigma_2 w(t,h,H))$, we obtain
\begin{align*}
\hat B_{t+h|t}^{(1)}&=\Sigma_{Y,Z} \Sigma_Z^{-1} z\\
&=(\sigma_1^2 w(t,h,H_1),\rho\sigma_1\sigma_2 w(t,h,H))\cdot \left(\begin{array}{c}\frac{B_t^{(1)}}{\sigma_1^2(1-\rho^2)t^{2H_1}}-\frac{\rho B_t^{(2)}}{\sigma_1\sigma_2(1-\rho^2)t^{2H}}\\ \frac{B_t^{(2)}}{\sigma_2^2(1-\rho^2)t^{2H_2}}-\frac{\rho B_t^{(1)}}{\sigma_1\sigma_2(1-\rho^2)t^{2H}}\end{array}\right)\,.
\end{align*} 
Computing this product proves \eqref{forecast}. The MSFE equals the conditional variance:
\begin{align*}&\E\Big[\big(\hat B_{t+h|t}^{(1)}-B_{t+h}^{(1)}\big)^2\Big]=\Sigma_Y-\Sigma_{Y,Z} \Sigma_Z^{-1} \Sigma_{Y,Z}^{\top}\\
&=\sigma_1^2\bigg((t+h)^{2H_1}-\frac{1}{1-\rho^2}\cdot\bigg( \frac{(w(t,h,H_1))^2}{t^{2H_1}}+\rho^2\bigg(\frac{2 w(t,h,H_1) w(t,h,H)}{t^{2H}}-\frac{(w(t,h,H))^2}{t^{2H_2}}\bigg)\bigg)\,.\hfill\qed
\end{align*}

\subsection{Proof of Proposition \ref{forecastprop2}}
With similar notation as in the proof of Proposition \ref{forecastprop}, we apply \eqref{condexp} with $Y=B_{t+h}^{(1)}$, and $Z=(B_t^{(1)},\ldots, B_t^{(d)})^{\top}$. The main step is to see that the $(d\times d)$ inverse of the covariance matrix $\Sigma_Z$ has entries
\[(\Sigma_Z^{-1})_{ij}=(\Sigma^{-1})_{ij}\,t^{-(H_i+H_j)}~,\,1\le i,j\le d\,.\]
First, $\det(\Sigma_Z)=\det(\Sigma)\cdot t^{2\textstyle\sum_{k=1}^dH_k}$ is verified by induction. For $d=3$, this follows readily with the rule of Sarrus. Then, a Laplace expansion, e.g., with respect to the first row, yields the induction step. Hence, the determinant depends in general only on the sum of all Hurst exponents. This yields for the inverse for $1\le i,j\le d$ that
\begin{align*}\big(\Sigma^{-1}_Z\big)_{ij}&=\frac{1}{\det(\Sigma_Z)}\operatorname{adj}(\Sigma_Z)=\frac{(-1)^{i+j}}{\det(\Sigma_Z)}\det\big((\Sigma_Z)_{-j,-i}\big)\\
&=\frac{(-1)^{i+j}}{\det(\Sigma)}t^{-2\textstyle\sum_{k=1}^dH_k}\det\big((\Sigma)_{-j,-i}\big)\,t^{\textstyle\sum_{k=1}^dH_k(1+\1(k\notin\{H_i,H_j\})}\\
&=(\Sigma^{-1})_{ij}\,t^{-(H_i+H_j)}\,,
\end{align*}
where we write $(\Sigma_Z)_{-j,-i}$ for the $((d-1)\times (d-1))$ matrix obtained from $\Sigma_Z$ by deleting the $j$th row and $i$th column, and $(\Sigma)_{-i,-j}$ analogously. We obtain the result from
\begin{align*}
\hat B_{t+h|t}^{(1)}&=\Sigma_{Y,Z} \Sigma_Z^{-1} \cdot (B_t^{(1)},\ldots, B_t^{(d)})^{\top}\,,\end{align*}
with $\Sigma_{Y,Z}=(\Sigma_{11}w(t,h,H_1),\ldots, \Sigma_{1d}w(t,h,(H_1+H_d)/2))$.\hfill\qed\\

\subsection{Proof of Proposition \ref{msedim}}
We use analogous notation as in the previous proofs. Based on similar linear algebra as above, we can prove inductively that $\Sigma^{-1}$ equals
\[\frac{1}{1+(d-2)\rho-(d-1)\rho^2}\left(\begin{array}{ccccc}1+(d-2)\rho&-\rho &-\rho&\ldots&-\rho\\ -\rho&1+(d-2)\rho&-\rho&\ldots&-\rho\\-\rho&\ldots&\ddots&\ldots&-\rho\\-\rho&\ldots&\ldots&1+(d-2)\rho&-\rho\\-\rho&-\rho&\ldots&-\rho&1+(d-2)\rho\end{array}\right).\]
Computing 
\begin{align*}
\big(w(t,h,H_1),\rho w(t,h,(H_1+H)/2),\ldots,\rho w(t,h,(H_1+H)/2)\big)\Sigma_Z^{-1}\left(\begin{array}{c}w(t,h,H_1)\\\rho w(t,h,(H_1+H)/2)\\ \vdots\\ \rho w(t,h,(H_1+H)/2)\end{array}\right)
\end{align*}
and re-using that $\big(\Sigma^{-1}_Z\big)_{ij}=(\Sigma^{-1})_{ij}\,t^{-(H_i+H_j)}$ yields the result.
\hfill\qed\\

\subsection{Proof of Proposition \ref{unifracprop}}
Since for $H=H_1=H_2$, it holds that 
\[\E\big[B_sB_t^{\top}\big]=w(t,s-t,H)\cdot \left(\begin{array}{cc}\sigma_1^2 & \rho\sigma_{1}\sigma_2 \\ \rho\sigma_{1}\sigma_2& \sigma_2^2\end{array}\right)\,,\]
the bivariate time-reversible fBm satisfies in this case the equality in distribution $B_t\stackrel{d}{=}A\cdot (W_t^{(1)},W_t^{(2)})^{\top}$, where $W_t^{(1)}$ and $W_t^{(2)}$ are two independent univariate fBms with Hurst exponents $H$, and
\[A= \left(\begin{array}{cc}\sigma_1 & 0 \\ \rho\sigma_2& \sqrt{1-\rho^2}\sigma_2\end{array}\right)~\mbox{with}~AA^{\top}=\left(\begin{array}{cc}\sigma_1^2 & \rho\sigma_{1}\sigma_2 \\ \rho\sigma_{1}\sigma_2& \sigma_2^2\end{array}\right)\,.\]
Since the covariance function uniquely determines the distribution of a centered Gaussian process, the equality is implied by that of the covariance functions. Let us point out that for $H_1\ne H_2$, it is impossible to illustrate $(B_t)$ as a linear transformation of independent fBms, unless all correlations between components of $(B_t)$ directly vanish.

The main step of the proof is now to show that for any $c,d,x,y\in\R$, we have that
\begin{align}\label{h1}(\sigma_1^2\,c,\rho\sigma_1\sigma_2\,c)A^{-\top}\left(\begin{array}{cc}d & 0 \\ 0& d\end{array}\right)A^{-1}\left(\begin{array}{c}x \\ y\end{array}\right)=c\cdot d\cdot x\,.\end{align}
We use the standard notation $A^{-\top}$ for $(A^{-1})^{\top}=(A^{\top})^{-1}$. Since the inverse
\begin{align*}A^{-1}=\left(\begin{array}{cc}\frac{1}{\sigma_1} & 0 \\ -\frac{\rho}{\sigma_1\sqrt{1-\rho^2}}& \frac{1}{\sigma_2\sqrt{1-\rho^2}}\end{array}\right)\,,\end{align*}
of the triangular matrix is easily determined, we obtain that
\begin{align*}&(\sigma_1^2\,c,\rho\sigma_1\sigma_2\,c)A^{-\top}\left(\begin{array}{cc}d & 0 \\ 0& d\end{array}\right)A^{-1}\left(\begin{array}{c}x \\ y\end{array}\right)=\\
&\quad=(\sigma_1^2\,c,\rho\sigma_1\sigma_2\,c)\left(\begin{array}{cc}\frac{1}{\sigma_1} & -\frac{\rho}{\sigma_1\sqrt{1-\rho^2}} \\[.25cm] 0& \frac{1}{\sigma_2\sqrt{1-\rho^2}}\end{array}\right)\left(\begin{array}{cc}\frac{d}{\sigma_1} & 0\\[.25cm] -\frac{d\rho}{\sigma_1\sqrt{1-\rho^2}}& \frac{d}{\sigma_2\sqrt{1-\rho^2}}\end{array}\right)\left(\begin{array}{c}x \\ y\end{array}\right)\\
&\quad=(\sigma_1^2\,c,\rho\sigma_1\sigma_2\,c)\left(\begin{array}{c}\Big(\frac{d}{\sigma_1^2}+\frac{d\rho^2}{\sigma_1^2(1-\rho^2)}\Big)x -\frac{d\rho}{\sigma_1\sigma_2(1-\rho^2)}y\\[.25cm]  -\frac{d\rho}{\sigma_1\sigma_2(1-\rho^2)}x+\frac{d}{\sigma_2^2(1-\rho^2)}y\end{array}\right)\\
&\quad=cd\Big(1+\frac{\rho^2}{1-\rho^2}\Big)x-\frac{cd\rho\sigma_1}{\sigma_2(1-\rho^2)}y-\frac{cd\rho^2}{1-\rho^2}x+\frac{cd\rho\sigma_1}{\sigma_2(1-\rho^2)}y\\
&\quad=cdx\,.
\end{align*}
Consider a forecast based on observations at two time points $t_1$ and $t_2$, being more general than equidistant ones. The covariance matrix $\mathfrak{W}$ of $(W_{t_1}^{(1)},W_{t_1}^{(2)},W_{t_2}^{(1)},W_{t_2}^{(2)})^{\top}$ contains due to the independence of the components four $(2\times 2)$-diagonal blocks:
\[\mathfrak{W}=\left(\begin{array}{cccc}(t_1)^{2H} & 0&w(t_1,t_2,H)&0 \\ 0&(t_1)^{2H} & 0&w(t_1,t_2,H)\\ w(t_1,t_2,H)&0&(t_2)^{2H}&0\\0&w(t_1,t_2,H)&0&(t_2)^{2H}\end{array}\right)\,,\]
and its inverse  $\mathfrak{W}^{-1}$ is of similar block-diagonal form. This and its explicit entries are determined in Lemma \ref{lemmatrix}, but the explicit entries are not required to conclude. Based on the linear transformation
\[(B_{t_1}^{(1)},B_{t_1}^{(2)},B_{t_2}^{(1)},B_{t_2}^{(2)})^{\top}\stackrel{d}{=}\left(\begin{array}{cc}A&0\\0&A\end{array}\right)(W_{t_1}^{(1)},W_{t_1}^{(2)},W_{t_2}^{(1)},W_{t_2}^{(2)})^{\top}\] 
the covariance matrix of $(B_{t_1}^{(1)},B_{t_1}^{(2)},B_{t_2}^{(1)},B_{t_2}^{(2)})^{\top}$ is given by 
\[\left(\begin{array}{cc}A&0\\0&A\end{array}\right)\mathfrak{W}\left(\begin{array}{cc}A^{\top}&0\\0&A^{\top}\end{array}\right)\,.\]
Writing  $\mathfrak{W}^{-1}$ with three $(2\times 2)$-diagonal blocks $D_1$, $D_2$ and $D_{12}$, the inverse covariance matrix of $(B_{t_1}^{(1)},B_{t_1}^{(2)},B_{t_2}^{(1)},B_{t_2}^{(2)})^{\top}$ yields
\[\left(\begin{array}{cc}A^{-\top}&0\\ 0&A^{-\top}\end{array}\right)\left(\begin{array}{cc}D_1&D_{12}\\ D_{12}&D_2\end{array}\right)\left(\begin{array}{cc}A^{-1}&0\\ 0&A^{-1}\end{array}\right)=\left(\begin{array}{cc}A^{-\top}D_1A^{-1}&A^{-\top}D_{12}A^{-1}\\ A^{-\top}D_{12}A^{-1}&A^{-\top}D_2A^{-1}\end{array}\right)\,.\]
Computing the optimal forecast based on the conditional expectation \eqref{condexp}, \eqref{h1} applies to all four blocks in the same way, such that the forecast does not contain the other component. The same $(2\times 2)$-diagonal block structure readily generalizes to arbitrarily many observation times, such that an algebraic induction step only requires an iterated application of \eqref{h1}. The induction hence yields the general result.\hfill\qed\\

To prove Proposition \ref{unifracgen}, we require the following algebraic lemma on inverses of block matrices of a specific form.
\begin{lem}\label{lemmatrix}
A positive definite, symmetric matrix $\mathfrak{W}\in\R^{nd\times nd}$ of the form
\begin{align*}\mathfrak{W}=\left(\begin{array}{cccc}w_{1,1}I_d&w_{1,2}I_d&\ldots&w_{1,n}I_d\\ \vdots& w_{2,2}I_d&\ldots&\vdots\\ \vdots&&\ddots&\\w_{1,n}I_d&w_{2,n}I_d&\ldots&w_{n,n}I_d \end{array}\right),~w_{i,j}\in\R,1\le i,j\le n,\end{align*}
where $I_d$ denotes the $(d\times d)$ identity matrix, has an inverse of the same form, i.e.\ there exist $\tilde w_{i,j}\in\R,1\le i,j\le n$, such that
\begin{align*}\mathfrak{W}^{-1}=\left(\begin{array}{cccc}\tilde w_{1,1}I_d&\tilde w_{1,2}I_d&\ldots&\tilde w_{1,n}I_d\\ \vdots& \tilde w_{2,2}I_d&\ldots&\vdots\\ \vdots&&\ddots&\\ \tilde w_{1,n}I_d&\tilde w_{2,n}I_d&\ldots&\tilde w_{n,n}I_d \end{array}\right).\end{align*}
\end{lem}

We prove the lemma by algebraic induction. For $n=2$, it is simple to verify by matrix multiplication that
\begin{align*}\left(\begin{array}{cc}w_{1,1}I_d&w_{1,2}I_d\\ w_{1,2}I_d&w_{2,2}I_d\end{array}\right)=\frac{1}{w_{1,1}w_{2,2}-w_{1,2}^2}\left(\begin{array}{cc}w_{2,2}I_d&-w_{1,2}I_d\\ -w_{1,2}I_d&w_{1,1}I_d\end{array}\right)\,,\end{align*}
what serves as the induction basis and yields an explicit formula for the one-period setting. If a positive definite, symmetric $\mathfrak{W}\in\R^{nd\times nd}$ is of the above form, write $\mathfrak{W}\in\operatorname{BD}(nd)$. Assume as induction hypothesis that for $\mathfrak{W}_{n-1}\in\operatorname{BD}((n-1)d)$, $n\ge 3$ arbitrary, it holds that  $\mathfrak{W}_{n-1}^{-1}\in\operatorname{BD}((n-1)d)$. To prepare the induction step, write for $\mathfrak{W}_n\in\operatorname{BD}(nd)$:
\begin{align*}\mathfrak{W}_n=\left(\begin{array}{cc}\mathfrak{W}_{n-1}&C\\ C^{\top}&w_{n,n}I_d\end{array}\right)\,,\end{align*}
with $w_{n,n}\in\R$, $C^{\top}=\big(w_{1,n}I_d,\ldots, w_{n-1,n}I_d\big)\in\R^{d\times (n-1)d}$. A general block inversion formula, see Sec.\ 9.3.1 in \cite{matrix}, yields
\begin{align*}\mathfrak{W}_n^{-1}=\left(\begin{array}{cc}\big(\mathfrak{W}_{n-1}-Cw_{n,n}^{-1}I_dC^{\top}\big)^{-1}&-\mathfrak{W}_{n-1}^{-1}C\big(w_{n,n}I_d-C^{\top}\mathfrak{W}_{n-1}^{-1}C\big)^{-1}\\ -\big(w_{n,n}I_d-C^{\top}\mathfrak{W}_{n-1}^{-1}C\big)^{-1}C^{\top}\mathfrak{W}_{n-1}^{-1}&\big(w_{n,n}I_d-C^{\top}\mathfrak{W}_{n-1}^{-1}C\big)^{-1}\end{array}\right)\,.\end{align*}
Since $C$ and $C^{\top}$ consist of $(d\times d)$-diagonal blocks in that $I_d$ is multiplied with real scalars, it follows that $C^{\top}\mathfrak{W}_{n-1}^{-1}C$ is a $(d\times d)$-diagonal matrix of the same form, such that $w_{n,n}I_d-C^{\top}\mathfrak{W}_{n-1}^{-1}C$ and its inverse are as well of this form. Since
\begin{align*}CC^{\top}=\left(\begin{array}{ccc}w_{1,n}^2I_d&\ldots&w_{1,n}w_{n-1,n}I_d\\ &\ddots&\\w_{1,n}w_{n-1,n}I_d&\ldots&w_{n-1,n}^2I_d\end{array}\right)\in\operatorname{BD}((n-1)d)\,,\end{align*}
$\mathfrak{W}_{n-1}-w_{n,n}^{-1}CC^{\top}$, and its inverse, are as well in $\operatorname{BD}((n-1)d)$. $C$ containing $(d\times d)$-diagonal blocks in that $I_d$ is multiplied with real scalars, and since $\mathfrak{W}_{n-1}^{-1}\in\operatorname{BD}((n-1)d)$, $\mathfrak{W}_{n-1}^{-1}C\big(w_{n,n}I_d-C^{\top}\mathfrak{W}_{n-1}^{-1}C\big)^{-1}$ preserves this structure, such that we conclude the induction step $\mathfrak{W}_{n}^{-1}\in\operatorname{BD}((n-1)d)$.

\subsection{Proof of Proposition \ref{unifracgen}}
In the general, $d$-dimensional setting with a mfBm $B_t=(B_t^{(1)},\ldots,B_t^{(d)})^{\top}$, with all Hurst exponents being $H\in(0,1)$, it holds that
\[\E\big[B_sB_t^{\top}\big]=w(t,s-t,H)\cdot \Sigma\,,\]
with the static, positive definite variance-covariance matrix $\Sigma\in\R^{d\times d}$ of $(B_1^{(1)},\ldots,B_1^{(d)})^{\top}$. By the spectral theorem, there is a diagonal matrix $\Lambda\in\R^{d\times d}$, with strictly positive entries on the diagonal, and an orthogonal matrix $O\in\R^{d\times d}$, such that
$\Sigma=O\Lambda O^{\top}$ and we set $\Sigma^{1/2}=O\Lambda^{1/2}O^{\top}$, where $\Lambda^{1/2}$ is obtained from $\Lambda$ taking square roots entry-wise. Similar as in the proof of Proposition \ref{unifracprop}, \eqref{mcov} yields an illustration
\begin{align*}B_t\stackrel{d}{=}\Sigma^{1/2}\cdot W_t\,,\end{align*}
with $W_t=(W_t^{(1)},\ldots,W_t^{(d)})^{\top}$ being a vector of $d$ independent fBms with equal Hurst exponents $H$. Considering $t$ observations in a vector $(B_{\Delta}^{(1)},\ldots,B_{\Delta}^{(d)},\ldots,B_{t\Delta}^{(1)},\ldots,B_{t\Delta}^{(d)})^{\top} \in\R^{td}$, this vector has the variance-covariance matrix $\mathfrak{C}\in\R^{td\times td}$ of the form
\begin{align*}\mathfrak{C}=BD_{\Sigma^{1/2}}\cdot \left(\begin{array}{cccc}\Delta^{2H}I_d&w(\Delta,2\Delta,H)&\ldots&w(\Delta,t\Delta,H)\\ w(\Delta,2\Delta,H)&(2\Delta)^{2H}I_d&\ldots&\vdots\\\vdots& &\ddots&\vdots\\w(\Delta,t\Delta,H)&\ldots&\ldots&(t\Delta)^{2H}I_d\end{array}\right)\cdot BD_{\Sigma^{1/2}}\,,\end{align*}
where $I_d$ denotes the $(d\times d)$ identity matrix and $BD_{\Sigma^{1/2}}$ the block-diagonal matrix
\begin{align*}BD_{\Sigma^{1/2}}=\left(\begin{array}{cccc}\Sigma^{1/2}&0&\ldots&0\\ 0&\Sigma^{1/2}&0\ldots&\vdots\\\vdots& &\ddots&0\\0&\ldots&0&\Sigma^{1/2}\end{array}\right)\,.\end{align*}
The inverse $\mathfrak{C}^{-1}\in\R^{td\times td}$, entering the conditional expectation of $B_{t\Delta+h}^{(1)}$, $h>0$, given the $t$ observations at times $j\Delta,1\le j\le t$, is with Lemma \ref{lemmatrix} hence of the form
\begin{align*}\mathfrak{C}^{-1}=\left(\begin{array}{cccc}\Sigma^{-1}c_{1,1}&\Sigma^{-1}c_{1,2}&\ldots&\Sigma^{-1}c_{1,t}\\ \Sigma^{-1}c_{1,2}&\Sigma^{-1}c_{2,2}&\ldots&\vdots\\\vdots& &\ddots&\\\Sigma^{-1}c_{1,t}&\ldots&\ldots&\Sigma^{-1}c_{t,t}\end{array}\right)\,,\end{align*}
with real scalars $c_{i,j}$, $1\le i,j\le t$, which depend on $\Delta$, $H$ and the observation times, but whose explicit values are not important.
With algebraic induction as in the proof of Proposition \ref{unifracprop}, it suffices to conclude block-wise that
\begin{align}\label{h2}c\cdot\Sigma_{1\bullet}\cdot\Sigma^{-1}\cdot d \cdot\left(\begin{array}{c}x_1\\ x_2\\\vdots\\ x_d\end{array}\right)=c\cdot d\cdot x_1\,,\end{align}
for arbitrary $c,d\in\R$, with $\Sigma_{1\bullet}\in\R^{1\times d}$ being the first row of $\Sigma$, entering the conditional expectation of $B_{t\Delta+h}^{(1)}$ given $(B_{\Delta}^{(1)},\ldots,B_{\Delta}^{(d)},\ldots,B_{t\Delta}^{(1)},\ldots,B_{t\Delta}^{(d)})^{\top}$. \eqref{h2} is readily obtained, since
\[\Sigma\cdot \Sigma^{-1}=I_d=\left(\begin{array}{c}\Sigma_{1\bullet}\cdot \Sigma^{-1}\\ \vdots\\ \vdots\end{array}\right)\,.\]
Similar as in the proof of Proposition \ref{unifracprop}, the main ingredient is that $\mathfrak{C}^{-1}$ only contains block matrices of the form $\Sigma^{-1}\cdot d$, $d\in\R$. The equidistant observation scheme, $t_j=j\Delta,1\le j\le t$, is not important for this proof which generalizes to arbitrary observation times.
\hfill\qed
%%%%%%%%%%%%%%%%%%%%%%%%%%%%%%%%%%%%%%%%%%%%%%%%%%%%%%%%%%%%%%%%%%%%%%%%%%%%%%%%%%%%%%%%%%%%%%%%%%%%%
\section{Monte Carlo Study}
\label{sec:monte_carlo}
\subsection{Finite-sample performance of proposed estimators}
We first examine the finite-sample performance of the proposed estimators. Since our estimators are derived either component-wise or pairwise, we consider the bivariate time-reversible fBm without loss of generality. Two empirically realistic parameter settings in the rough region are considered: $\{H_1,H_2,\sigma_1^2,\sigma_2^2,\rho_{12}\}=\{0.1, 0.4,1,1, 0\}$, $\{0.1, 0.4,1,1, 0.4\}$. We set the sample size to $n = 500$, $1000$, $\Delta=1/52$, $1/250$, and conduct $1000$ replications.

Tables \ref{finite_per_a}-\ref{finite_per_b} report the bias, the standard deviation, and the root mean squared error (RMSE) of the estimators across replications, with the standard error implied by our asymptotic theory in parentheses. The proposed method performs well and the empirical standard deviation is accurately predicted by the asymptotic theory.

We next evaluate the finite-sample performance of our proposed estimators in comparison with those from \cite{amblard2011identification}. As our estimators are consistent for general mfBm, we can evaluate their performance when $\eta_{1,2}\ne 0$. We focus on the bivariate case, as \cite{amblard2011identification} noted that estimating $\eta_{1,2}$ was "very difficult" with their method. If their estimator for $\eta_{1,2}$ performs poorly in the bivariate case, it is unlikely to improve in higher dimensions, whereas our estimators are dimension-independent. Four parameter configurations are considered: $\{H_1,H_2,\sigma_1^2,\sigma_2^2,\rho_{12}, \eta_{1,2}\}=\{0.1, 0.4,1,1, 0.2, 0\}$, $\{0.1, 0.4,1,1, 0.4, 0\}$, $\{0.1, 0.4,1,1, 0.2, 0.5\}$, $\{0.1, 0.4,1,1, 0.4, 0.5\}$. The sample size is set to $n=500, 1000$, and the sample interval $\Delta$ to $1/250$. For clarity, BYZ denote our estimators and AC those from \cite{amblard2011identification}. The AC estimators are implemented using Coeurjolly's \texttt{SimEstFBM.R} package with its default configuration.

\begin{table}[H]
\centering
\caption{ Bias, standard error (asymptotic standard errors in parentheses) and RMSE of our method: $H_1=0.1$, $H_2=0.4$, $\sigma_1^2=1$,  $\sigma_2^2=1$, $\rho=0$ and $\eta_{1,2}=0$ (denoted by $\eta$ in the table).}
\scalebox{0.75}{
\begin{tabular}{cccccccc}
   \hline \hline
                                    &  & $H_1$           & $H_2$           & $\sigma_1^2$      & $\sigma_2^2$      & $\rho$          & $\eta$          \\
                                    \hline
 \multirow{4}{*}{$n=500,\Delta=1/52$}   & Bias                 & 0.0013  & -0.0012 & 0.0567 & 0.0222 & 0.0015  & 0.0003  \\
                                    & \multirow{2}{*}{Std} & 0.0441  & 0.0356  & 0.3449 & 0.2934 & 0.0482  & 0.1192  \\
                                    &                      &   (0.0431)      &   (0.0351)      &   (0.3404)     & (0.2774)       &     (0.0472)    &   (0.1137)      \\
                                    & RMSE                 & 0.0441  & 0.0357  & 0.3496 & 0.2943 & 0.0483  & 0.1192  \\\hline
\multirow{4}{*}{$n=1000,\Delta=1/52$}  & Bias                 & -0.0009 & 0.0001  & 0.0144 & 0.0156 & -0.0005 & -0.0041 \\
                                    & \multirow{2}{*}{Std} & 0.0293  & 0.0248  & 0.2246 & 0.1977 & 0.0334  & 0.0788  \\
                                    &                      &     (0.0305)    &   (0.0248)      &    (0.2407)    &    (0.1962)    &    (0.0334)     &    (0.0804)     \\
                                    & RMSE                 & 0.0293  & 0.0248  & 0.2250 & 0.1984 & 0.0334  & 0.0789 \\\hline
\multirow{4}{*}{$n=500,\Delta=1/250$}  & Bias                 & -0.0014 & -0.0025 & 0.0837 & 0.0445 & -0.0011 & -0.0017 \\
                                    & \multirow{2}{*}{Std} & 0.0430  & 0.0353  & 0.5132 & 0.4075 & 0.0480  & 0.1109  \\
                                    &                      &     (0.0431)    &   (0.0351)      &   (0.4756)     &   (0.3876)     &   (0.0472)      &     (0.1137)    \\
                                    & RMSE                 & 0.0430  & 0.0354  & 0.5203 & 0.4101 & 0.0480  & 0.1110  \\\hline
\multirow{4}{*}{$n=1000,\Delta=1/250$} & Bias                 & -0.0005 & 0.0001  & 0.0399 & 0.0366 & -0.0015 & 0.0022  \\
                                    & \multirow{2}{*}{Std} & 0.0301  & 0.0246  & 0.3285 & 0.2800 & 0.0333  & 0.0759  \\
                                    &                      &      (0.0305)   &     (0.0248)    & (0.3363)       &    (0.2741)    &  (0.0334)       &   (0.0804)      \\
                                    & RMSE                 & 0.0301  & 0.0246  & 0.3309 & 0.2824 & 0.0334  & 0.0760  \\
                                    \hline \hline
\end{tabular}}
\label{finite_per_a}
\end{table}

%This improvement is notable given \cite{amblard2011identification}'s own observation that estimating $\eta_{1,2}$ was ``very difficult to estimate, at least with the method adopted here''. Our approach successfully overcomes this estimation challenge.

Table \ref{sec:finite_per_c} presents the bias, standard deviation, and RMSE of both estimators across 1000 replications. Our estimator for $\rho$ shows performance comparable to that of \cite{amblard2011identification}, while our estimator for $\eta_{1,2}$ demonstrates substantially improved accuracy, particularly when $\eta_{1,2} = 0$. For instance, when $\rho = 0.4$, $\eta_{1,2} = 0$, and $n = 500$, the RMSE ratio for $\rho$ is 100.3\%, indicating nearly identical performance, whereas the ratio for $\eta_{1,2}$ is 34.8\%, clearly favoring our estimator. Real-data results echo this scenario where $\eta_{1,2} = 0$, providing further support for the practical relevance of our estimation approach. The results in Online Supplement \ref{generalestimation} show that our optimal GMM estimator for $\rho$ achieves moderately higher accuracy. 
 
\begin{table}[H]
\centering
\caption{Bias, standard error (asymptotic standard errors in parentheses) and RMSE of our method: $H_1=0.1$, $H_2=0.4$, $\sigma_1^2=1$,  $\sigma_2^2=1$, $\rho=0.4$ and $\eta_{1,2}=0$ (denoted by $\eta$ in the table).}
\scalebox{0.75}{
\begin{tabular}{cccccccc}
\hline \hline
                                          &  & $H_1$           & $H_2$           & $\sigma_1^2$      & $\sigma_2^2$      & $\rho$          & $\eta$          \\
                                       \hline
\multirow{4}{*}{$n=500,\Delta=1/52$}   & Bias                 & -0.0029 & -0.0011 & 0.0214 & 0.0206 & 0.0006  & 0.0024  \\
                                    & \multirow{2}{*}{Std} & 0.0420  & 0.0341  & 0.3285 & 0.2738 & 0.0402  & 0.1019  \\
                                    &                      &      (0.0431)   &   (0.0351)      & (0.3404)       &     (0.2774)   &   (0.0394)      &      (0.1036)   \\
                                    & RMSE                 & 0.0421  & 0.0341  & 0.3294 & 0.2747 & 0.0402  & 0.1019  \\\hline
\multirow{4}{*}{$n=1000,\Delta=1/52$}  & Bias                 & -0.0007 & -0.0011 & 0.0157 & 0.0084 & -0.0006 & -0.0014 \\
                                    & \multirow{2}{*}{Std} & 0.0300  & 0.0250  & 0.2225 & 0.1990 & 0.0284  & 0.0746  \\
                                    &                      &     (0.0305)    &   (0.0248)      &  (0.2407)      &    (0.1962)    &     (0.0279)    &  (0.0733)       \\
                                    & RMSE                 & 0.0300  & 0.0250  & 0.2230 & 0.1991 & 0.0284  & 0.0746  \\\hline
\multirow{4}{*}{$n=500,\Delta=1/250$}  & Bias                 & -0.0008 & -0.0005 & 0.0868 & 0.0726 & 0.0014  & 0.0041  \\
                                    & \multirow{2}{*}{Std} & 0.0421  & 0.0353  & 0.4956 & 0.4242 & 0.0388  & 0.1080  \\
                                    &                      &   (0.0431)      &   (0.0351)      &     (0.4756)   &      (0.3877)  &   (0.0394)      &     (0.1036)    \\
                                    & RMSE                 & 0.0421  & 0.0354  & 0.5034 & 0.4306 & 0.0389  & 0.1081  \\\hline
\multirow{4}{*}{$n=1000,\Delta=1/250$} & Bias                 & -0.0032 & 0.0003  & 0.0149 & 0.0387 & -0.0013 & 0.0000  \\
                                    & \multirow{2}{*}{Std} & 0.0306  & 0.0250  & 0.3275 & 0.2889 & 0.0280  & 0.0744  \\
                                    &                      &      (0.0305)       &   (0.0248)      &  (0.3363)      &    (0.2741)    &    (0.0279)     &   (0.0733)      \\
                                    & RMSE                 & 0.0308  & 0.0250  & 0.3279 & 0.2915 & 0.0280  & 0.0744 \\
                                    \hline\hline
\end{tabular}}
\label{finite_per_b}
\end{table}

%Notably, when $\rho=0$, an interesting contrast emerges: while the estimator from \cite{amblard2011identification} exhibits smaller variance, our estimator achieves significantly reduced bias. 
For both parameters ($\rho$ and $\eta_{1,2}$), we find that when their true values are near zero—a scenario explicitly excluded in the construction of \cite{amblard2011identification}'s estimators—their estimators exhibit substantial bias. Although their correlation estimator maintains reasonable RMSE performance, their asymmetry estimates suffer from severe bias. Overall, our estimators deliver smaller biases and markedly improved statistical inference for $\eta_{1,2}$ compared to \cite{amblard2011identification}.

\begin{table}[H]
\centering
\caption{Bias, standard error (asymptotic standard errors in parentheses), and RMSE of 2 methods for different combinations of $\rho$ and $\eta_{1,2}$ (denoted by $\eta$ in the table), with $H_1=0.1$, $H_2=0.4$, $\sigma_1^2=1$ and $\sigma_2^2=1$.}
\scalebox{0.75}{
\begin{tabular}{c|c|cc|cc|cc|cc}
\hline \hline
          &      & \multicolumn{4}{c|}{$n=500$}    &  \multicolumn{4}{c}{$n=1000$}   \\ \hline
          &      & \multicolumn{2}{c|}{BYZ}    &  \multicolumn{2}{c|}{AC} & \multicolumn{2}{c|}{BYZ}    &  \multicolumn{2}{c}{AC}  \\ \hline
                                &      &$\rho$   & $\eta$     &   $\rho$     & $\eta$  &$\rho$   & $\eta$     &   $\rho$     & $\eta$   \\
                                \hline
				 & Bias & -0.0002& 0.0019& 0.0045 &0.3078 & -0.0002&-0.0041  & -0.0025&0.2068   \\
{$\rho=0.2$}  & Std  & 0.0456&0.1146 &0.0477 &0.1462 & 0.0322 & 0.0764   &  0.0320 & 0.0971                  \\
$\eta=0$& RMSE &  0.0456 & 0.1146   & 0.0479& 0.3408 & 0.0322 &0.0765   & 0.0321 &0.2284                         \\\hline
          & Bias & -0.0022&0.0005    & -0.0026&0.2843  & -0.0012&-0.0018&  -0.0014&0.1936\\
$\rho=0.4$& Std  & 0.0381 & 0.1093   & 0.0380 &0.1407 &  0.0276 &0.0735 &   0.0269 &0.0950                                      \\
$\eta=0$  & RMSE & 0.0382& 0.1093  &  0.0381 &0.3173&  0.0276 &0.0735  & 0.0270 &0.2157                            \\\hline
         & Bias & 0.0004&0.0116 &-0.0033  &-0.0541 & 0.0000&0.0012  & -0.0019&-0.0864   \\
{$\rho=0.2$} & Std  & 0.0444&0.1357 &0.0436 &0.1946 & 0.0314 & 0.0894   &  0.0295 & 0.1479                  \\
$\eta=0.5$ & RMSE & 0.0444  & 0.1362   &0.0437 & 0.2020 & 0.0314 &0.0894   & 0.0295 &0.1713                           \\\hline
          & Bias &0.0017&0.0051   & -0.0011&-0.0573 & 0.0006&0.0037 & -0.0013&-0.0852\\
$\rho=0.4$& Std  & 0.0372 &0.1279   & 0.0358& 0.1835 & 0.0269 & 0.0874  & 0.0253 & 0.1468\\
$\eta=0.5$& RMSE & 0.0372& 0.1280   &  0.0358& 0.1924& 0.0269 & 0.0875  &  0.0253 & 0.1697\\            \hline \hline                              
\end{tabular}}
\label{sec:finite_per_c}
\end{table}

\subsection{Size and power of the test}
\label{sec:power}

To examine the size and the power of the proposed test from Corollary \ref{etatest}, we consider the following parameter configurations, which are empirically relevant: \( H_1 = 0.1 \), \( H_2 = 0.4 \), \( \rho = 0.4 \), \( \Delta = 1/250 \), \( \sigma_1^2 = 1 \), and \( \sigma_2^2 = 1 \). The true value for \( \eta_{1,2}\) varies over the interval \([0, 0.65]\). The sample sizes are set to \( 500 \) and \( 1000 \), and the number of replications is \( 5000 \). The test has good size when $\eta_{1,2}=0$ with the nominal sizes close to the real sizes in finite sample. Moreover, testing at both 1\% and 5\% significance levels, we find the procedure maintains good finite-sample properties (Table \ref{sec:finite_per_d}). Consistent with theoretical expectations, the test's power grows monotonically as the true $\eta_{1,2}$ value deviates further from zero and gets larger for a larger sample size.

\begin{table}[H]
\centering
\caption{Size and power of the test for $\eta_{1,2}$ (denoted by $\eta$ in the table).}
\scalebox{0.75}{
\begin{tabular}{ccccccccc}
\hline \hline
       & $\eta=0$  & $\eta=0.1$ & $\eta=0.2$ & $\eta=0.3$ & $\eta=0.4$ & $\eta=0.5$ & $\eta=0.6$ & $\eta=0.65$ \\
       \hline 
       \multicolumn{9}{c}{\( 1\% \) significance level } \\
       \hline
$n=500$  &   0.0120  & 0.0616  & 0.2608  & 0.6214  & 0.8852  & 0.9856&0.9986  & 0.9998   \\
$n=1000$ & 0.0122 & 0.1074  & 0.5556  & 0.9230  & 0.9978  & 1.0000  & 1.0000  & 1.0000  \\
       \hline 
       \multicolumn{9}{c}{\( 5\% \) significance level } \\
       \hline
$n=500$  &   0.0582  & 0.1610  & 0.4806  & 0.8126  &  0.9634  & 0.9972  & 1.0000  &1.0000 \\
$n=1000$ & 0.0526 & 0.2742  & 0.7736  & 0.9772  &  0.9996  & 1.0000  & 1.0000  & 1.0000  \\
       \hline
\hline \hline
\end{tabular}}
\label{sec:finite_per_d}
\end{table}

\subsection{Finite-sample performance of optimal forecasts}
This subsection examines potential improvements in out-of-sample forecasting accuracy through mfBm modeling. While Section \ref{sec:4.1} established theoretical foundations through Propositions \ref{forecastprop}-\ref{msedim} under simplified conditions, we now conduct numerical simulations to assess forecasting performance in empirically realistic settings. We design four Monte Carlo experiments. In all experiments, we generate forecasts based on the first 500 observations ($n = 500$) sampled at frequency $\Delta=1/250$. Through 10,000 Monte Carlo replications, we compute the root MSFE (RMSFE). Each replication simulates 505 observations to accommodate forecast horizons $h \in\{1,\ldots,5\}$. To isolate the pure forecasting gains from estimation effects, we assume that $\Sigma$ and $H$ are known in these simulations. 

The first Monte Carlo experiment is designed to investigate how forecast gains vary with inter-component correlation. Using simulated data from a bivariate time-reversible fBm with Hurst exponents $H_1=0.1$, $H_2=0.4$, we evaluate three distinct correlation structures for $\Sigma$ (i.e., $\rho=0,0.4,0.8$):
 \[ \begin{bmatrix}
1&0\\
0&1
\end{bmatrix}, \begin{bmatrix}
1&0.4\\
0.4&1
\end{bmatrix}\, \text{and}\,\begin{bmatrix}
1&0.8\\
0.8&1
\end{bmatrix}. \]

The results, presented in Table \ref{simu_com_1}, reveal two key findings. First, incorporating the second (correlated) component in bivariate fBm (bfBm) yields significant improvements over univariate fBm forecasting. Second, these gains increase monotonically with the absolute correlation strength. As expected, forecasts coincide when $\rho=0$ due to component independence. For $\rho=0.4$, we observe moderate efficiency gains, while $\rho=0.8$ demonstrates the potential for substantial RMSFE reduction (up to 15.4\% in our simulations).

\begin{table}[H]
\centering
\caption{RMSFEs of $h$-day-ahead forecasts for $B^{(i)}_{(t+h)\Delta}$, with and without taking the other component into account. The theoretical RMSFEs are in the brackets.}
	\scalebox{0.65}{
\begin{tabular}{c|ccccc|ccccc}
\hline
\hline
&$h=1$&$h=2$&$h=3$&$h=4$&$h=5$&$h=1$&$h=2$&$h=3$&$h=4$&$h=5$ \\
\hline\hline
 &\multicolumn{5}{c}{$i=1$} &\multicolumn{5}{|c}{$i=2$} \\\hline
&\multicolumn{10}{c}{$Corr=0$}\\\hline
 \multirow{2}{*}{fBm\&bfBm} & .4845 & .5084 & .5237 & .5389 & .5501& .1093 & .1441 & .1683 & .1881 & .2040 \\           
                      & (.4802) & (.5077) & (.5254) & (.5387) & (.5495)& (.1085) & (.1430) & (.1681) & (.1886) & (.2061) \\ 
\hline
&\multicolumn{10}{c}{$Corr=0.4$}\\\hline
 \multirow{2}{*}{fBm} & .4832 & .5119 & .5239 & .5370 & .5551& .1096 & .1433 & .1689 & .1899 & .2080 \\           
 & (.4802) & (.5077) & (.5254) & (.5387) & (.5495) & (.1085) & (.1430) & (.1681) & (.1886) & (.2061) \\ \hline
 \multirow{2}{*}{bfBm}  & .4794 & .5080 & .5203 & .5337 & .5517& .1088 & .1424 & .1680 & .1888 & .2066 \\           
 & (.4756) & (.5035) & (.5213) & (.5348) & (.5456) & (.1075) & (.1417) & (.1666) & (.1869) & (.2043)\\ \hline
&\multicolumn{10}{c}{$Corr=0.8$}\\\hline
   \multirow{2}{*}{fBm}  & .4839 & .5044 & .5220 & .5370 & .5492& .1085 & .1420 & .1683 & .1883 & .2067 \\           
  &(.4802) & (.5077) & (.5254) & (.5387) & (.5495)& (.1085) & (.1430) & (.1681) & (.1886) & (.2061) \\\hline
\multirow{2}{*}{bfBm} & .4286 & .4502 & .4698 & .4830 & .4940& .0946 & .1224 & .1441 & .1593 & .1734 \\          
  &(.4246) & (.4526) & (.4700) & (.4827) & (.4927)& (.0953) & (.1242) & (.1443) & (.1602) & (.1734) \\
\hline \hline
\end{tabular}}
\label{simu_com_1}
\end{table}

The second experiment is designed to examine how gains depend on the differences between Hurst exponents. We fix $\sigma_1^2=\sigma_2^2=1$, $\rho=0.4$, and allow $(H_1,H_2)=$ (0.1, 0.1), (0.1, 0.2) and (0.1, 0.4). The RMSFE is computed from 10,000 replications. Results are given in Table \ref{simu_com_2}. Consistent with our theoretical insights, gains become more pronounced for increasing differences between the idiosyncratic Hurst exponents. 

\begin{table}[H]
\centering
\caption{RMSFEs of $h$-day-ahead forecasts for $B_{(t+h)\Delta}^{(i)}$, with and without taking the second component into account. The theoretical RMSFEs are in the brackets.}
	\scalebox{0.65}{
\begin{tabular}{c|ccccc|ccccc}
\hline
\hline
&$h=1$&$h=2$&$h=3$&$h=4$&$h=5$&$h=1$&$h=2$&$h=3$&$h=4$&$h=5$ \\
\hline\hline
 &\multicolumn{5}{c}{$i=1$} &\multicolumn{5}{|c}{$i=2$} \\\hline
 &\multicolumn{10}{c}{$H_1=0.1, H_2=0.1$}\\\hline
 \multirow{2}{*}{fBm\&bfBm} & .4754 & .5103 & .5348 & .5396 & .5533 & .4847 & .5143 & .5321 & .5356 & .5469\\           
 &  (.4802) & (.5077) & (.5254) & (.5387) & (.5495)& (.4802) & (.5077) & (.5254) & (.5387) & (.5495) \\ \hline
 & \multicolumn{10}{c}{$H_1=0.1, H_2=0.2$}\\\hline
 \multirow{2}{*}{fBm}  & .4818 & .5098 & .5226 & .5375 & .5495 & .3019  &  .3425  &  .3670  &  .3904  &  .4059\\           
&  (.4802) & (.5077) & (.5254) & (.5387) & (.5495) &(.2999) &   (.3411) &   (.3682)  & (.3890)   & (.4061) \\\hline
\multirow{2}{*}{bfBm}  & .4810 & .5093 & .5217 & .5370 & .5492& .3013 &   .3420 &   .3665  &  .3900   & .4055 \\           
&  (.4795) & (0.5071) & (.5249) & (.5382) & (.5490)  & (.2995)  &   (.3407)  &   (.3679)  &   (.3887)  &   (.4058)\\ \hline
&  \multicolumn{10}{c}{$H_1=0.1, H_2=0.4$}\\\hline
  \multirow{2}{*}{fBm}   & .4832 & .5119 & .5239 & .5370 & .5551& .1096 & .1433 & .1689 & .1899 & .2080 \\           
& (.4802) & (.5077) & (.5254) & (.5387) & (.5495)& (.1085) & (.1430) & (.1681) & (.1886) & (.2061) \\\hline
 \multirow{2}{*}{bfBm}  & .4794 & .5080 & .5203 & .5337 & .5517& .1088 & .1424 & .1680 & .1888 & .2066 \\          
& (.4756) & (.5035) & (.5213) & (.5348) & (.5456)& (.1075) & (.1417) & (.1666) & (.1869) & (.2043) \\
\hline \hline
\end{tabular}}
\label{simu_com_2}
\end{table}

The third experiment is designed to examine gains when including more components of mfBm. Assume there are three components with  $H_1=0.1$, $H_2=H_3=0.4$, and 
\[\Sigma_1= \begin{bmatrix}
1&0.4&0.4\\
0.4&1&0\\
0.4&0&1\\
\end{bmatrix}.\]

We evaluate three forecasting approaches: univariate fBm, bfBm, and three-dimensional mfBm (mfBm3). The forecasting results are reported in Table \ref{simu_com_3}. As expected from the theoretical insights, including more correlated components further increases the forecasting performance. By further incorporating an additional component that is correlated only with the first component, with $H_4=0.4$, such that
\[\Sigma_2=\begin{bmatrix}
1&w_{0.4}^{\top}\\
w_{0.4}&I_3\\
\end{bmatrix},~\text{with}~I_3~\text{the $(3\times 3)$ identity matrix}~\text{and}~w_{0.4}=(0.4,0.4,0.4)^{\top},\]  
the forecasting error will be reduced accordingly.

\begin{table}[H]
\centering
\caption{RMSFEs of $h$-day-ahead forecasts for $B_{(n+h)\Delta}^{(1)}$, with and without taking other components into account. The theoretical RMSFEs are in the brackets.}
\scalebox{0.7}{
\begin{tabular}{c|ccccc|ccccc}
\hline
\hline
$h$&$1$&$2$&$3$&$4$&$5$&$1$&$2$&$3$&$4$&$5$ \\
\hline
 \hline
 &\multicolumn{5}{c|}{$\Sigma=\Sigma_1$}&\multicolumn{5}{c}{$\Sigma=\Sigma_2$}\\\hline
  \multirow{2}{*}{fBm} & .4784 & .5025 & .5213 & .5361 & .5458& .4833 & .5081 & .5293 & .5359 & .5463 \\          
 & (.4802) & (.5077) & (.5254) & (.5387) & (.5495)& (.4802) & (.5077) & (.5254) & (.5387) & (.5495) \\\hline
 \multirow{2}{*}{bfBm} & .4740 & .4982 & .5177 & .5328 & .5427 & .4783 & .5040 & .5248 & .5318 & .5421\\          
 & (.4756) & (.5035) & (.5213) & (.5348) & (.5456) & (.4756) & (.5035) & (.5213) & (.5348) & (.5456) \\\hline
 \multirow{2}{*}{mfBm3} & .4656 & .4919 & .5108 & .5266 & .5352& .4697 & .4970 & .5176 & .5249 & .5351 \\          
 & (.4686) & (.4969) & (.5150) & (.5286) & (.5396)& (.4686) & (.4969) & (.5150) & (.5286) & (.5396) \\\hline
\multirow{2}{*}{mfBm4}  &&&&&& .4586 & .4852 & .5053 & .5136 & .5225 \\          
&&&&&& (.4563) & (.4851) & (.5035) & (.5173) & (.5284) \\
\hline \hline
\end{tabular}}
\label{simu_com_3}
\end{table}

The fourth experiment considers a setting that explicitly accounts for estimation errors. 
We examine two scenarios: (1) independence and (2) moderate correlation with $\rho = 0.4$. 
The parameters are set to $H_1 = 0.2$, $H_2 = 0.4$, $\sigma_1^2 = 1$, and $\sigma_2^2 = 1$. 
The value of $H_1$ is increased to avoid negative estimates of $H$ when applying the estimators in 
\eqref{sigmaest}, \eqref{Hest}, \eqref{rhohat}, and \eqref{etahat}. The results reported in Table \ref{simu_com_4} indicate that, compared to the known-parameter case, estimation errors reduce forecasting performance. Nevertheless, also in the presence of estimation errors, the bivariate fBm yields more accurate forecasts when the correlation is nonzero.

\begin{table}[H]
\centering
\caption{RMSFEs of $h$-day-ahead forecasts for $B^{(i)}_{(t+h)\Delta}$, with taking estimation errors into account. The label Known indicates that the parameters are assumed to be known, whereas the label Unknown indicates that the parameters need be estimated.}
	\scalebox{0.65}{
\begin{tabular}{c|c|ccccc|ccccc}
\hline
\hline
&&$h=1$&$h=2$&$h=3$&$h=4$&$h=5$&$h=1$&$h=2$&$h=3$&$h=4$&$h=5$ \\
\hline\hline
 &&\multicolumn{5}{c}{$i=1$} &\multicolumn{5}{|c}{$i=2$} \\\hline
&&\multicolumn{10}{c}{$Corr=0$}\\\hline
 \multirow{2}{*}{fBm} &Known&.2952  &  .3394    &.3668  &  .3851  &  .4043&   .1086 &   .1444  &  .1692&    .1915  &  .2078 \\       
     &Unknown&  .2963 &   .3412  &  .3685&    .3868 &   .4058&     .1085  &  .1446   & .1695  &  .1918   & .2081\\    
                   & Theoretical  & .2999 &   .3411  &  .3682  &  .3890  &  .4061&  .1085 &    .1430  &   .1681 &    .1886   &  .2061\\ \hline
\multirow{2}{*}{bfBm} & Known& .2952  &  .3394    &.3668  &  .3851  &  .4043& .1086 &   .1444  &  .1692&    .1915  &  .2078 \\        
     &Unknown&  .2963 &   .3412  &  .3685&    .3868 &   .4059&   .1085  &  .1446   & .1695  &  .1918   & .2081 \\       
                    &  Theoretical &.2999 &   .3411  &  .3682  &  .3890  &  .4061& .1085 &    .1430  &   .1681 &    .1886   &  .2061 \\ 
\hline
&&\multicolumn{10}{c}{$Corr=0.4$}\\\hline
 \multirow{2}{*}{fBm} &Known&    .2999  &  .3406  &  .3680 &   .3868  &  .4049 &    .1091  &  .1420 &   .1690  &  .1877  &  .2066\\    
 &Unknown&  .3008  &  .3421    &.3691  &  .3883   & .4063 &    .1091 &   .1420 &   .1692 &   .1878  &  .2068\\  
 &Theoretical &   .2999 &   .3411  &  .3682   & .3890  &  .4061 & .1085 & .1430 & .1681 & .1886 & .2061 \\ \hline
 \multirow{2}{*}{bfBm}  &Known & .2984   & .3394&    .3666  &  .3860   & .4036&   .1087 &   .1416&    .1684  &  .1871  &  .2059 \\
 & Unknown &  .2995  &  .3411  &  .3677  &  .3877  &  .4052   & .1089 &   .1418&    .1687  &  .1875  &  .2063 \\         
 &Theoretical & .2990  &   .3400  &   .3671  &   .3879  &   .4050 &  .1082&    .1426 &   .1676 &   .1880  &  .2055\\ \hline
\hline \hline
\end{tabular}}
\label{simu_com_4}
\end{table}

In sum, we draw the following conclusions about efficiency gains from mfBm based on the analytical and simulation results.
(1) If all Hurst exponents are identical, the forecasts based on fBm and mfBm coincide. Hence, there is no efficiency gain.
(2) If all correlations are zero, i.e., $\Sigma$ is diagonal, the forecasts based on fBm and mfBm coincide. Hence, there is no efficiency gain. (3) In a scenario that is not covered by the previous two cases, the forecast efficiency improves compared to the univariate model. Larger differences between Hurst exponents lead to larger efficiency gains. Larger absolute correlations between components lead to larger efficiency gains.  (4) Including more correlated components yields further improvements of the forecasting performance. (5) Estimation errors reduce forecasting performance; however, the advantages of mfBm still remain when $\rho \neq 0$ and Hurst exponents differ.

\subsection{Effects of measurement errors} 
\label{sec:error_on_eta_rho}
RV is an observable proxy for a latent volatility process, and the Hurst exponent estimates are downward biased due to measurement error. Although we find that the estimated correlation is around 0.3 and the asymmetry parameter is close to zero from Tables \ref{dj30_corr_7}, \ref{dj30_corr30} and \ref{dj30_eta30}, these estimates may not accurately reflect the latent volatility process. To see this more clearly, consider a simple setting in which the latent integrated volatility process follows a bivariate fBm, and each component is contaminated by i.i.d.\ noise. In this case,
\[
\sqrt{n}\left(\log RV_t - \log IV_t\right) \xrightarrow{d} \mathcal{N}(0,2),
\]
as shown in Theorem 2.1 of \citet{Fukasawa-Takabatake-Westphal-2022}. This setting corresponds to interpreting parameters of integrated volatility using realized volatility as a proxy. We conduct a simulation study. We set $H_1 = 0.1$, $H_2 = 0.4$, and $\sigma_1^2 = \sigma_2^2 = 1$. We consider four cases for $(\rho,\eta)$: (1) $\rho = 0$, $\eta = 0$; (2) $\rho = 0.4$, $\eta = 0$; (3) $\rho = 0$, $\eta = 0.5$; and (4) $\rho = 0.4$, $\eta = 0.5$. For $5$-minute realized variance computed from market data with six trading hours, we have $n = 72$. The number of replications is 1{,}000, and the sample sizes are 500 and 1{,}000. We report the simulation results in Table \ref{tab:bias_rho_eta}. The results indicate that positive $\eta$ and $\rho$ are subject to downward bias. For example, when the sample size is 500, the downward bias is approximately $-0.24$ for $\rho=0.4$ and $-0.36$ for $\eta=0.5$. The intuition is as follows. Consider our estimators of $\rho$ and $\eta$ in (\ref{rhohat}) and (\ref{etahat}): the i.i.d. noise does not affect the numerator, but it inflates the denominator. As a result, the estimates tend to be biased toward zero.

\begin{table}[H]
\centering
\caption{Bias of BYZ methods for different combinations of $\rho$ and $\eta_{1,2}$ (denoted by $\eta$ in the table) under model misspecification, with $H_1=0.1$, $H_2=0.4$, $\sigma_1^2=1$ and $\sigma_2^2=1$.}
\scalebox{0.75}{
\begin{tabular}{c|cc|cc|cc|cc}
\hline \hline
              & \multicolumn{4}{c|}{$n=500$}    &  \multicolumn{4}{c}{$n=1000$}   \\ \hline
              & \multicolumn{2}{c|}{bfBm}    &  \multicolumn{2}{c|}{bfBm+noise} & \multicolumn{2}{c|}{bfBm}    &  \multicolumn{2}{c}{bfBm+noise}  \\ \hline
                                   &$\rho$   & $\eta$     &   $\rho$     & $\eta$  &$\rho$   & $\eta$     &   $\rho$     & $\eta$   \\
                                \hline
 $\rho=0$, $\eta=0$          & -0.0006&-0.0009 &  -0.0024&-0.0004& 0.0027&0.0037  & 0.0010& 0.0016   \\\hline
   $\rho=0.4$, $\eta=0$        & -0.0022&0.0005    & \textbf{-0.2430}&0.0008  & -0.0012&-0.0018&\textbf{-0.2436}&0.0011\\\hline
 $\rho=0$,  $\eta=0.5$      & -0.0010&0.0040   & 0.0011&\textbf{-0.3691} & 0.0009&0.0050& -0.0015&\textbf{-0.3687}\\ \hline
$\rho=0.4$,  $\eta=0.5$    &0.0017&0.0051   & \textbf{-0.2405}&\textbf{-0.3679} & 0.0006&0.0037 & \textbf{-0.2418}&\textbf{-0.3680}\\         \hline \hline                              
\end{tabular}}
\label{tab:bias_rho_eta}
\end{table}

\section{Additional Empirical Results} 
\label{sec:add_empirics}

\subsection{Parameter estimates for Dow Jones 30 stocks}
\label{sec:est_eta_rho}

Table \ref{dj30_corr30}  reports the point estimates of $H$ and $\rho$ for  Dow Jones 30 stocks  from January 5, 2005, to January 14, 2025 in alphabetical order. We also report the point estimates of $\eta$ in Table \ref{dj30_eta30}  for these stocks. Table \ref{dj30_corr30} suggests that there are differences among the estimated Hurst exponents and that the estimated correlation parameters are always positive and far away from zero. We expect mfBm should offer forecasting improvements over univariate fBm. Table \ref{dj30_eta30} indicates that the asymmetry parameters are close to zero, indicating the usefulness of time-reversible mfBm.
 
\begin{table}[H]
\centering
\caption{ Estimates of $H$ and $\rho$ for Dow Jones 30 stocks}                                                                                                                                                                                                                                                                                   
\rotatebox{90}{   
\scalebox{0.44}{  
\begin{tabular}{l|c|llllllllllllllllllllllllllllll}
\hline
\hline
 &     H   &   AAPL & ALD & AMGN & AXP & BA & BEL & CAT & CHV & CRM & CSCO & DIS & GS & HD & IBM & INTC & JNJ & JPM & KO & MCD  &MMM&MRK&MSFT&NIKE&PG&SPC&UNH&V&WAG&WMT&XOM     \\
 \hline
AAPL & 0.2609 & 1 &       &       &       &       &       &       &       &       &       &       &       &       &       &       &       &       &       &       &       &       &       &       &       &       &       &       &       &       &       \\
 ALD& 0.2050 & 0.3902 & 1 &       &       &       &       &       &       &       &       &       &       &       &       &       &       &       &       &       &       &       &       &       &       &       &       &       &       &       &       \\
AMGN & 0.1645 & 0.3074 & 0.3205 & 1 &       &       &       &       &       &       &       &       &       &       &       &       &       &       &       &       &       &       &       &       &       &       &       &       &       &       &       \\
AXP & 0.2058 & 0.3721 & 0.4011 & 0.2893 & 1 &       &       &       &       &       &       &       &       &       &       &       &       &       &       &       &       &       &       &       &       &       &       &       &       &       &       \\
 BA& 0.2153 & 0.3602 & 0.3850 & 0.2696 & 0.3405 & 1 &       &       &       &       &       &       &       &       &       &       &       &       &       &       &       &       &       &       &       &       &       &       &       &       &       \\
BEL & 0.1776 & 0.3248 & 0.3474 & 0.2642 & 0.3456 & 0.2850 & 1 &       &       &       &       &       &       &       &       &       &       &       &       &       &       &       &       &       &       &       &       &       &       &       &       \\
CAT & 0.2138 & 0.3894 & 0.4246 & 0.2807 & 0.4080 & 0.3594 & 0.3135 & 1 &       &       &       &       &       &       &       &       &       &       &       &       &       &       &       &       &       &       &       &       &       &       &       \\
CHV & 0.2248 & 0.4170 & 0.4262 & 0.3200 & 0.4229 & 0.3701 & 0.3758 & 0.4434 & 1 &       &       &       &       &       &       &       &       &       &       &       &       &       &       &       &       &       &       &       &       &       &       \\
CRM & 0.1918 & 0.3473 & 0.2889 & 0.2113 & 0.2985 & 0.2669 & 0.2462 & 0.2806 & 0.299 & 1 &       &       &       &       &       &       &       &       &       &       &       &       &       &       &       &       &       &       &       &       &       \\
 CSCO& 0.2285 & 0.4611 & 0.4131 & 0.3309 & 0.4102 & 0.3772 & 0.3567 & 0.4072 & 0.4565 & 0.3381 & 1 &       &       &       &       &       &       &       &       &       &       &       &       &       &       &       &       &       &       &       &       \\
 DIS& 0.1989 & 0.3468 & 0.3646 & 0.2795 & 0.3450 & 0.3162 & 0.3412 & 0.3429 & 0.4023 & 0.2546 & 0.3876 & 1 &       &       &       &       &       &       &       &       &       &       &       &       &       &       &       &       &       &       &       \\
GS & 0.2515 & 0.4591 & 0.4127 & 0.2980 & 0.4639 & 0.3610 & 0.3429 & 0.4203 & 0.4667 & 0.3287 & 0.4594 & 0.3772 & 1 &       &       &       &       &       &       &       &       &       &       &       &       &       &       &       &       &       &       \\
HD & 0.2125 & 0.4016 & 0.4157 & 0.2855 & 0.3994 & 0.3422 & 0.3374 & 0.3843 & 0.4309 & 0.2858 & 0.4170 & 0.3680 & 0.4511 & 1 &       &       &       &       &       &       &       &       &       &       &       &       &       &       &       &       &       \\
 IBM& 0.2149 & 0.4096 & 0.4035 & 0.2754 & 0.3802 & 0.3499 & 0.3378 & 0.393 & 0.4201 & 0.2959 & 0.4391 & 0.3641 & 0.4216 & 0.3863 & 1 &       &       &       &       &       &       &       &       &       &       &       &       &       &       &       &       \\
INTC & 0.2246 & 0.4358 & 0.3857 & 0.3035 & 0.3879 & 0.3481 & 0.3135 & 0.3766 & 0.4398 & 0.2982 & 0.4696 & 0.3487 & 0.4033 & 0.3863 & 0.4059 & 1 &       &       &       &       &       &       &       &       &       &       &       &       &       &       &       \\
JNJ & 0.1634 & 0.3148 & 0.3654 & 0.3015 & 0.3554 & 0.3177 & 0.3627 & 0.3139 & 0.3818 & 0.2461 & 0.3663 & 0.3314 & 0.4104 & 0.3784 & 0.3416 & 0.3227 & 1 &       &       &       &       &       &       &       &       &       &       &       &       &       &       \\
 JPM& 0.2516 & 0.4479 & 0.4469 & 0.3282 & 0.5171 & 0.3953 & 0.4024 & 0.4682 & 0.4801 & 0.3110 & 0.4904 & 0.4163 & 0.6394 & 0.5020 & 0.4259 & 0.4289 & 0.4378 & 1 &       &       &       &       &       &       &       &       &       &       &       &       &       \\
KO & 0.1887 & 0.3554 & 0.3912 & 0.2885 & 0.3704 & 0.3406 & 0.3748 & 0.3808 & 0.4317 & 0.2890 & 0.4035 & 0.3734 & 0.4209 & 0.4069 & 0.3854 & 0.3608 & 0.3951 & 0.4586 & 1 &       &       &       &       &       &       &       &       &       &       &       &       \\
MCD & 0.1909 & 0.3145 & 0.3346 & 0.2732 & 0.3254 & 0.2725 & 0.3085 & 0.3355 & 0.3681 & 0.2204 & 0.3317 & 0.3212 & 0.3343 & 0.3583 & 0.3231 & 0.2966 & 0.2948 & 0.3834 & 0.3407 & 1 &       &       &       &       &       &       &       &       &       &       &       \\
 MMM& 0.1993 & 0.3627 & 0.4370 & 0.2827 & 0.3710 & 0.3223 & 0.3967 & 0.4165 & 0.4183 & 0.2729 & 0.3890 & 0.3605 & 0.4077 & 0.3814 & 0.3783 & 0.3406 & 0.3787 & 0.4228 & 0.3946 & 0.3386 & 1 &       &       &       &       &       &       &       &       &       &       \\
 MRK& 0.1894 & 0.2698 & 0.3156 & 0.2943 & 0.2931 & 0.2510 & 0.3068 & 0.3112 & 0.3294 & 0.2367 & 0.3034 & 0.2893 & 0.2878 & 0.3119 & 0.2983 & 0.2676 & 0.3456 & 0.3437 & 0.3226 & 0.2564 & 0.3263 & 1 &       &       &       &       &       &       &       &       &       \\
 MSFT& 0.2234 & 0.5051 & 0.4800 & 0.3690 & 0.4390 & 0.3807 & 0.3808 & 0.4430 & 0.4836 & 0.3571 & 0.5357 & 0.4304 & 0.4836 & 0.4448 & 0.4451 & 0.5102 & 0.3841 & 0.5031 & 0.4359 & 0.3539 & 0.4098 & 0.3126 & 1 &       &       &       &       &       &       &       &       \\
NIKE & 0.2069 & 0.3279 & 0.3277 & 0.2362 & 0.3141 & 0.2795 & 0.3006 & 0.3145 & 0.3430 & 0.2652 & 0.3308 & 0.3022 & 0.3382 & 0.3766 & 0.3273 & 0.3212 & 0.2898 & 0.3729 & 0.3252 & 0.2811 & 0.3233 & 0.2675 & 0.3899 & 1 &       &       &       &       &       &       &       \\
 PG& 0.1824 & 0.3371 & 0.3918 & 0.2891 & 0.3635 & 0.3074 & 0.3531 & 0.3501 & 0.4147 & 0.2485 & 0.3753 & 0.3441 & 0.3678 & 0.3811 & 0.3664 & 0.3623 & 0.3498 & 0.4223 & 0.4387 & 0.3139 & 0.3613 & 0.3353 & 0.4108 & 0.3168 & 1 &       &       &       &       &       &       \\
SPC & 0.1404 & 0.2900 & 0.3400 & 0.2663 & 0.3824 & 0.2780 & 0.3454 & 0.3183 & 0.3628 & 0.2323 & 0.3283 & 0.2997 & 0.3986 & 0.3530 & 0.3232 & 0.2961 & 0.3289 & 0.4174 & 0.3635 & 0.2628 & 0.3484 & 0.2700 & 0.3666 & 0.2743 & 0.3550 & 1 &       &       &       &       &       \\
 UNH& 0.1804 & 0.2643 & 0.2909 & 0.2501 & 0.2454 & 0.2250 & 0.2595 & 0.2606 & 0.3057 & 0.1966 & 0.2820 & 0.2476 & 0.3194 & 0.2939 & 0.2756 & 0.2495 & 0.2951 & 0.2955 & 0.3190 & 0.2557 & 0.2531 & 0.2396 & 0.3036 & 0.2353 & 0.2564 & 0.2598 & 1 &       &       &       &       \\
 V& 0.2225 & 0.3957 & 0.3887 & 0.2833 & 0.4311 & 0.3290 & 0.3342 & 0.3668 & 0.4072 & 0.3404 & 0.3847 & 0.3552 & 0.4052 & 0.3960 & 0.3602 & 0.3405 & 0.3390 & 0.4594 & 0.3394 & 0.2976 & 0.3547 & 0.3014 & 0.4632 & 0.3274 & 0.3378 & 0.3229 & 0.2789 & 1 &       &       &       \\
WAG & 0.1995 & 0.2687 & 0.2976 & 0.2134 & 0.2852 & 0.2538 & 0.2294 & 0.2819 & 0.3019 & 0.1928 & 0.3025 & 0.2624 & 0.2901 & 0.3213 & 0.2940 & 0.2643 & 0.2396 & 0.3272 & 0.2841 & 0.2315 & 0.2647 & 0.2466 & 0.2934 & 0.2436 & 0.2799 & 0.2544 & 0.2104 & 0.2429 & 1 &       &       \\
WMT & 0.1977 & 0.3293 & 0.3484 & 0.2800 & 0.3480 & 0.2933 & 0.3164 & 0.3176 & 0.3982 & 0.2364 & 0.3981 & 0.3018 & 0.3550 & 0.4269 & 0.3408 & 0.3244 & 0.3303 & 0.3928 & 0.3798 & 0.2975 & 0.3489 & 0.2652 & 0.3706 & 0.3130 & 0.3707 & 0.3253 & 0.2494 & 0.3038 & 0.3049 & 1 &       \\
XOM & 0.2311 & 0.4285 & 0.4385 & 0.3301 & 0.4143 & 0.3665 & 0.3889 & 0.4451 & 0.7700 & 0.3108 & 0.4574 & 0.4088 & 0.4812 & 0.4426 & 0.4281 & 0.4494 & 0.3974 & 0.5051 & 0.4431 & 0.3517 & 0.4281 & 0.3468 & 0.4911 & 0.3536 & 0.4301 & 0.3885 & 0.3269 & 0.4110 & 0.3267 & 0.4179 & 1\\
\hline
\hline
\end{tabular}}}
\label{dj30_corr30}    
\end{table}

\begin{table}[H]
\centering
\caption{Estimates of $\eta_{1,2}$ for Dow Jones 30 stocks}
\rotatebox{90}{   
\scalebox{0.44}{  
\begin{tabular}{l|llllllllllllllllllllllllllllll}
\hline
\hline
 & AAPL & ALD & AMGN & AXP & BA & BEL & CAT & CHV & CRM & CSCO & DIS & GS & HD & IBM & INTC & JNJ & JPM & KO & MCD  &MMM&MRK&MSFT&NIKE&PG&SPC&UNH&V&WAG&WMT&XOM \\
 \hline
 AAPL  & 0         &          &          &          &          &          &          &          &          &          &          &          &          &          &          &          &          &          &          &          &          &          &          &          &          &          &          &          &         &    \\
ALD & 0.0950    &  0       &          &          &          &          &          &          &          &          &          &          &          &          &          &          &          &          &          &          &          &          &          &          &          &          &          &          &         &    \\
AMGN  & 0.0123  & -0.0080   &    0     &          &          &          &          &          &          &          &          &          &          &          &          &          &          &          &          &          &          &          &          &          &          &          &          &          &         &    \\
AXP   & 0.0556  & -0.0593 & 0.0276  &   0       &          &          &          &          &          &          &          &          &          &          &          &          &          &          &          &          &          &          &          &          &          &          &          &          &         &    \\
BA     & 0.0899  & -0.0937 & -0.0239 & -0.0336 &    0      &          &          &          &          &          &          &          &          &          &          &          &          &          &          &          &          &          &          &          &          &          &          &          &         &    \\
BEL     & 0.0710    & -0.0653 & -0.0151 & 0.0010    & -0.0759 &     0   &          &          &          &          &          &          &          &          &          &          &          &          &          &          &          &          &          &          &          &          &          &          &         &    \\
CAT     & 0.0980    & 0.0077  & -0.0318 & -0.0248 & 0.0665  & 0.0338  &    0      &          &          &          &          &          &          &          &          &          &          &          &          &          &          &          &          &          &          &          &          &          &         &    \\
 CHV    & 0.0461  & -0.0583 & -0.0291 & -0.0540   & 0.0183  & -0.0083 & -0.0038 &  0        &          &          &          &          &          &          &          &          &          &          &          &          &          &          &          &          &          &          &          &          &         &    \\
CRM     & 0.0839  & -0.0468 & -0.0338 & -0.0195 & 0.0190    & 0.0185  & 0.0160    & 0.0159  &   0       &          &          &          &          &          &          &          &          &          &          &          &          &          &          &          &          &          &          &          &         &    \\
  CSCO   & 0.1513  & -0.0308 & -0.0453 & 0.0027  & 0.0311  & -0.0336 & -0.0725 & 0.0117  & -0.0052 &    0      &          &          &          &          &          &          &          &          &          &          &          &          &          &          &          &          &          &          &         &    \\
  DIS   & 0.0597  & -0.0484 & -0.0185 & -0.0184 & -0.0286 & -0.0241 & -0.0283 & -0.0654 & -0.0205 & -0.0420   &    0      &          &          &          &          &          &          &          &          &          &          &          &          &          &          &          &          &          &         &    \\
 GS    & 0.1132  & -0.0542 & -0.0235 & 0.0048  & 0.0307  & -0.0057 & -0.0348 & 0.0098  & -0.0022 & -0.0201 & 0.066    &    0     &          &          &          &          &          &          &          &          &          &          &          &          &          &          &          &          &         &    \\
  HD   & 0.0216  & -0.0976 & -0.0588 & -0.0577 & -0.0327 & -0.0072 & -0.0890   & -0.0347 & -0.0307 & -0.0469 & -0.0311 & -0.0214 &    0    &          &          &          &          &          &          &          &          &          &          &          &          &          &          &          &         &    \\
 IBM    & 0.0176  & -0.1240   & -0.0810   & -0.1142 & -0.0188 & -0.0226 & -0.0466 & -0.0262 & -0.0116 & -0.0501 & -0.0045 & -0.0312 & -0.0269 &    0      &          &          &          &          &          &          &          &          &          &          &          &          &          &          &         &    \\
   INTC  & 0.1117  & -0.0275 & 0.0279  & -0.0599 & 0.0402  & -0.0258 & -0.0429 & -0.0163 & 0.0034  & 0.0406  & 0.0692  & 0.0090    & 0.0711  & 0.0632  &   0      &          &          &          &          &          &          &          &          &          &          &          &          &          &         &    \\
   JNJ  & 0.0818  & -0.0588 & -0.0084 & -0.0043 & -0.0056 & -0.0085 & -0.0251 & 0.0125  & -0.0248 & -0.0188 & -0.0042 & -0.0409 & 0.0098  & -0.0248 & -0.0739 &    0    &          &          &          &          &          &          &          &          &          &          &          &          &         &    \\
JPM     & 0.0484  & -0.0539 & -0.0440   & -0.0062 & 0.0451  & -0.0119 & -0.0519 & 0.0382  & -0.0067 & -0.0004 & 0.0284  & -0.0385 & 0.0390    & 0.0531  & 0.0164  & 0.0696  &  0        &          &          &          &          &          &          &          &          &          &          &          &         &    \\
 KO    & 0.0784  & -0.0593 & -0.0215 & -0.0446 & -0.0379 & 0.0332  & -0.0528 & 0.0234  & -0.0225 & -0.0044 & 0.0484  & -0.0349 & 0.0286  & 0.0113  & -0.0110   & 0.0292  & 0.0155  &    0      &          &          &          &          &          &          &          &          &          &          &         &    \\
 MCD    & 0.0677  & -0.0219 & -0.0323 & -0.0147 & 0.0466  & -0.0021 & -0.0355 & 0.0120    & -0.0048 & 0.0484  & 0.0324  & 0.0187  & 0.0391  & 0.0023  & 0.0005  & -0.0135 & 0.0504  & 0.0004  &   0     &          &          &          &          &          &          &          &          &          &         &    \\
  MMM   & 0.0388  & -0.1179 & -0.0755 & -0.0316 & -0.0720  & -0.0471 & -0.0111 & -0.0346 & -0.0147 & -0.0826 & 0.0458  & -0.0226 & -0.0030   & 0.0111  & -0.0841 & -0.0425 & 0.0011  & -0.0524 & -0.0318 &     0     &          &          &          &          &          &          &          &          &         &    \\
 MRK    & -0.0021 & -0.0242 & -0.0126 & -0.0428 & 0.0097  & 0.0135  & -0.0705 & -0.0214 & -0.0047 & -0.0108 & 0.0284  & -0.0259 & 0.0142  & 0.0276  & -0.0514 & -0.0497 & -0.0172 & -0.0080   & -0.0275 & 0.0463  &   0       &          &          &          &          &          &          &          &         &    \\
   MSFT  & 0.0654  & -0.0692 & 0.0025  & 0.0065  & 0.0426  & 0.0158  & -0.0151 & 0.0020    & -0.0151 & -0.0058 & 0.0125  & -0.0434 & 0.0520    & 0.0854  & -0.0052 & 0.0205  & 0.0108  & 0.0313  & -0.0200   & 0.0964  & 0.0296  &      0    &          &          &          &          &          &          &         &    \\
  NIKE   & 0.0780    & 0.0132  & -0.0129 & -0.0162 & 0.0273  & 0.0501  & 0.0382  & 0.0435  & 0.0328  & 0.0304  & 0.0378  & 0.0348  & 0.0665  & 0.1001  & 0.0107  & 0.0371  & 0.0605  & 0.0173  & 0.0400    & 0.0500    & 0.0418  & 0.0467  &     0     &          &          &          &          &          &         &    \\
  PG   & 0.0249  & -0.0145 & -0.0539 & -0.0221 & 0.0068  & 0.0159  & -0.0640   & -0.0099 & -0.0103 & -0.0089 & 0.0336  & -0.0258 & 0.0497  & 0.0470    & -0.0281 & 0.0561  & 0.0199  & 0.0066  & 0.0120    & 0.0611  & 0.0319  & -0.0476 & -0.0131 &    0      &          &          &          &          &         &    \\
 SPC    & 0.0388  & -0.0942 & -0.0228 & -0.0237 & -0.0685 & -0.0112 & -0.0048 & 0.0208  & 0.0209  & -0.0279 & 0.0081  & 0.0236  & 0.0204  & 0.0289  & -0.0168 & 0.0546  & 0.0018  & 0.0251  & -0.0007 & 0.0489  & 0.0032  & -0.0417 & -0.0146 & 0.0418  &   0       &          &          &          &         &    \\
 UNH    & 0.0892  & -0.0156 & -0.0206 & -0.0496 & -0.0056 & 0.0069  & 0.0181  & 0.0163  & -0.0150   & 0.0286  & 0.0492  & 0.0437  & 0.0216  & 0.0206  & -0.0162 & 0.0441  & 0.0562  & 0.0409  & 0.0056  & 0.0608  & 0.0051  & 0.0038  & -0.0114 & -0.0035 & -0.0166 &    0      &          &          &         &    \\
 V    & 0.0246  & -0.0779 & -0.0201 & -0.0018 & 0.0188  & 0.0019  & -0.0068 & -0.0052 & 0.0324  & -0.0453 & 0.0549  & -0.0099 & 0.0782  & 0.0724  & -0.0977 & -0.0122 & 0.0244  & -0.0116 & 0.0331  & 0.0702  & -0.0122 & 0.0267  & -0.0245 & -0.0373 & 0.0170    & -0.0032 &       0  &          &         &    \\
WAG     & 0.1016  & -0.0447 & -0.0342 & -0.0189 & 0.0008  & 0.0403  & -0.0006 & 0.0141  & -0.0018 & -0.0224 & 0.037    & -0.0114 & -0.0193 & 0.0186  & -0.0281 & 0.0213  & 0.0144  & 0.0254  & -0.0122 & 0.0336  & 0.0134  & -0.0068 & -0.0356 & -0.0059 & -0.0191 & -0.0516 & -0.0096 &     0     &         &    \\
  WMT   & 0.0762  & -0.0593 & -0.0280   & -0.0256 & -0.0327 & -0.0022 & -0.0390   & 0.0241  & 0.0050    & 0.0080    & 0.0481  & -0.0093 & 0.0131  & 0.0447  & -0.0521 & -0.0050   & -0.0178 & -0.0304 & -0.0360   & -0.0120   & -0.0112 & -0.0371 & -0.0642 & 0.0476  & 0.0033  & -0.0311 & 0.0313  & -0.0496 &   0      &    \\
   XOM  & 0.1168  & -0.0511 & -0.0173 & -0.0658 & 0.0234  & -0.0172 & -0.0281 & -0.0588 & 0.0068  & -0.0166 & 0.0530    & -0.0254 & 0.0243  & 0.0512  & -0.0217 & 0.0217  & 0.0031  & -0.0291 & 0.0072  & 0.0361  & 0.0291  & -0.0081 & -0.0385 & -0.0195 & -0.0272 & -0.0317 & -0.0008 & -0.0026 & 0.0054 &0\\
   \hline
\hline
\end{tabular}}}   
\label{dj30_eta30}   
\end{table}

\subsection{HAR and Its Variants}
\subsubsection{HAR model}
The HAR model proposed by \cite{corsi2009} has a simple linear specification:  
\begin{equation}
RV_{t+h} = \beta_0 + \beta_1 RV_t + \beta_2 RV_{t|t-4} + \beta_3 RV_{t|t-21} + \epsilon_t,  \label{eq:HAR}
\end{equation}  
where \( RV_{t} \) is the daily realized volatility and \( RV_{t-j|t-k} = \frac{1}{k-j+1} \sum_{i=j}^{k} RV_{t-i} \) with \( j \leq k \), and \( \epsilon_t \) is the disturbance term. Based on this definition, \( RV_{t|t-4} \) and \( RV_{t|t-21} \) represent the weekly and monthly average realized volatility, respectively. The linear HAR model can be estimated using the ordinary least squares (OLS) method.
\subsubsection{Vector HAR model}
The HAR model only handles a univariate time series. If we extend it to a multivariate setting and incorporate information from other time series for forecasting, a natural approach is to stack the realized volatilities and formulate a vector autoregressive model as follows (referred to as the Vector HAR model or VHAR for short):
\[
\mathbf{RV}_{t+h} = \mathbf{\beta_0} + \mathbf{\Phi}_1 \mathbf{RV}_{t} + \mathbf{\Phi}_2 \mathbf{RV}_{t|t-4} + \mathbf{\Phi}_3 \mathbf{RV}_{t|t-21} + \mathbf{\epsilon_t},
\]
where
\[
\mathbf{RV}_t = \begin{pmatrix} RV_t^1 \\ RV_t^2\\... \\ RV_t^J \end{pmatrix}, ~
\mathbf{RV}_{t|t-4} = \begin{pmatrix} RV_{t|t-4}^1 \\ RV_{t|t-4}^2\\... \\ RV_{t|t-4}^J \end{pmatrix}, ~
\mathbf{RV}_{t|t-21} = \begin{pmatrix} RV_{t|t-21}^1 \\ RV_{t|t-21}^2\\... \\ RV_{t|t-21}^J \end{pmatrix}, ~
\mathbf{C} = \begin{pmatrix} \beta_0^1 \\ \beta_0^2\\... \\ \beta_0^J \end{pmatrix}, \quad
\mathbf{\epsilon}_t = \begin{pmatrix} \epsilon_{t}^1 \\ \epsilon_t^2\\... \\ \epsilon_t^J \end{pmatrix},
\]
$J$ represents the number of realized volatilities and the coefficient matrices are defined as ($i=1,2,3$):
\[
\mathbf{\Phi}_i = \begin{pmatrix}
\phi_{11,i} & \phi_{12,i} & ... &\phi_{1J,i} \\
\phi_{21,i} & \phi_{22,i} &...& \phi_{2J,i} \\
...&...&...&...\\
\phi_{J1,i} & \phi_{J2,i} & ...&\phi_{JJ,i}
\end{pmatrix}, \quad
\]

Assuming that the covariance matrix of the error term, \( \mathbf{\epsilon}_t \), is a diagonal matrix allows estimation and forecasting to be performed row by row. To forecast the first element in \( \mathbf{RV}_t \), denoted by \( RV_{t}^1 \), the regression model can be expressed as follows and estimated using OLS:
\begin{align}
RV_{t+h}^1=&\beta _{0}^1+\phi_{11,1}RV_{t}^1+\phi_{11,2}RV_{t|t-4}^1+\phi_{11,3}RV_{t|t-21}^2\notag \\
+&\sum_{j=2}^J\left\{\phi_{1j,1}RV_{t}^j+\phi_{1j,2}RV_{t|t-4}^j+\phi_{1j,3}RV_{t|t-21}^j\right\}+\epsilon_{t}^1.  \label{eq:vector-HAR}
\end{align}
The additional terms in the vector HAR model, compared to the standard HAR model,  
\[
\sum_{j=2}^J \left\{\phi_{1j,1}RV_{t}^j + \phi_{1j,2}RV_{t|t-4}^j + \phi_{1j,3}RV_{t|t-21}^j \right\}
\]  
represent information from other time series.

\subsubsection{Vector HAR model with a single common factor}
The vector HAR model with a single common factor (VHARF) is specified below, according to (6) in \cite{Bauwens2023} and (1.1) in \cite{Ding2025}. This formulation effectively imposes equal weights across components in \eqref{eq:vector-HAR}:
\begin{align}
RV_{t+h}^1 =& \beta_0^1 + \phi_{11,1}RV_t^1 + \phi_{11,2}RV_{t|t-4}^1 + \phi_{11,3}RV_{t|t-21}^1 \notag \\
+& \bar{\phi}_1 \overline{RV}_t + \bar{\phi}_2 \overline{RV}_{t|t-4} + \bar{\phi}_3 \overline{RV}_{t|t-21} + \varepsilon_t^1, \label{eq:vector-HARc}
\end{align}
where $\overline{RV}_t = \frac{1}{J-1}\sum_{j=2}^J RV_t^j$, $\overline{RV}_{t|t-4} = \frac{1}{J-1}\sum_{j=2}^J RV_{t|t-4}^j$ and $\overline{RV}_{t|t-21} = \frac{1}{J-1}\sum_{j=2}^J RV_{t|t-21}^j$.

To evaluate this approach, we forecast AAPL's realized volatility using the same set of information sources (AAPL, ALD, AMGN, AXP, and BA) across our five progressively expanded model specifications. The common factor adapts dynamically to each specification. Table \ref{emp: HARC} reports MSFEs of HAR and 4 VHARF models. While the common factor models enhance the performance relative to their respective VHAR specifications, they perform worse than the univariate HAR model and, more importantly, the corresponding mfBm models.

\begin{table}[H]
\centering
\caption{MSFEs  of VHARF models for $h$-day-ahead forecasts of RV, along with p-values from model confidence set (MCS) tests,  between January 3, 2007 to January 14, 2025}
\scalebox{0.6}{ 
\begin{tabular}{ccccccccccccccccccc}
\hline \hline
&  \multicolumn{8}{c}{MSFE $\times$ 100}  & & \multicolumn{8}{c}{MCS p-value} \\
\hline
              &       $h=1  $    &  $h=2 $     &  $h=3  $    &  $h=4 $     &  $h=5 $     &  $h=10 $    &  $h=15 $    &  $h=20$&&  $h=1  $    &  $h=2 $     &  $h=3  $    &  $h=4 $     &  $h=5 $     &  $h=10 $    &  $h=15 $    &  $h=20$  \\
                         \hline
HAR   & 0.5157 & 0.6513 & 0.7369 & 0.7870 & 0.8678 & 1.1109 & 1.2468 & 1.3072&&1.0000 & 1.0000 & 1.0000 & 1.0000 & 1.0000 & 1.0000 & 1.0000 & 1.0000 \\
VHARF2  & 0.5316 & 0.6703 & 0.7701 & 0.8292 & 0.9307 & 1.2834 & 1.4560 & 1.5053&&0.1446 & 0.1664 & 0.0958 & 0.0730 & 0.0946 & 0.1972 & 0.2186 & 0.2222 \\
VHARF3 & 0.5300 & 0.6695 & 0.7691 & 0.8249 & 0.9278 & 1.2629 & 1.3993 & 1.4299 &&0.1446 & 0.1664 & 0.0958 & 0.0730 & 0.0946 & 0.1972 & 0.2186 & 0.2222\\
VHARF4 & 0.5293 & 0.6691 & 0.7743 & 0.8348 & 0.9372 & 1.2306 & 1.3949 & 1.4871 &&0.1446 & 0.1664 & 0.0958 & 0.0730 & 0.0688 & 0.1662 & 0.1718 & 0.2222\\
VHARF5 & 0.5323 & 0.6712 & 0.7773 & 0.8378 & 0.9454 & 1.2351 & 1.4075 & 1.4925&&0.1446 & 0.1664 & 0.0854 & 0.0584 & 0.0946 & 0.1972 & 0.2186 & 0.2222\\
\hline \hline
\end{tabular}}
\label{emp: HARC}
\end{table}

\subsubsection{Vector log-HAR model}
We replace $RV$ in (\ref{eq:vector-HAR}) with $\log RV$ and obtain the corresponding vector log-HAR (VLHAR) models. The model can be readily estimated by OLS, and the corresponding forecasts can be easily constructed using the expectation implied by the log-normal distribution. Table \ref{emp: LHAR} reports MSFEs of HAR and 4 VLHAR models. While the  VLHAR models enhance the performance relative to their respective VHAR specifications, they perform worse than the univariate log-HAR model and, more importantly, the corresponding mfBm models.

\begin{table}[H]
\centering
\caption{MSFEs of VLHAR models for $h$-day-ahead forecasts of RV, along with p-values from model confidence set (MCS) tests, between January 3, 2007 to January 14, 2025}
\scalebox{0.6}{ 
\begin{tabular}{ccccccccccccccccccc}
\hline \hline
&  \multicolumn{8}{c}{MSFE $\times$ 100}  & & \multicolumn{8}{c}{MCS p-value} \\
\hline
              &       $h=1  $    &  $h=2 $     &  $h=3  $    &  $h=4 $     &  $h=5 $     &  $h=10 $    &  $h=15 $    &  $h=20$   && $h=1  $    &  $h=2 $     &  $h=3  $    &  $h=4 $     &  $h=5 $     &  $h=10 $    &  $h=15 $    &  $h=20$ \\
                         \hline
LHAR    & 0.5003 & 0.6389 & 0.7263 & 0.7774 & 0.8510 & 1.0568 & 1.1918 & 1.2721 && 1.0000 & 1.0000 & 1.0000 & 1.0000 & 1.0000 & 1.0000 & 1.0000 & 1.0000\\
VLHAR2  & 0.5090 & 0.6495 & 0.7425 & 0.7915 & 0.8723 & 1.1162 & 1.2874 & 1.4055 &&0.2292 & 0.2532 & 0.1940 & 0.3522 & 0.1412 & 0.1462 & 0.1670 & 0.1332\\
VLHAR3 & 0.5136 & 0.6563 & 0.7565 & 0.8066 & 0.8954 & 1.1598 & 1.3485 & 1.4600&& 0.2292 & 0.0840 & 0.0830 & 0.0930 & 0.1154 & 0.1462 & 0.1670 & 0.1332\\
VLHAR4 & 0.5174 & 0.6619 & 0.7675 & 0.8184 & 0.9079 & 1.1716 & 1.3644 & 1.4771&& 0.1992 & 0.0678 & 0.0830 & 0.0930 & 0.1154 & 0.1462 & 0.1670 & 0.1332\\
VLHAR5 & 0.5263 & 0.6699 & 0.7769 & 0.8349 & 0.9293 & 1.1905 & 1.3842 & 1.4928&&0.1376 & 0.0048 & 0.0124 & 0.0062 & 0.0128 & 0.0208 & 0.0394 & 0.0354\\
\hline \hline
\end{tabular}}
\label{emp: LHAR}
\end{table}

\subsection{Magnificent 7 stocks}
\label{sec:m6}
\begin{figure}[t]
\centering
\makebox[\textwidth]{
	\includegraphics[width=15cm]{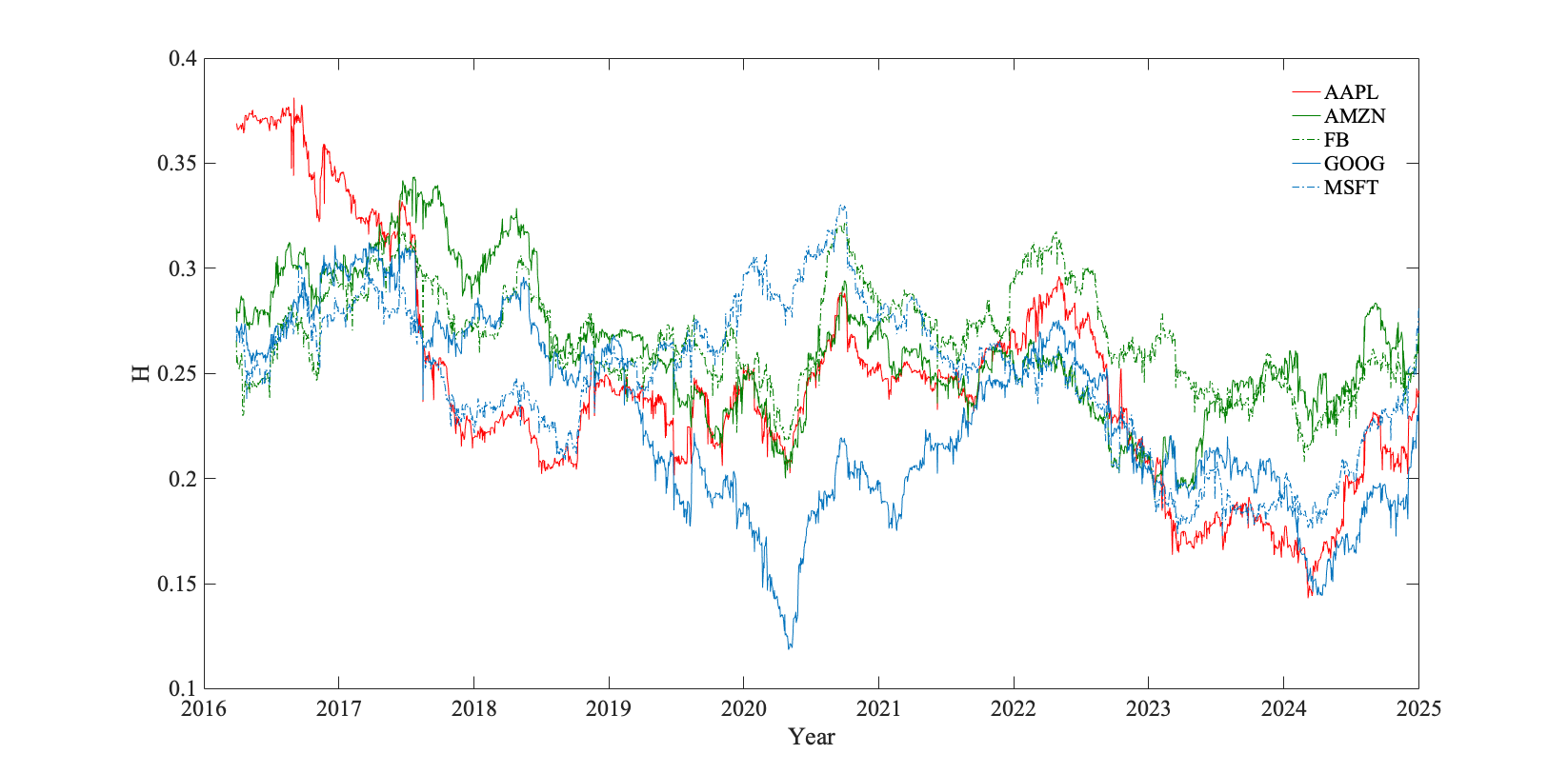}}
	\vspace{-0.3cm}
	\caption{Rolling window estimates of $H$ for log realized volatilities.}
\label{fig:roll_h_m}
\end{figure}

We also consider five stocks from the Magnificent 7 --- AAPL, AMZN, FB, GOOG, and MSFT --- and pay special attention to the forecasting performance of alternative models for AAPL. Figure \ref{fig:roll_h_m} reports the two-year  rolling window estimates of the Hurst exponent $H$ for log realized volatilities, showing that the estimates are close across assets. We consider five model specifications by progressively expanding the asset set: fBm (AAPL), bfBm (AAPL and AMZN), mfBm3 (AAPL, AMZN, and FB), mfBm4 (AAPL, AMZN, FB, and GOOG), and mfBm5 (AAPL, AMZN, FB, GOOG, and MSFT). The RMSEs and MCS p-values (for each model class) of the competing models are reported in Table \ref{emp: mse_m6}. Overall, the mfBm delivers more accurate forecasts than the univariate fBm, as evidenced by both the RMSE results and the MCS p-values. In contrast, for the vector HAR specifications—the vector HAR, the vector HAR with a single common factor and the vector log-HAR models—the univariate version consistently performs best.

\begin{table}[H]
\centering
\caption{MSFEs of alternative models for $h$-day-ahead forecasts of RV, along with p-values from model confidence set (MCS) tests for each model class, between March 29, 2016 to January 14, 2025.}
\scalebox{0.6}{ 
\begin{tabular}{ccccccccccccccccccc}
\hline \hline
&  \multicolumn{8}{c}{MSFE $\times$ 100}  & & \multicolumn{8}{c}{MCS p-value} \\
\hline
              &       $h=1  $    &  $h=2 $     &  $h=3  $    &  $h=4 $     &  $h=5 $     &  $h=10 $    &  $h=15 $    &  $h=20$ && $h=1  $    &  $h=2 $     &  $h=3  $    &  $h=4 $     &  $h=5 $     &  $h=10 $    &  $h=15 $    &  $h=20$   \\
                                  \hline
                         \multicolumn{18}{c}{MfBm}      \\
\hline
fBm   & 0.2793 & 0.3754 & 0.4484 & 0.5072 & 0.5574 & 0.7139 & 0.8382 & 0.9277&&0.6086 & 0.7242 & 0.8884 & 0.5896 & 0.4706 & 0.3962 & 0.3340 & 0.2852 \\
bfBm  & 0.2788 & 0.3747 & 0.4473 & 0.5060 & 0.5563 & 0.7133 & 0.8369 & 0.9270 &&0.6086 & 0.7242 & 0.9690 & 0.7982 & 0.6258 & 0.5078 & 0.4002 & 0.3368\\
mfBm3 & 0.2782 & 0.3743 & 0.4470 & 0.5055 & 0.5557 & 0.7139 & 0.8380 & 0.9286 && 1.0000 & 1.0000 & 1.0000 & 0.7982 & 0.6370 & 0.5918 & 0.4838 & 0.3636 \\
mfBm4 & 0.2820 & 0.3793 & 0.4486 & 0.5055 & 0.5545 & 0.6990 & 0.8117 & 0.8913&&0.4934 & 0.5618 & 0.8884 & 0.7982 & 0.6370 & 0.5918 & 0.4838 & 0.3636  \\
mfBm5 & 0.2820 & 0.3785 & 0.4471 & 0.5018 & 0.5489 & 0.6917 & 0.8000 & 0.8774&&0.6086 & 0.7242 & 1.0000 & 1.0000 & 1.0000 & 1.0000 & 1.0000 & 1.0000\\
\hline
\multicolumn{18}{c}{Vector HAR model}      \\
\hline
HAR    & 0.2876 & 0.3814 & 0.4575 & 0.5204 & 0.5842 & 0.7698 & 0.8024 & 0.8502 &&1.0000 & 1.0000 & 1.0000 & 1.0000 & 1.0000 & 1.0000 & 1.0000 & 1.0000 \\
VHAR2   & 0.2909 & 0.3922 & 0.4765 & 0.5485 & 0.6303 & 0.8854 & 0.9025 & 0.9520 &&0.3014 & 0.3284 & 0.1494 & 0.1794 & 0.2672 & 0.2746 & 0.3512 & 0.4608\\
VHAR3 & 0.2932 & 0.3983 & 0.4864 & 0.5611 & 0.6441 & 0.8826 & 0.9606 & 1.0669 &&0.3014 & 0.3284 & 0.1494 & 0.1794 & 0.2672 & 0.2746 & 0.3512 & 0.4608\\
VHAR4  & 0.2952 & 0.4016 & 0.4936 & 0.5727 & 0.6575 & 0.9521 & 1.1467 & 1.3803&&0.3014 & 0.3284 & 0.1494 & 0.1794 & 0.2672 & 0.2746 & 0.3512 & 0.4608 \\
VHAR5 & 0.2991 & 0.4073 & 0.5024 & 0.5920 & 0.6937 & 1.1794 & 1.6927 & 2.5475&&0.1056 & 0.2162 & 0.1298 & 0.0958 & 0.0576 & 0.0574 & 0.0494 & 0.0462\\
\hline
\multicolumn{18}{c}{Vector HAR model with a single common factor}       \\
\hline
HAR    & 0.2876 & 0.3814 & 0.4575 & 0.5204 & 0.5842 & 0.7698 & 0.8024 & 0.8502 &&1.0000 & 1.0000 & 1.0000 & 1.0000 & 1.0000 & 1.0000 & 1.0000 & 1.0000\\
VHARF2   & 0.2909 & 0.3922 & 0.4765 & 0.5485 & 0.6303 & 0.8854 & 0.9025 & 0.9520&&0.5240 & 0.4436 & 0.3050 & 0.2604 & 0.3086 & 0.2598 & 0.2630 & 0.2398 \\
VHARF3 & 0.2898 & 0.3918 & 0.4760 & 0.5475 & 0.6245 & 0.8823 & 0.9298 & 0.9964&&0.7310 & 0.4436 & 0.3050 & 0.2604 & 0.3086 & 0.2598 & 0.2630 & 0.2398 \\
VHARF4 & 0.2899 & 0.3914 & 0.4755 & 0.5474 & 0.6227 & 0.8800 & 0.9268 & 0.9881 &&0.5240 & 0.4436 & 0.3050 & 0.2024 & 0.3086 & 0.2480 & 0.2630 & 0.2398\\
VHARF5 & 0.2888 & 0.3891 & 0.4701 & 0.5396 & 0.6115 & 0.8537 & 0.8893 & 0.9562&&0.7310 & 0.4436 & 0.3050 & 0.2604 & 0.3086 & 0.2598 & 0.2630 & 0.2398\\
\hline
\multicolumn{18}{c}{Vector log-HAR model}       \\
\hline
LHAR   & 0.2833 & 0.3782 & 0.4484 & 0.5076 & 0.5609 & 0.7111 & 0.7734 & 0.8268 &&1.0000 & 1.0000 & 1.0000 & 1.0000 & 1.0000 & 1.0000 & 1.0000 & 1.0000 \\
VLHAR2 & 0.2875 & 0.3878 & 0.4627 & 0.5227 & 0.5801 & 0.7540 & 0.8189 & 0.8878&&0.8472 & 0.4750 & 0.4886 & 0.4668 & 0.4996 & 0.5762 & 0.5968 & 0.5936 \\
VLHAR3 & 0.2868 & 0.3864 & 0.4596 & 0.5164 & 0.5698 & 0.7347 & 0.8164 & 0.9086 &&0.8472 & 0.4750 & 0.4886 & 0.6474 & 0.7842 & 0.5762 & 0.5968 & 0.5936\\
VLHAR4& 0.2841 & 0.3799 & 0.4547 & 0.5148 & 0.5704 & 0.7488 & 0.8397 & 0.9369&&0.8472 & 0.7342 & 0.4886 & 0.6474 & 0.7842 & 0.5762 & 0.5968 & 0.5936 \\
VLHAR5 & 0.2888 & 0.3936 & 0.4800 & 0.5565 & 0.6307 & 0.9054 & 1.0053 & 1.0578&&0.8472 & 0.4750 & 0.2456 & 0.2358 & 0.2336 & 0.2238 & 0.2142 & 0.2106\\
\hline \hline
\end{tabular}}
\label{emp: mse_m6}
\end{table}

\subsection{Model confidence set test}
\label{sec:mcs}
The model confidence set (MCS) provides a confidence set containing the best models with a probability
greater than or equal to a specified level (e.g., 25\%). It also assigns a
p-value to each individual model, allowing for a comprehensive evaluation of
their statistical significance.

Suppose we have a set of competing models indexed by $\mathbb{M}%
_{0}=\left\{ {i=1,\ldots ,M}\right\} $. The loss function ($L_{i,t}$) can be the squared forecast error  for model $i$ at time $t$. We calculate the relative
performance as $d_{{ij,t}}=L_{i,t}-L_{j,t}$ for all $i,j\in \mathbb{M}_{0}$%
. The MCS procedure involves an iterative process to identify the best model
set. For iteration $s$, the null and alternative hypotheses are as follows: 
\begin{equation*}
H_{0,\mathbb{M}_{s}}:\E(d_{ij,t})=0\,\text{for\thinspace\ all}\,i,j\in 
\mathbb{M}_{s}\subset \mathbb{M}_{0},
\end{equation*}%
and the alternative 
\begin{equation*}
H_{A,\mathbb{M}_{s}}:\E(d_{ij,t})\neq 0\,\text{for\thinspace\ some}\,i,j\in 
\mathbb{M}_{s}.
\end{equation*}

We perform a model equivalence test using the $T_{\max,M}$ statistic with a
bootstrapped implementation (block length of 20 and $5,000$ replications),
as recommended by \cite{hansen2011}. This statistic measures the maximum
absolute difference between the empirical distribution functions of the two
models, offering insights into their overall dissimilarity.  If $H_{0,%
\mathbb{M}_s}$ is not rejected, the best model confidence set is $\mathbb{M%
}_s$. Otherwise, we apply an elimination rule to remove models from $%
\mathbb{M}_s$ according to the guidelines specified in \citep{hansen2011}
and repeat the test.

Let $P_{H_{0,\mathbb{M}_{s}}}$ denote the p-value associated with the null
hypothesis $H_{0,\mathbb{M}_{s}}$, and let $e_{\mathbb{M}_{s}}$ represent
the model eliminated from the set $\mathbb{M}_{s}$ when $H_{0,\mathbb{M}%
_{s}}$ is rejected. The MCS p-value for model $e_{\mathbb{M}_{s}}$ is
defined as

\begin{equation*}
\hat{p}_{e_{\mathbb{M}_{s}}} = \max_{k \leq s} P_{H_{0,\mathbb{M}_{k}}},
\end{equation*}
where $\mathbb{M}_{1} \supset \mathbb{M}_{2} \supset \ldots \supset 
\mathbb{M}_{s}$.

\subsection{QLIKE}
Our theoretical results on optimal forecasting are developed under the MSFE criterion, as reported in the previous tables. We also consider the QLIKE loss function \citep{patton2011, bollerslev2016}, defined as
\begin{eqnarray*}
\frac{RV_{t+h}}{\widehat{RV}_{t+h}}-\log \frac{RV_{t+h}}{\widehat{RV}_{t+h}}-1,
\end{eqnarray*}%
where $h$ denotes the forecasting horizon. The corresponding MCS p-values  for each model class are also reported. Unlike absolute forecast errors, both MSFE and QLIKE yield inference that is invariant to the choice of measurement units. We consider two sets of assets, \{AAPL, ALD, AMGN, AXP, and BA\} and \{AAPL, AMZN, FB, GOOG, and MSFT\}. As before, we examine five model specifications by progressively expanding the asset set.

From Tables \ref{emp: qlike_dj} and \ref{emp: qlike_m6}, we find that, consistent with the theoretical predictions and the empirical results based on MSFE, the mfBm delivers smaller QLIKE losses than the univariate fBm. The corresponding MCS p-values further support this conclusion. In contrast, under the other modeling frameworks, the univariate specification consistently performs best within each model class, as indicated by both the QLIKE criterion and the MCS p-values. Furthermore, across all model specifications, the mfBm generally delivers the most accurate forecasts in terms of both MSFE and QLIKE.

\begin{table}[H]
\centering
\caption{QLIKEs of alternative models for $h$-day-ahead forecasts of RV, along with p-values from model confidence set (MCS) tests for each model class,  between January 3, 2007 to January 14, 2025. The asset set consists of AAPL, ALD, AMGN, AXP, and BA.}
\scalebox{0.6}{ 
\begin{tabular}{ccccccccccccccccccc}
\hline \hline
&  \multicolumn{8}{c}{QLIKE}  & & \multicolumn{8}{c}{MCS p-value} \\
\hline
              &       $h=1  $    &  $h=2 $     &  $h=3  $    &  $h=4 $     &  $h=5 $     &  $h=10 $    &  $h=15 $    &  $h=20$ && $h=1  $    &  $h=2 $     &  $h=3  $    &  $h=4 $     &  $h=5 $     &  $h=10 $    &  $h=15 $    &  $h=20$   \\
                                  \hline
                         \multicolumn{18}{c}{MfBm}       \\
\hline
fBm   & 0.0280 & 0.0388 & 0.0448 & 0.0482 & 0.0523 & 0.0649 & 0.0752 & 0.0817 &&0.7192 & 0.6710 & 1.0000 & 0.8030 & 0.0950 & 0.0252 & 0.0448 & 0.0248\\
bfBm  & 0.0279 & 0.0388 & 0.0448 & 0.0481 & 0.0521 & 0.0644 & 0.0746 & 0.0809&&1.0000 & 1.0000 & 0.9054 & 0.8182 & 0.3336 & 0.2536 & 0.4018 & 0.4066 \\
mfBm3 & 0.0280 & 0.0388 & 0.0448 & 0.0481 & 0.0521 & 0.0644 & 0.0743 & 0.0807 &&0.2020 & 0.5500 & 0.4768 & 0.8182 & 0.3336 & 0.3256 & 0.6268 & 0.4710\\
mfBm4 & 0.0280 & 0.0388 & 0.0449 & 0.0481 & 0.0520 & 0.0643 & 0.0742 & 0.0806 && 0.2020 & 0.5044 & 0.3292 & 0.8182 & 0.3336 & 0.3256 & 0.6268 & 0.4710 \\
mfBm5 & 0.0280 & 0.0388 & 0.0448 & 0.0481 & 0.0520 & 0.0643 & 0.0740 & 0.0804&&0.7192 & 0.5500 & 0.7052 & 1.0000 & 1.0000 & 1.0000 & 1.0000 & 1.0000\\
\hline
\multicolumn{18}{c}{Vector HAR model}       \\
\hline
HAR   & 0.0284 & 0.0387 & 0.0445 & 0.0480 & 0.0522 & 0.0660 & 0.0748 & 0.0799 && 1.0000 & 1.0000 & 1.0000 & 1.0000 & 1.0000 & 1.0000 & 1.0000 & 1.0000 \\
VHAR2  & 0.0289 & 0.0393 & 0.0455 & 0.0497 & 0.0556 & 0.0825 & 0.1023 & 0.1604&&0.0020 & 0.0870 & 0.0795 & 0.0735 & 0.1350 & 0.7220 & 0.5855 & 0.2320 \\
VHAR3 &0.0294 & 0.0403 & 0.0469 & 0.0521 & 0.0585 & 0.1275 & 0.1312 & 0.1593&&0      & 0.0020 & 0.0160 & 0.0735 & 0.1350 & 0.7220 & 0.5855 & 0.2320 \\
VHAR4 & 0.0301 & 0.0419 & 0.0495 & 0.0554 & 0.0628 & 0.2009 & 0.2896 & 0.4238 &&0      & 0.0020 & 0.0040 & 0.0105 & 0.1120 & 0.6735 & 0.3875 & 0.2320 \\
VHAR5 & 0.0310 & 0.0437 & 0.0525 & 0.0598 & 0.0740 & 0.1931 & 0.2247 & 0.2137&&0      & 0      & 0.0005 & 0.0105 & 0.1120 & 0.7220 & 0.3875 & 0.2320\\
\hline
\multicolumn{18}{c}{Vector HAR model with a single common factor}      \\
\hline
HAR   & 0.0284 & 0.0387 & 0.0445 & 0.0480 & 0.0522 & 0.0660 & 0.0748 & 0.0799 &&1.0000 & 1.0000 & 1.0000 & 1.0000 & 1.0000 & 1.0000 & 1.0000 & 1.0000 \\
VHARF2  & 0.0289 & 0.0393 & 0.0455 & 0.0497 & 0.0556 & 0.0825 & 0.1023 & 0.1604&&0.2360 & 0.3400 & 0.3180 & 0.3515 & 0.3505 & 0.5475 & 0.6345 & 0.4770 \\
VHARF3 &0.0289 & 0.0395 & 0.0456 & 0.0496 & 0.0549 & 0.1390 & 0.0992 & 0.1564 &&0.2360 & 0.0465 & 0.2405 & 0.3515 & 0.3505 & 0.5475 & 0.6345 & 0.4770 \\
VHARF4 & 0.0289 & 0.0394 & 0.0454 & 0.0495 & 0.0551 & 0.0806 & 0.1064 & 0.1374 &&0.2360 & 0.2700 & 0.3180 & 0.3515 & 0.2440 & 0.5475 & 0.6345 & 0.4770 \\
VHARF5 & 0.0288 & 0.0393 & 0.0456 & 0.0495 & 0.0553 & 0.0871 & 0.1066 & 0.1724&&0.2360 & 0.3400 & 0.2995 & 0.3515 & 0.3505 & 0.5475 & 0.6345 & 0.4770\\
\hline
\multicolumn{18}{c}{Vector log-HAR model}       \\
\hline
LHAR   &0.0277 & 0.0381 & 0.0439 & 0.0475 & 0.0516 & 0.0640 & 0.0729 & 0.0779&&1.0000 & 1.0000 & 1.0000 & 1.0000 & 1.0000 & 1.0000 & 1.0000 & 1.0000 \\
VLHAR2   & 0.0279 & 0.0382 & 0.0441 & 0.0477 & 0.0522 & 0.0655 & 0.0761 & 0.0822&&0.1450 & 0.5675 & 0.6575 & 0.6105 & 0.3070 & 0.3255 & 0.2855 & 0.2525 \\
VLHAR3 & 0.0280 & 0.0385 & 0.0445 & 0.0485 & 0.0533 & 0.0676 & 0.0789 & 0.0856&&0.1450 & 0.2615 & 0.3165 & 0.1490 & 0.0565 & 0.0360 & 0.0285 & 0.0445 \\
VLHAR4 & 0.0282 & 0.0391 & 0.0456 & 0.0498 & 0.0547 & 0.0699 & 0.0823 & 0.0896 &&0.0205 & 0.0335 & 0.0315 & 0.0275 & 0.0250 & 0.0320 & 0.0285 & 0.0230 \\
VLHAR5 & 0.0285 & 0.0397 & 0.0464 & 0.0508 & 0.0560 & 0.0719 & 0.0835 & 0.0899&&0.0015 & 0.0015 & 0.0045 & 0.0045 & 0.0055 & 0.0050 & 0.0075 & 0.0085\\
\hline \hline
\end{tabular}}
\label{emp: qlike_dj}
\end{table}

\begin{table}[H]
\centering
\caption{QLIKEs of alternative models for $h$-day-ahead forecasts of RV, along with p-values from model confidence set (MCS) tests for each model class,  between March 29, 2016 to January 14, 2025. The asset set consists of AAPL, AMZN, FB, GOOG, and MSFT.}
\scalebox{0.6}{ 
\begin{tabular}{ccccccccccccccccccc}
\hline \hline
&  \multicolumn{8}{c}{QLIKE}  & & \multicolumn{8}{c}{MCS p-value} \\
\hline
              &       $h=1  $    &  $h=2 $     &  $h=3  $    &  $h=4 $     &  $h=5 $     &  $h=10 $    &  $h=15 $    &  $h=20$&& $h=1  $    &  $h=2 $     &  $h=3  $    &  $h=4 $     &  $h=5 $     &  $h=10 $    &  $h=15 $    &  $h=20$   \\
                         \hline
                         \multicolumn{18}{c}{MfBm}       \\
\hline
fBm   & 0.0234 & 0.0317 & 0.0370 & 0.0412 & 0.0451 & 0.0562 & 0.0647 & 0.0722&&0.8588 & 0.7628 & 0.2552 & 0.0892 & 0.2140 & 0.2928 & 0.2472 & 0.2962 \\
bfBm  & 0.0233 & 0.0316 & 0.0369 & 0.0410 & 0.0450 & 0.0562 & 0.0647 & 0.0723&&0.8588 & 0.9860 & 0.7158 & 0.5666 & 0.5920 & 0.4644 & 0.2618 & 0.2614 \\
mfBm3 & 0.0233 & 0.0316 & 0.0369 & 0.0410 & 0.0450 & 0.0563 & 0.0647 & 0.0724 &&1.0000 & 1.0000 & 0.7444 & 0.6074 & 0.5920 & 0.4980 & 0.3118 & 0.2614\\
mfBm4 & 0.0234 & 0.0317 & 0.0369 & 0.0409 & 0.0450 & 0.0557 & 0.0638 & 0.0708&&0.8588 & 0.7628 & 0.7444 & 0.6074 & 0.5920 & 0.4980 & 0.3118 & 0.2962 \\
mfBm5 & 0.0234 & 0.0316 & 0.0368 & 0.0408 & 0.0446 & 0.0553 & 0.0632 & 0.0699&&0.8588 & 0.9860 & 1.0000 & 1.0000 & 1.0000 & 1.0000 & 1.0000 & 1.0000\\
\hline
\multicolumn{18}{c}{Vector HAR model}    \\
\hline
HAR  & 0.0238 & 0.0328 & 0.0394 & 0.0445 & 0.0490 & 0.0640 & 0.0704 & 0.0768 &&1.0000 & 1.0000 & 1.0000 & 1.0000 & 1.0000 & 1.0000 & 1.0000 & 1.0000 \\
VHAR2  & 0.0239 & 0.0331 & 0.0401 & 0.0456 & 0.0510 & 0.1031 & 0.0771 & 0.0843&& 0.7800 & 0.3230 & 0.0925 & 0.1605 & 0.2065 & 0.1825 & 0.5205 & 0.1260\\
VHAR3 &0.0243 & 0.0344 & 0.0427 & 0.0501 & 0.0606 & 0.1487 & 0.1215 & 0.1767&&0.0390 & 0.0465 & 0.0925 & 0.1605 & 0.2065 & 0.1825 & 0.5205 & 0.1260\\
VHAR4 & 0.0247 & 0.0355 & 0.0445 & 0.0555 & 0.1088 & 0.0921 & 0.1541 & 0.1975&&0.0295 & 0.0465 & 0.0925 & 0.1605 & 0.2065 & 0.1700 & 0.5205 & 0.1260 \\
VHAR5 & 0.0251 & 0.0365 & 0.0456 & 0.0551 & 0.1027 & 0.1113 & 0.2255 & 0.4699&&0.0055 & 0.0070 & 0.0135 & 0.0160 & 0.1165 & 0.1825 & 0.5205 & 0.1215\\
\hline
\multicolumn{18}{c}{Vector HAR model with a single common factor}      \\
\hline
HAR    & 0.0238 & 0.0328 & 0.0394 & 0.0445 & 0.0490 & 0.0640 & 0.0704 & 0.0768&&0.9700 & 1.0000 & 1.0000 & 1.0000 & 1.0000 & 1.0000 & 1.0000 & 1.0000\\
VHARF2   & 0.0239 & 0.0331 & 0.0401 & 0.0456 & 0.0510 & 0.1031 & 0.0771 & 0.0843&&0.9500 & 0.6215 & 0.6250 & 0.2405 & 0.1345 & 0.3080 & 0.3075 & 0.2980   \\
VHARF3 & 0.0238 & 0.0331 & 0.0400 & 0.0454 & 0.0503 & 0.0676 & 0.0747 & 0.0838&& 1.0000 & 0.6215 & 0.6250 & 0.2405 & 0.1345 & 0.3080 & 0.3075 & 0.2980 \\
VHARF4 & 0.0239 & 0.0332 & 0.0400 & 0.0456 & 0.0505 & 0.0671 & 0.0748 & 0.0840&&0.9700 & 0.5955 & 0.6250 & 0.2370 & 0.1345 & 0.3080 & 0.3075 & 0.2980 \\
VHARF5  &0.0239 & 0.0333 & 0.0400 & 0.0455 & 0.0503 & 0.0662 & 0.0735 & 0.0826&&0.9680 & 0.5905 & 0.6250 & 0.2405 & 0.1345 & 0.3080 & 0.3075 & 0.2980\\
\hline
\multicolumn{18}{c}{Vector log-HAR model}     \\
\hline
LHAR   & 0.0236 & 0.0323 & 0.0384 & 0.0431 & 0.0474 & 0.0598 & 0.0668 & 0.0733 &&1.0000 & 1.0000 & 1.0000 & 1.0000 & 1.0000 & 1.0000 & 1.0000 & 1.0000  \\
VLHAR2  &0.0236 & 0.0325 & 0.0390 & 0.0440 & 0.0488 & 0.0624 & 0.0706 & 0.0790  &&0.8435 & 0.5845 & 0.1330 & 0.0900 & 0.1190 & 0.1455 & 0.1780 & 0.1985\\
VLHAR3 &0.0237 & 0.0329 & 0.0395 & 0.0445 & 0.0493 & 0.0634 & 0.0739 & 0.0837  &&0.4645 & 0.3790 & 0.1330 & 0.0900 & 0.1190 & 0.1455 & 0.1780 & 0.1985\\
VLHAR4 &0.0238 & 0.0330 & 0.0399 & 0.0452 & 0.0504 & 0.0664 & 0.0790 & 0.0914&&0.4645 & 0.3790 & 0.1330 & 0.0705 & 0.0690 & 0.0750 & 0.0785 & 0.0945  \\
VLHAR5 & 0.0242 & 0.0341 & 0.0419 & 0.0483 & 0.0551 & 0.0859 & 0.1060 & 0.1420&&0.0450 & 0.0390 & 0.0515 & 0.0605 & 0.0690 & 0.0750 & 0.0785 & 0.0945\\
\hline \hline
\end{tabular}}
\label{emp: qlike_m6}
\end{table}

\section{GMM for Correlation}
\label{generalestimation}
\subsection{Optimal GMM}
With lagged increments $\Delta_{k,{{{\ell}}}} B^{(p)}=B^{(p)}_{k\Delta}-B^{(p)}_{(k-{{\ell}})\Delta},\ell\le k\le n$, consider a class of estimators 
\begin{align}\label{rhohatlag}
\hat\rho_{\text{lag}}&=\hat\rho_{\ell}=\frac{\sum_{k=\ell}^n\Delta_{k,\ell} B^{(1)}\cdot \Delta_{k,\ell} B^{(2)}}{\sqrt{\sum_{k=\ell}^n\big(\Delta_{k,\ell} B^{(1)}\big)^2\cdot \sum_{k=\ell}^n\big(\Delta_{k,\ell} B^{(2)}\big)^2}}\,,
\end{align}
generalizing the empirical correlation from Eq.\ (8a) in the main paper, such that $\hat\rho_{1}=\hat\rho$ coincides with our standard estimator. The lag $\ell$ is a fixed positive integer $\ell\in\N$. The expectations of squared increments $\Delta_{k,\ell} B^{(p)}$, with lag $\ell$, contain an additional factor $\ell^{2H_p}$, and the expectation 
\[\E[\Delta_{k,\ell} B^{(1)}\cdot\Delta_{k,\ell} B^{(2)}]=\ell^{H_1+H_2}\Delta^{H_1+H_2}\rho\sigma_1\sigma_2\,,\]
the additional factor $\ell^{H_1+H_2}$, see \eqref{covlag}. This makes the statistics \eqref{rhohatlag} self-normalized and yields consistent estimators for any $1\le \ell=\KLEINO(n)$. 
\begin{prop}\label{propapp}For any lag $1\le \ell=\KLEINO(n)$, the estimator $\hat\rho_{\text{lag}}$ is for a time-reversible, bivariate fBm with $\max(H_1,H_2)<3/4$, $|\rho|<1$, $H_1+H_2\ne 1$, as $n\rightarrow \infty$, we have, 
\begin{align}\label{cltrholag}\hat\rho_{\text{lag}}\stackrel{p}{\rightarrow}\rho,\; \sqrt{n}\big(\hat\rho_{\text{lag}}-\rho\big)\stackrel{d}{\rightarrow}\mathcal{N}\Big(0,\text{AVAR}_{\hat\rho_{\text{lag}}}\Big)\,,\end{align}
with $\text{AVAR}_{\hat\rho_{\text{lag}}}$ contained in \eqref{acov} and strictly increasing in $\ell$.
\end{prop}
It is not surprising that the asymptotic variance $\hat\rho_{\text{lag}}$ increases with $l$; see \eqref{avar}. Nevertheless, an optimal combination of different estimators $\hat\rho_{\ell},1\le \ell\le L$, can potentially improve upon our MM estimator $\hat\rho_{1}=\hat\rho$. Given the Gaussianity of mfBm, the optimal combination is linear and the optimization problem has an explicit solution, which is given in the next theorem.
\begin{theo}\label{theoapp}Denote $\mathfrak{C}$ the asymptotic variance-covariance matrix of $(\hat\rho_{1},\ldots,\hat\rho_{\text{L}})^{\top}$, which is determined by \eqref{avar}, and $\1$ the $L$-dimensional vector with all entries equal to 1. Let the conditions of Proposition 4.1 be satisfied. Then the estimator 
\begin{align}\label{optest}\hat\rho^{opt}=\frac{\1^{\top}\mathfrak{C}^{-1}(\hat\rho_{1},\ldots,\hat\rho_{L})^{\top}}{\1^{\top}\mathfrak{C}^{-1}\1}\end{align}
minimizes the mean squared error (MSE) asymptotically among all possible combinations of $\hat\rho_{\text{1}},\ldots,\hat\rho_{\text{L}}$, attains a smaller asymptotic variance than $\hat\rho_{1}$ given by 
\begin{align}\lim_{n\to\infty}n\cdot\var(\hat\rho^{opt})=\text{AVAR}_{\hat\rho^{opt}}=\frac{1}{\1^{\top}\mathfrak{C}^{-1}\1}\,,\end{align}
and satisfies the central limit theorem 
\begin{align}\label{cltrholagopt}\hat\rho^{opt}\stackrel{p}{\rightarrow}\rho,\; \sqrt{n}\big(\hat\rho^{opt}-\rho\big)\stackrel{d}{\rightarrow}\mathcal{N}\Big(0,\big(\1^{\top}\mathfrak{C}^{-1}\1\big)^{-1}\Big)\,.\end{align}
\end{theo}
%The solution of the minimization problem of the MSE by an optimal combination of estimators is closely related to the derivation of the global minimum variance portfolio in finance. That the problem has an explicit solution turns out to be beneficial for the implementation and practical performance and the similarity of the formulas to well-known results from portfolio optimization is neat to easily remember the results.
The solution of minimizing the asymptotic MSE by an optimal combination of estimators is closely related to the derivation of the global minimum-variance portfolio in finance. The existence of an explicit solution is advantageous for implementation and practical performance. Moreover, the close analogy to well-known results from portfolio optimization provides an intuitive interpretation and facilitates recall of the resulting formulas. 

Clearly, $\mathfrak{C}$ and $\mathfrak{C}^{-1}$ depend on unknown parameters, so that $\hat\rho^{opt}$ is an infeasible oracle estimator. To obtain a feasible version, we adopt a two-step procedure. In the first step, feasible consistent pilot estimators are computed. In the second step, these estimates are plugged into the expression for the optimal weights, yielding a feasible counterpart of $\hat\rho^{opt}$.

The efficiency of GMM estimator critically depends on the choice of moment conditions. In this paper, we deliberately refrain from incorporating more general covariance relations between shifted increments at different lags, which would compromise the desirable self-normalization property and lead to a considerably more involved optimization problem. The resulting numerical procedure would be computationally demanding and its efficiency gains remain unclear. Our simulation experiments suggest that, in practice, it is difficult to implement a general GMM specification that delivers significant efficiency gains relative to our simpler pair- and entry-wise estimators.
%The simulation study below suggests that, in practice, it is difficult to implement a general GMM specification that delivers efficiency gains relative to our simpler estimators.

%As our simulations demonstrate, it is difficult to implement a general GMM such that efficiency gains compared to our simple estimators can be realized. 
%The estimator $\hat\rho^{opt}$ is of GMM-type, since it is based on moment equalities. We do not use more general covariance equations between shifted increments at different lags, since this would manipulate the sound self-normalization property and yield a complicated optimization problem whose numerical solution would take considerable computational time while it is unclear if any efficiency gains can be realized.
\subsection{Estimation based on second-order increments}
Using the notation $\Delta_{k}^{(2)} B^{(p)}=B^{(p)}_{k\Delta}-2B^{(p)}_{(k-1)\Delta}+B^{(p)}_{(k-2)\Delta},2\le k\le n$, for second-order increments, one can consider an empirical correlation statistic
\begin{align}\label{rhohatso}
\hat\rho_1^{(2)}&=\frac{\sum_{k=2}^n\Delta_{k}^{(2)} B^{(1)}\cdot \Delta_{k}^{(2)} B^{(2)}}{\sqrt{\sum_{k=2}^n\big(\Delta_{k}^{(2)} B^{(1)}\big)^2\cdot \sum_{k=2}^n\big(\Delta_{k}^{(2)} B^{(2)}\big)^2}}\,,
\end{align}
and similarly with larger lags. Without providing a formal proof, it is not difficult to conjecture that the estimator \eqref{rhohatso} satisfies a central limit theorem at the optimal rate $n^{-1/2}$ without requiring the restriction $\max(H_1,H_2)<3/4$ imposed in Proposition \ref{propapp}, provided that the remaining assumptions hold. At the same time, for all $H_1$ and $H_2$ such that $\max(H_1,H_2)<3/4$, the asymptotic variance of the estimator \eqref{rhohatso} will exceed that of our standard estimator $\hat\rho_1$. Since our primary interest lies in rough volatility data, we therefore focus on methods based on first-order increments. In the next subsection, we provide a quantitative comparison of the variances of our standard estimator and the estimator $\hat\rho_1^{(2)}$ in simulations.
%Without proof, it is not difficult to conjecture that the estimator \eqref{rhohatso} satisfies a central limit theorem at optimal rate $n^{-1/2}$, without the condition $\max(H_1,H_2)<3/4$ required in Proposition \ref{propapp}, if the other assumptions are satisfied. At the same time, for all $H_1$ and $H_2$ with $\max(H_1,H_2)<3/4$, the asymptotic variance of estimator \eqref{rhohatso} will be larger than that of our standard estimator $\hat\rho_1$. Since our main interest in in rough volatility data, we propose methods based on first-order increments. We shall quantitatively compare the different variances of our standard estimator and estimator $\hat\rho^{(2)}$ in the next subsection.
\subsection{Comparison of correlation estimators in simulations}
We compare in a Monte Carlo simulation the variances of different estimators for the parameter $\rho=0{.}4$. We simulate $n=1{,}000$ high-frequency observations on $[0,1]$ of a time-reversible, bivariate fBm with Hurst exponents $H_1=0{.}1$ and $H_2=0{.}4$, and with $\sigma_1=1$ and $\sigma_2=1$. 

This scenario has also been considered in the Monte Carlo study of Section \ref{sec:monte_carlo}. It is particularly challenging for empirical correlations at larger lags, since the variances of the normalized empirical quadratic variations of the second component dominate the remaining terms in the asymptotic variance. In such settings, it is well known that, for empirical correlations, approximating the asymptotic variance via the delta method may lead to rather poor finite-sample performance. In line with this, we observe that the variance of $\hat\rho_{\text{lag}}$ is overestimated by the theoretical expression for larger lags, whereas the central limit theorem for $\hat\rho_1$ delivers accurate confidence intervals for $\rho$, as seen in Section \ref{sec:monte_carlo}. But from the perspective of potential efficiency gains when estimating  $\rho$ based on a combination of $\hat\rho_1,\ldots,\hat\rho_L$, this is in fact encouraging.

\begin{table}[t]
\begin{center}
\begin{tabular}{|c|cccc|}
   \hline 
& $\hat\rho_1 $ & $\hat\rho_1^{(2)}$  & $\hat\rho^{opt}$  & $\hat\rho^{opt,boot}$\\
\hline
  Mean  & 0.400  & 0.407 & 0.400& 0.398  \\
  Variance & 0.77  & 1.16 & 0.65& 0.72  \\
 RMSE & 0.88  & 1.08  & 0.81 & 0.85 \\\hline
\end{tabular}
\end{center}
\caption{\label{Tab1}Monte Carlo averages (Mean), empirical variances of Monte Carlo estimates (Variance) normalized with the rate factor $n$, and the resulting normalized root mean squared error (RMSE) across different estimation methods.}
\end{table}

\begin{figure}[t]
\begin{center}
\includegraphics[width=14cm]{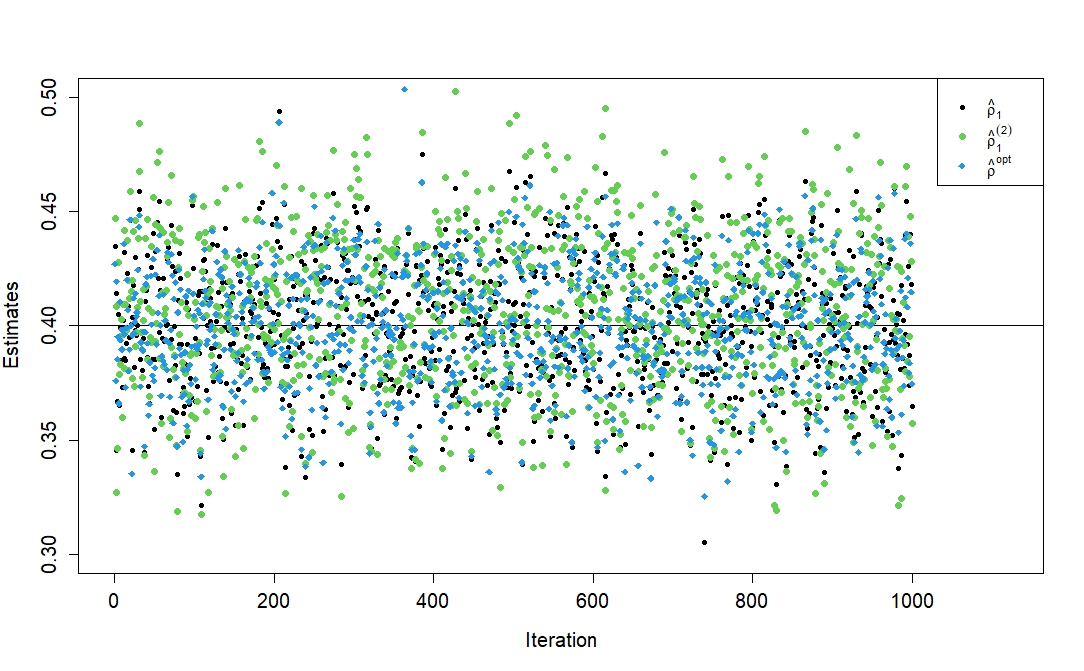}
\end{center}
\caption{\label{opt_plot}Realized estimates of $\hat\rho_{1}$, $\hat\rho_1^{(2)}$ and $\hat\rho_1^{opt}$ in 1{,}000 Monte Carlo iterations.}
\end{figure}

We implement two feasible versions of the optimal estimator	\eqref{optest}. The two-step estimator for which we plug in estimated parameters $H_1$, $H_2$, $\sigma_1$, $\sigma_2$ and $\rho$, using the estimators (6) and (7) from the paper, and $\hat\rho_1$ as pilot estimator, in $\mathfrak{C}$ to estimate the optimal weight matrix. This estimator is denoted by $\hat\rho^{opt}$ in Table \ref{Tab1} and Figure \ref{opt_plot}. To get an accurate finite-sample fit of the variance-covariance structure of $\hat\rho_1,\ldots,\hat\rho_L$, we implement additionally a wild bootstrap procedure. The method requires additional computation time, but the implementation is very simple. After estimating parameters $H_1$, $H_2$, $\sigma_1$ and $\sigma_2$ with the estimators (6) and (7) from the paper, and $\rho$ with $\hat\rho_1$ as pilot estimator, we sample $B=1{,}000$ Monte Carlo bootstrap iterations from our model with the estimated parameters. Based on these inner Monte Carlo iterations within each replication of the outer Monte Carlo study, we estimate $\mathfrak{C}$ by the empirical variance-covariance matrix of the correlation estimates with different lags computed from the $B$ bootstrap samples. Its numerical inverse is then used to construct a feasible version of $\hat\rho^{opt}$, which we denote by $\hat\rho^{opt,boot}$.
Similar wild bootstrap procedures are popular and often perform well in high-frequency statistics, see, for instance, \cite{boot} and \cite{dovonon2019bootstrapping}.

In Table \ref{Tab1} the averages, variances and RMSEs of the estimates across different methods are given based on $1{,}000$ Monte-Carlo iterations. By the sound self-normalization property, both estimators $\hat\rho_1$ and $\hat\rho^{opt}$ are almost perfectly unbiased in practice, while $\hat\rho_1^{(2)}$ has a small positive bias. The estimator $\hat\rho^{opt}$ performs well and has a variance which is more than 15\% smaller than that of $\hat\rho_1$. While the standard two-step estimator performs slightly better than expected, the bootstrap approach does not. Its variance is only about 7\% smaller than that of $\hat\rho_1$ and it does not improve upon the performance of the two-step estimator. This effect appears to be driven by the influence of the pilot estimators in the inner Monte Carlo approximation. Given the considerable additional computation time required by the bootstrap, $\hat\rho^{opt}$ should therefore be preferred.

%While the standard two-step estimator performs slightly better than expected, the bootstrap does not. It attains a variance which is approximately only 7\% smaller than that of $\hat\rho_1$ and shows no improvements compared to the two-step estimator. Since it requires considerable additional computation time, $\hat\rho^{opt}$ should be preferred.

\begin{figure}[t]
\begin{center}
\includegraphics[width=14cm]{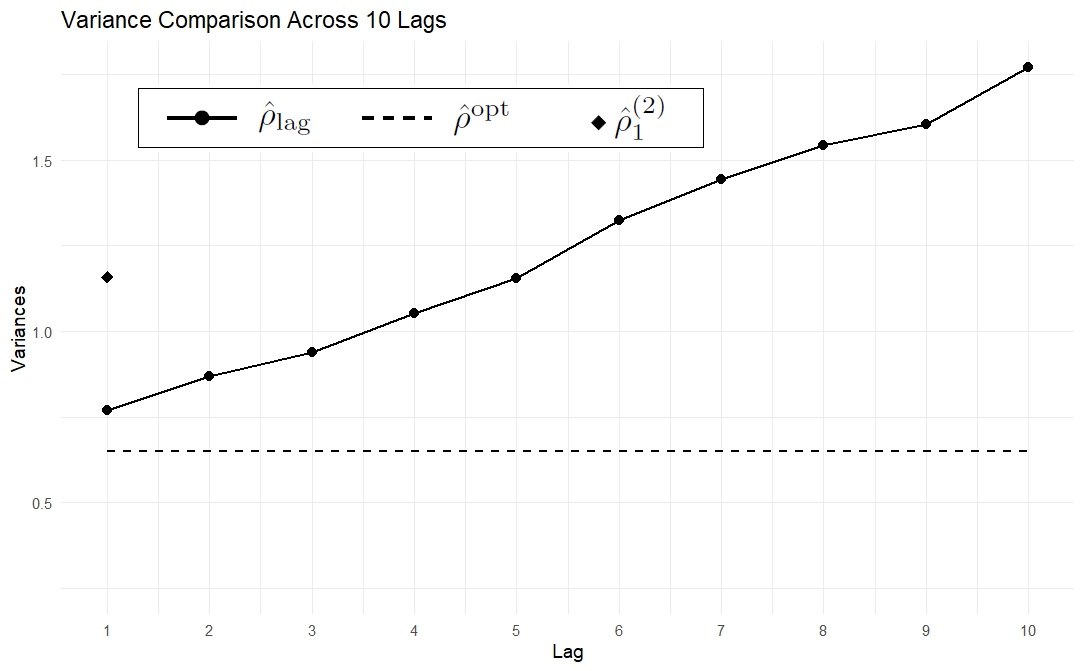}
\end{center}
\caption{\label{comp_plot}Normalized variances of estimators $\hat\rho_{\text{lag}},1\le\text{lag}\le 10$, $\hat\rho_1^{(2)}$, and ${\hat\rho}^{opt}$.}
%\caption{\label{comp_plot}Normalized variances of estimators $\hat\rho_{\text{lag}},1\le\text{lag}\le 10$, $\hat\rho_1^{(2)}$, and the bootstrap version of ${\hat\rho}^{opt}$.}
\end{figure}

In Figure \ref{comp_plot}, we compare the empirical variances based on $1{,}000$ outer Monte-Carlo iterations of the estimators $\hat\rho_{\text{lag}},1\le\text{lag}\le 10$, the estimator $\hat\rho_1^{(2)}$ from \eqref{rhohatso} based on second-order increments, and $\hat\rho^{opt}$, all of them normalized with the rate factor $n$. Figure \ref{opt_plot} shows estimates of the three methods across the 1{,}000 Monte Carlo iterations. As expected, the variances of $\hat\rho_{\text{lag}},1\le\text{lag}\le 10$, increase for larger lags. The variance of $\hat\rho_1^{(2)}$ is more than 50\% larger than that of our standard estimator $\hat\rho_1$. Therefore, one should prefer estimates based on first-order increments in a rough regime, while it is possible to exploit methods using second-order increments to test for roughness first. Finally, $\hat\rho^{opt}$ performs well in practice and has a considerably smaller variance than $\hat\rho_1$. %Occasionally, it is possible that the estimate of $\mathfrak{C}$ is non-invertible. With $B=1{,}000$ Monte Carlo bootstrap iterations that occurred, however, only in ? outer iterations.

We conjecture that maximum likelihood estimation could yield improved and optimal estimates. However, the corresponding theory has not yet been developed for mfBm. As already in the univariate case, we expect that it would be difficult to compute or approximate MLE, primarily due to the need to invert the variance–covariance matrix of the observations, which would be computationally demanding. By contrast, all empirical correlation estimators considered here can be computed without any significant computation time.    
\subsection{Proofs}
Based on the covariance function of mfBm, an elementary computation yields the following generalization of Eq.\ (18) from Section \ref{sec:proof_theorem} for covariances between lagged increments $\Delta_{k,{{{\ell_1}}}} B^{(p)}=B^{(p)}_{k\Delta}-B^{(p)}_{(k-{{\ell_1}})\Delta}$ of the time-reversible mfBm:
\begin{align}\label{covlag}&\cov(\Delta_{k,{{{\ell_1}}}}  B^{(p)}\negthinspace ,\negthinspace \Delta_{j,{{{\ell_2}}}} B^{(q)}) =\negthinspace \Delta^{\negthinspace H_p+H_q}\sigma_{\negthinspace p}\sigma_{\negthinspace q}\rho_{p,q}\,\gamma_{p,q}^{{{\ell_1,\ell_2}}}(k -j),\\
&\notag\mbox{with}~\gamma_{p,q}^{{{\ell_1,\ell_2}}}(r)=\tfrac12(|r\negthinspace+\negthinspace{{\ell_1}}|^{H_p+H_q}+|r\negthinspace-\negthinspace{{\ell_2}}|^{H_p+H_q}-|r|^{H_p+H_q}-|r\negthinspace+\negthinspace{{\ell_1\negthinspace-\negthinspace\ell_2}}|^{H_p+H_q})\,,
%&\mbox{and}~\tilde \gamma_{p,q}^{{\dr{\ell_1,\ell_2}}}(r)=\tfrac12({\scriptstyle{\operatorname{sgn}(r+{\dr{\ell_1}})}}|r+{\dr{\ell_1}}|^{H_p+H_q}+{\scriptstyle{\operatorname{sgn}(r-{\dr{\ell_2}})}}|r-{\dr{\ell_2}}|^{H_p+H_q}-\ldots~,
%{\scriptstyle{\operatorname{sgn}(r)}}|r|^{H_p+H_q})-{\scriptstyle{\operatorname{sgn}(r+\ell_1-\ell_2)}}|r+\ell_1-\ell_2|^{H_p+H_q}).
\end{align}
where $\gamma_{p,q}^{{{\ell,\ell}}}(0)=\ell^{H_p+H_q}$.

For components $p,q,r,s\in\{1,2\}$, and lags $\ell_1,\ell_2\in\{1,\ldots,L\}$, consider the covariances of suitably normalized empirical covariations
\begin{align*}
&\cov\Big(\frac{\Delta^{-H_p-H_q}}{\sqrt{n}}\sum_{k=\ell_1}^n\Delta_{k,\ell_1}B^{(p)}\Delta_{k,\ell_1}B^{(q)},\frac{\Delta^{-H_r-H_s}}{\sqrt{n}}\sum_{k=\ell_2}^n\Delta_{k,\ell_2}B^{(r)}\Delta_{k,\ell_2}B^{(s)}\Big)\\
&=\frac{\Delta^{-H_p-H_q-H_r-H_s}}{n}\sum_{\substack{\ell_1\le k\le n\\ \ell_2\le l\le n}}\cov\big(\Delta_{k,\ell_1}B^{(p)}\Delta_{k,\ell_1}B^{(q)},\Delta_{l,\ell_2}B^{(r)}\Delta_{l,\ell_2}B^{(s)}\big)\\
&\quad=\frac{\Delta^{-H_p-H_q-H_r-H_s}}{n}\sum_{\substack{\ell_1\le k\le n\\ \ell_2\le l\le n}}\Big(\cov\big(\Delta_{k,\ell_1}B^{(p)},\Delta_{l,\ell_2}B^{(r)}\big)\cov\big(\Delta_{k,\ell_1}B^{(q)},\Delta_{l,\ell_2}B^{(s)}\big)\\
&\hspace*{4.2cm}+\cov\big(\Delta_{k,\ell_1}B^{(p)},\Delta_{l,\ell_2}B^{(s)}\big)\cov\big(\Delta_{k,\ell_1}B^{(q)},\Delta_{l,\ell_2}B^{(r)}\big)\Big)\\
&\quad=\frac{\sigma_p\sigma_r\sigma_q\sigma_s}{n}\sum_{\substack{\ell_1\le k\le n\\ \ell_2\le l\le n}}\Big(\rho_{p,r}\rho_{q,s}\gamma_{p,r}^{\ell_1,\ell_2}(k-l)\gamma_{q,s}^{\ell_1,\ell_2}(k-l)+\rho_{p,s}\rho_{q,r}\gamma_{p,s}^{\ell_1,\ell_2}(k-l)\gamma_{q,r}^{\ell_1,\ell_2}(k-l)\Big)\,.
%&\quad=\sigma_p^2\sigma_q^2\Big(\gamma_p(0)\gamma_q(0)+2\sum_{r=1}^n\frac{n-r}{n}\gamma_p(r)\gamma_q(r)+\rho_{p,q}^2\big(\gamma_{p,q}^2(0)+2\sum_{r=1}^n\frac{n-r}{n}\gamma^2_{p,q}(r)\big)\Big)\,,
\end{align*}
The identities use Isserlis' theorem from Eq.\ (20) in Section \ref{sec:proof_theorem} and \eqref{covlag}. Dominated convergence yields that
\begin{align}&\notag\lim_{n\to\infty} \cov\Big(\frac{\Delta^{-H_p-H_q}}{\sqrt{n}}\sum_{k=\ell_1}^n\Delta_{k,\ell_1}B^{(p)}\Delta_{k,\ell_1}B^{(q)},\frac{\Delta^{-H_r-H_s}}{\sqrt{n}}\sum_{k=\ell_2}^n\Delta_{k,\ell_2}B^{(r)}\Delta_{k,\ell_2}B^{(s)}\Big)\\
&\label{avar}=\sigma_p\sigma_r\sigma_q\sigma_s\big(\rho_{p,r}\rho_{q,s}\gamma_{p,r}^{\ell_1,\ell_2}(0)\gamma_{q,s}^{\ell_1,\ell_2}(0)+\rho_{p,s}\rho_{q,r}\gamma_{p,s}^{\ell_1,\ell_2}(0)\gamma_{q,r}^{\ell_1,\ell_2}(0)\big)\\
&\notag\quad +\sigma_p\sigma_r\sigma_q\sigma_s\Big(\sum_{j=1}^{\infty}\big(\rho_{p,r}\rho_{q,s}\gamma_{p,r}^{\ell_1,\ell_2}(j)\gamma_{q,s}^{\ell_1,\ell_2}(j)+\rho_{p,s}\rho_{q,r}\gamma_{p,s}^{\ell_1,\ell_2}(j)\gamma_{q,r}^{\ell_1,\ell_2}(j)\big)\\
&\notag\hspace*{2.25cm}+\sum_{j=1}^{\infty}\big(\rho_{p,r}\rho_{q,s}\gamma_{p,r}^{\ell_1,\ell_2}(-j)\gamma_{q,s}^{\ell_1,\ell_2}(-j)+\rho_{p,s}\rho_{q,r}\gamma_{p,s}^{\ell_1,\ell_2}(-j)\gamma_{q,r}^{\ell_1,\ell_2}(-j)\big)\Big)\,.
\end{align}
Note that while the symmetry $\gamma_{p,q}^{\ell_1,\ell_1}(-r)=\gamma_{p,q}^{\ell_1,\ell_1}(r)$ holds for all $p,q$, in general $\gamma_{p,q}^{\ell_1,\ell_2}(-r)=\gamma_{p,q}^{\ell_2,\ell_1}(r)$, such that we cannot use equality of the two series expressions in the limit.
These entries define an asymptotic covariance matrix of vectors $\big(QV^{(1)}_{\ell_1},QV^{(2)}_{\ell_1},QC_{\ell_1}\big)^{\top}$ and $\big(QV^{(1)}_{\ell_2},QV^{(2)}_{\ell_2},QC_{\ell_2}\big)^{\top}$, writing
\begin{subequations}
\begin{align}QV^{(1)}_{\ell}&=\frac{\Delta^{-2H_1}}{\sqrt{n}}\sum_{k=\ell}^n\big(\Delta_{k,\ell}B^{(1)}\big)^2\,,\\
QV^{(2)}_{\ell}&=\frac{\Delta^{-2H_2}}{\sqrt{n}}\sum_{k=\ell}^n\big(\Delta_{k,\ell}B^{(2)}\big)^2\,,\\
\text{and}~~QC_{\ell}&=\frac{\Delta^{-H_1-H_2}}{\sqrt{n}}\sum_{k=\ell}^n \Delta_{k,\ell}B^{(1)}\Delta_{k,\ell}B^{(2)}\,,
\end{align}
\end{subequations}
for any lags $\ell_1,\ell_2\in\{1,\ldots, L\}$. Analogously as in Section \ref{sec:proof_theorem}, based on \cite{arcones} and Cram\'{e}r-Wold, there is a joint central limit theorem
\[
\sqrt{n}
\left(
\begin{pmatrix}
QV^{(1)}_{\ell_1}-\E[QV^{(1)}_{\ell_1}] \\[.1cm]
QV^{(2)}_{\ell_1}-\E[QV^{(2)}_{\ell_1}] \\[.1cm]
QC_{\ell_1}-\E[QC_{\ell_1}]
\end{pmatrix},
\,
\begin{pmatrix}
QV^{(1)}_{\ell_2}-\E[QV^{(1)}_{\ell_2}] \\[.1cm]
QV^{(2)}_{\ell_2}-\E[QV^{(2)}_{\ell_2}] \\[.1cm]
QC_{\ell_2}-\E[QC_{\ell_2}]
\end{pmatrix}
\right)
\;\xrightarrow{d}\;
\mathcal{N}(0,\AVAR_{\ell_1,\ell_2})\,,
\]
where the $3\times3$ matrix
\[
\AVAR_{\ell_1,\ell_2}
=
\Cov\!\left(
\begin{pmatrix}
QV^{(1)}_{\ell_1} \\[.1cm]
QV^{(2)}_{\ell_1} \\[.1cm]
QC_{\ell_1}
\end{pmatrix},
\begin{pmatrix}
QV^{(1)}_{\ell_2} \\[.1cm]
QV^{(2)}_{\ell_2} \\[.1cm]
QC_{\ell_2}
\end{pmatrix}
\right)
\]
is determined by the entries in \eqref{avar}. Since
\[
\hat\rho_{\ell}
=
g\!\left(
QV^{(1)}_{\ell},
QV^{(2)}_{\ell},
QC_{\ell}
\right),
\qquad
g(y_1,y_2,x)
=
\frac{x}{\sqrt{y_1 y_2}}\,,
\]
based on the joint central limit theorem and the gradient of $g$, which is
\[
\nabla g(y_1,y_2,x)
=
\Big(-\dfrac{x}{2 y_1^{3/2}\sqrt{y_2}},
-\dfrac{x}{2 y_2^{3/2}\sqrt{y_1}},
\dfrac{1}{\sqrt{y_1 y_2}}
\Big)^{\top}\,,
\]
the multivariate delta method with \(\psi_{\ell}=\nabla g(\sigma_1^2,\sigma_2^2,\rho\sigma_1\sigma_2)
\), 
yields that
\begin{align}\label{acov}
\lim_{n\to\infty}n\cdot \cov\!\left(
\hat\rho_{\ell_1},
\hat\rho_{\ell_2}
\right)=
\psi_{\ell_1}^{\top}
\AVAR_{\ell_1,\ell_2}
\psi_{\ell_2}\,.
\end{align}
An asymptotic lower bound for the MSE of linear estimators based on $\hat\rho_1,\ldots,\hat\rho_L$ is always a lower bound more generally for $\sigma(\hat\rho_1,\ldots,\hat\rho_L)$-measurable estimators, where in case of asymptotic normality we expect that both bounds coincide.
Writing $\hat\rho=\big(\hat\rho_1,\ldots,\hat\rho_L\big)^{\top}$, it holds by the multivariate delta method that
\[\sqrt{n}\big(\hat\rho-\rho\big)\;\xrightarrow{d}\;
\mathcal{N}(0,\mathfrak{C})\,,\]
where $\mathcal{C}$ is the $L\times L$ positive definite matrix with entries 
\begin{align}\label{avarc}\mathcal{C}_{\ell_1,\ell_2}=\lim_{n\to\infty}n\cdot \cov\!\left(
\hat\rho_{\ell_1},\hat\rho_{\ell_2}\right)
\end{align}
defined in \eqref{acov}. Consider the linear estimator
\[\hat\rho^{w}=w^{\top}\hat\rho\,,\]
with a weight vector $w$ satisfying the required constraint 
\[\sum_{j=1}^Lw_j=\1^{\top}\cdot w=\1^{\top}\cdot \mathfrak{C}^{-1/2}\mathfrak{C}^{1/2}\cdot w=1\,.\]
Minimizing the MSE within this class is closely related to the derivation of the global minimum-variance portfolio and usually conducted with Lagrange multipliers. We present a simpler proof. The Cauchy-Schwarz inequality yields that
\[|\1^{\top}\cdot\mathfrak{C}^{-1/2}\mathfrak{C}^{1/2}\cdot w|^2\le \|\1^{\top}\mathfrak{C}^{-1/2}\|^2\cdot \|\mathfrak{C}^{1/2}\cdot w \|^2=\|\1^{\top}\mathfrak{C}^{-1/2}\|^2\cdot\var(\hat{\rho}^w)\,,\]
where equality holds if 
\[\mathfrak{C}^{1/2}\cdot w=\lambda \mathfrak{C}^{-1/2}\1\,,\]
with proportionality constant $\lambda$, i.e.\ 
\[w=\lambda \mathfrak{C}^{-1}\1\,.\]
Since $\1^{\top}\cdot w=1$, we obtain
\[\lambda=\frac{1}{\1^{\top}\mathfrak{C}^{-1}\1}~~~\Rightarrow~~~w^{opt}=\frac{\mathfrak{C}^{-1}\1}{\1^{\top}\mathfrak{C}^{-1}\1}\,,\]
what yields the results in Theorem \ref{theoapp}.

\bibliographystyle{Chicago}
\bibliography{Bibliography-MM-MC}
%\bibliography{Bibliography-App}

\end{document}